\documentclass[fleqn,usenatbib]{mnras}

\usepackage{newtxtext,newtxmath}

\usepackage[T1]{fontenc}

\DeclareRobustCommand{\VAN}[3]{#2}
\let\VANthebibliography\thebibliography
\def\thebibliography{\DeclareRobustCommand{\VAN}[3]{##3}\VANthebibliography}

\usepackage{graphicx}	
\usepackage{amsmath}	
\usepackage{subcaption} 
\usepackage{ulem}
\usepackage{xspace}
\usepackage[final]{changes}
\usepackage{float}

\newcommand{\hbt}{\textsc{hbt+}\xspace}
\newcommand{\hbth}{\textsc{hbt-herons}\xspace}
\newcommand{\colibre}{\textsc{colibre}\xspace}

\title[Tidal evolution of COLIBRE satellites]{The tidal evolution of satellite galaxies in cosmological simulations: insights from \colibre}

\author[He et al.]{
Feihong He,$^{1,2,3}$\thanks{E-mail: fhtouma@sjtu.edu.cn (SJTU)}
Jiaxin Han,$^{1,2}$\thanks{E-mail: jiaxin.han@sjtu.edu.cn (SJTU)}
Joop Schaye,$^{4}$
Wenting Wang,$^{1,2}$
Zhaozhou Li,$^{5,6}$
Sylvia Ploeckinger,$^{7}$
\and Evgenii Chaikin,$^{4}$
Robert J. McGibbon,$^{4}$
Filip Hu\v{s}ko,$^{4}$
Matthieu Schaller,$^{4}$
Alejandro Benítez-Llambay,$^{8}$
\and Alexander J. Richings,$^{9,10}$
James W. Trayford,$^{11}$
Carlos S. Frenk,$^{12}$
and Fangzhou Jiang$^{3}$\
\\
$^{1}$Department of Astronomy, School of Physics and Astronomy, Shanghai Jiao Tong University, Shanghai, 200240, People’s Republic of China\\
$^{2}$State Key Laboratory of Dark Matter Physics, School of Physics and Astronomy, Shanghai Jiao Tong University, Shanghai 200240, China\\
$^{3}$Kavli Institute for Astronomy and Astrophysics, Peking University, Beijing, 100871, China\\
$^{4}$Leiden Observatory, Leiden University, PO Box 9513, 2300 RA, Leiden, The Netherlands\\
$^{5}$School of Astronomy and Space Science, Nanjing University, Nanjing 210093, China\\
$^{6}$Key Laboratory of Modern Astronomy and Astrophysics, Nanjing University, Ministry of Education, Nanjing 210093, China\\
$^{7}$Department of Astrophysics, University of Vienna, Türkenschanzstrasse 17, A-1180 Vienna, Austria\\
$^{8}$Dipartimento di Fisica G. Occhialini, Università degli Studi di Milano Bicocca, Piazza della Scienza, 3 I-20126 Milano MI, Italy\\
$^{9}$Centre for Data Science, Artificial Intelligence and Modelling, University of Hull, Cottingham Road, Hull, HU6 7RX, UK\\
$^{10}$E. A. Milne Centre for Astrophysics, University of Hull, Cottingham Road, Hull, HU6 7RX, UK\\
$^{11}$Institute of Cosmology and Gravitation, University of Portsmouth, Dennis Sciama Building, Burnaby Road, Portsmouth PO1 3FX, UK\\
$^{12}$Institute for Computational Cosmology, Department of Physics, University of Durham, South Road, Durham, DH1 3LE, UK\\
}

\date{Accepted XXX. Received YYY; in original form ZZZ}

\pubyear{\the\year{}}

\begin{document}
\label{firstpage}
\pagerange{\pageref{firstpage}--\pageref{lastpage}}
\maketitle

\begin{abstract}
We investigate the co-evolution of the stellar and dark matter mass of satellite galaxies using the \colibre cosmological hydrodynamical simulations with subhaloes resolved by the history-based \hbth subhalo finder. We identify a universal tidal track connecting stellar mass loss to subhalo mass loss characterized by two distinct phases, which can be well described by the two-parameter model. The initial phase consists primarily of dark matter stripping, whereas stellar stripping becomes significant only after the subhalo total bound mass fraction drops below a critical value ($\sim 0.057$). We find a bimodal mass loss rate distribution of subhaloes. In satellites with modest mass loss rates, the stellar mass is largely frozen. By contrast, the galaxy quickly becomes unresolved, along with the dark matter component for the extreme-mass-loss population, naturally explaining the lack of ``orphan'' galaxies in previous hydrodynamical simulations. Our model also predicts the formation condition for dark-matter-deficient galaxies (DMDGs), whose abundance peaks at $m_{*}\sim 10^{9.5}\,\rm{M}_{\odot}$. The abundance of DMDGs can be very sensitive to numerical effects, with \colibre resolving a much larger DMDG population than previous hydrodynamical simulations. We also estimate the influence of artificial disruption on the satellite stellar mass function, which can amount to 20 (50) per cent at $m_* \sim 10^{9} (10^{8}) \, \rm M_\odot$, given a baryonic mass resolution of $\sim 10^{6}\,\rm{M}_{\odot}$. 

\end{abstract}




\defcitealias{Han2016b}{Han16}
\defcitealias{vdB18b}{BO18}
\defcitealias{Green21}{Green21}
\defcitealias{Errani2021}{Errani21}
\defcitealias{Errani2022}{Errani22}
\defcitealias{He2025}{He25}
\defcitealias{Smith2016}{Smith16}

\section{Introduction}
In the standard paradigm of hierarchical structure formation, galaxies form and evolve within extended dark matter haloes (DM haloes)~\citep[e.g.][]{White1978}. Small haloes collapse early and are subsequently accreted onto more massive hosts, giving rise to a nested hierarchy of haloes and subhaloes \citep{Moore1998, Ghigna1998,Jiang2025}. Once a halo becomes a satellite (subhalo) within a deeper potential well, it experiences strong environmental effects. Non-linear processes such as tidal stripping, tidal heating, ram-pressure, and dynamical friction from the host reshape the structure of the subhalo, while repeated pericentric passages further transform the embedded galaxy~\citep[e.g.][]{Taylor2001, Taylor2004, Taylor2005b, Taylor2005}. As a result, satellites differ systematically from isolated field systems in their DM density profiles, sizes, morphologies, and (sub)halo-to-stellar mass relation.

Accurately modelling these tidal effects is essential for understanding satellite evolution, especially in the context of semi-analytical models (SAMs). These frameworks require analytic prescriptions that capture how subhaloes respond to tidal fields and how their structural properties evolve as mass is removed. A classical approach treats tidal stripping as an outside-in process in configuration space: material beyond the tidal radius is removed at each timestep \citep[e.g.][]{Zentner2005, Jiang2016}. However, this method relies on somewhat arbitrary specifications of the tidal radius and of how to redistribute or relax the remaining bound material, and its physical fidelity depends on these modelling choices \citep{Green2019, Green2021}.


While most theoretical and numerical studies have focused primarily on the evolution of dark-matter-only (DMO) subhaloes, real satellite galaxies consist of both DM and baryonic components, including stars and gas. Recent work has suggested that tidal stripping operates in a broadly similar fashion in energy space for both components: loosely bound material is removed first, and the most tightly bound material is stripped only under extreme mass loss conditions \citep[e.g.][]{Drakos2020, Errani2022, Arias2025}. Since stars form deep within the gravitational potential well of their host subhalo, they are generally much more resilient to stripping compared to the extended DM envelope.

This qualitative picture is supported by simulations. For example, \cite{Penarrubia2008} and \cite{Smith2013} showed that a subhalo typically must lose $80-90\%$ of its DM before even $10\%$ of its stellar mass is removed. Building on these insights, \cite{Smith2016} (hereafter \citetalias{Smith2016}) examined satellite evolution in hydrodynamical simulations and quantified the relation between DM mass loss and stellar mass loss. They proposed that the stellar mass remains largely intact during the early phase of evolution when DM stripping is mild; only after the subhalo has been heavily stripped does the stellar component begin to respond, tracking the continued loss of DM. On the other hand, \cite{Penarrubia2008}, \cite{Errani2018} and \cite{Carleton2019} used a suite of idealized numerical experiments to derive analogous relations connecting subhalo mass loss to the evolution of various satellite structural properties, including stellar mass and size. Their results demonstrated that the structural response of a satellite to tidal stripping is largely independent of its orbit and depends primarily on its initial internal structure. This evolutionary path is commonly referred to as a \textit{tidal track}.

It should be noted that the above results are mostly derived from the evolution history of resolved subhaloes and their galaxy components. However, the subhalo population in simulations is known to follow two different evolution channels: a modest mass loss population and an extreme mass loss population. \citet{Han2016b} found that about half of the subhaloes that have ever been accreted appear ``disrupted'' at $z=0$, a result that is robust against simulation resolution. This ``disrupted'' population was shown to correspond to a population of extreme mass loss~\citep[hereafter \citetalias{He2025}]{He2025} distinct from surviving ones, such that they remain unresolved under any realistic resolution in cosmological simulations. Given this bimodal evolution of the subhalo population, a critical open question is what happens to the stellar component for each population. Do they follow the same evolution, and, do we expect the galaxy to survive when its subhalo experiences extreme mass loss?

SAMs often assume that a satellite's stellar mass remains fixed after infall, also called frozen \citep[e.g.,][]{Moster2013, Reddick2013, Watson2013, Fu2022}. However, this assumption can strongly bias predictions for the satellite stellar mass function and for other statistics such as the galaxy correlation function. \cite{Fu2024} incorporated the \citetalias{Smith2016} stripping prescription into a semi-analytic framework and computed the resulting satellite stellar mass function. By comparing these predictions with observational measurements, they found that models including stellar mass loss and post-infall star formation provide a better match to the observed low-mass end of the satellite stellar mass function than frozen models.

Regarding the fate of satellites, a key uncertainty is whether galaxies can survive as ``orphans'' (a numerical placeholder for a missing subhalo) after their host subhaloes fall below the resolution limit \citep[e.g.][]{Springel01,Gao04,Guo2011, Henriques2015, SSI2025}. 
Historically, orphans were introduced in SAMs to compensate for the ``over-merging'' problem of subhaloes \citep{Moore1999, Klypin1999b, vandenBosch2018b}. While recent work by \citetalias{He2025} suggests that numerical disruption is no longer a major concern for the subhalo mass function in state-of-the-art DMO simulations, its impact on the satellite stellar mass function remains an open question. Simply discarding these lost systems could artificially suppress the low-mass stellar mass function, whereas keeping them all as long-lived orphans might overpredict it. Therefore, determining whether these ``lost'' orphans physically survive or wither requires a model that can extend their evolutionary tracks beyond the resolution limit for subhaloes.

A physically distinct but conceptually related phenomenon is the formation of Dark Matter Deficient Galaxies (DMDGs). Observations of ultra-diffuse galaxies such as NGC 1052-DF2 and DF4 have revealed systems with negligible DM content \citep{vanDokkum2018, vanDokkum2019}. Unlike the ``orphan'', a DMDG is a physical object characterized by an extreme stellar-to-halo mass ratio, likely formed through deep tidal stripping \citep[e.g.][]{Jackson2021, Ogiya2022, Yin2026}. To understand their origin and distribution, it is critical to properly model the stellar mass loss alongside the DM loss, especially for subhaloes experiencing extreme tidal stripping.

In this work, we analyse the \colibre cosmological hydrodynamical simulations \citep{colibre, Chaikin2025}, focusing on the L200m6 run, to characterise the tidal evolution of satellite galaxies. Our primary goal is to establish a quantitative connection between subhalo mass loss and stellar mass loss, and to investigate the ultimate evolutionary fate of satellites after infall. A key advantage of \colibre is its DM supersampling scheme, in which the simulation employs four times as many DM particles as baryonic particles. This design significantly reduces spurious energy transfer between particle species, leading to more reliable modelling of satellite structural evolution \citep{Ludlow2019, Ludlow2023}. Furthermore, \colibre features an advanced treatment of the interstellar medium (ISM). By moving away from an artificially pressurized equation of state and instead modelling a cold, molecular phase where star formation is restricted to gravitationally unstable regions, the simulation naturally yields more tightly self-bound stellar distributions. This physically motivated ISM model makes the embedded galaxies inherently more physical. Importantly, \colibre has been shown to successfully reproduce many relevant observational constraints, including the evolution of the galaxy stellar mass function and star formation rates \citep{Chaikin2025b}, as well as the evolution of various mass-size relations for active and passive galaxies \citep{Ludlow2026}. For structure identification and evolutionary tracing, \colibre employs the \hbth (sub)halo finder \citep{Victor2025}, an improved version of \hbt \citep{Han2012,Han2018} optimized for hydrodynamical simulations. These features make \colibre particularly well-suited for studying the detailed tidal evolution of satellites.

This paper is organised as follows. In Section~\ref {sec:data}, we describe the \colibre simulations and the construction of our subhalo catalogue. Section~\ref{sec:satellites} presents the mass evolution of satellites and their tidal tracks. In Section~\ref{sec:fate}, based on our tidal-track framework, we examine satellite evolutionary fates, including the role of artificial disruption and the emergence of dark-matter-deficient satellite galaxies. In Section~\ref{sec:discussion}, we discuss the dependence of tidal-track parameters on galaxy properties and explore the inner density evolution during stripping. We summarise our main findings in Section~\ref{sec:summary}.


\section{Simulation data}\label{sec:data} 

We use outputs from the \colibre suite of cosmological hydrodynamical simulations \citep{colibre, Chaikin2025}, which were run with the highly parallel gravity+hydrodynamics code {\textsc{swift}\xspace} \citep{swift}. \colibre adopts an updated galaxy-formation model that includes non-equilibrium hydrogen/helium chemistry \citep{Ploeckinger2025}, explicit dust physics \citep{Trayford2026}, star formation \citep{Nobels2024}, turbulent diffusion, a detailed subgrid model for chemical enrichment \citep{Correa2026}, pre-supernova stellar feedback \citep{ABL2026}, and prescriptions for supernova feedback \citep{Chaikin2023, colibre} and AGN feedback \citep{BS2009, colibre, Husko2026}. These subgrid prescriptions are calibrated to reproduce the present-day galaxy stellar mass function, galaxy mass--size relations, and black hole masses in massive galaxies \citep{Chaikin2025b}. To mitigate spurious numerical energy transfer from DM to baryons, \colibre ``supersamples'' the DM by a factor of four so that the dark and baryonic particle masses are comparable, thereby greatly enhancing the effective resolution for the baryonic component \citep{Ludlow2019, Ludlow2021, Ludlow2023}. The \colibre simulation suite comprises multiple box sizes and three mass resolutions (labelled m5–m7), with the L200m6 run (the simulation used in this work) corresponding to a $200$ cMpc cubed box and particle masses of $m_{\rm gas, p}\simeq1.84\times10^{6} \rm{M}_{\odot}$ and $m_{\rm CDM, p}\simeq2.42\times10^{6}\rm{M}_{\odot}$ (comoving softening $\epsilon_{\rm com}\simeq1.8$ ckpc; maximum proper softening $\simeq0.7$ pkpc). The initial conditions of the L200m6 simulation run contains $3008^{3}$ baryonic particles and $4\times3008^{3}$ CDM particles. We select the L200m6 run as the primary simulation for our analysis because it offers an optimal balance between cosmological volume and numerical resolution. The large box provides a statistically robust sample of satellite galaxies across a wide range of host halo masses, while the m6 resolution is the highest currently available for this volume within the \colibre suite.

Structure finding proceeds in a usual two-step manner: a Friends-of-Friends (FoF) group catalogue is generated from the particle distribution \citep{Press&Davis1982}, and substructure is then identified and tracked over time. For subhalo identification and merger-tree construction, we use \hbth \citep{Victor2025}, a history-based subhalo finder that is a successor to the \hbt algorithm \citep{Han2012, Han2018} and was developed specifically to improve robustness in hydrodynamical environments. \hbth builds subhaloes by tracing bound particles across snapshots rather than relying purely on instantaneous configuration or phase-space criteria. This tracking approach avoids misassignment of particles near host centres, produces self-consistent merger trees, and substantially improves the recovery of deeply embedded and heavily stripped subhaloes. These properties make \hbth particularly well suited for studies of satellite evolution and mass loss. To capture the detailed dynamical history, the \colibre simulation outputs 128 snapshots between $z=30$ and $z=0$. The time interval between snapshots is not constant, ranging from $\sim 0.02 \, \rm{Gyr}$ at high redshift ($z \gtrsim 10$) to $\sim 0.2 \, \rm{Gyr}$ at low redshift ($z \lesssim 1$). This output frequency provides the necessary temporal resolution to robustly trace the tidal evolution of satellites.

Throughout this paper, we adopt the convention of using uppercase letters (e.g., $M, R$) to denote the properties of the host halo, and lowercase letters (e.g., $m, l$) for those of the satellite galaxy or subhalo. When referring to the virial mass or radius of a host halo ($M_{\rm vir}, R_{\rm vir}$), we use the spherical overdensity definition corresponding to 200 times the critical density of the Universe ($200\rho_{\rm crit}$). \added{We use the term \textit{subhalo} to refer strictly to any self-bound DM substructure that inhabits a larger host halo. A subhalo that hosts a luminous stellar component is designated as a \textit{satellite galaxy}. Unless specified otherwise, we use the term ``satellite'' throughout this paper to denote the entire multi-component, self-bound system of interest, encompassing its DM, stellar, gas and black hole(BH) components. The total bound mass of a satellite is defined as $m_{\rm tot} = m_{\rm DM} + m_{*} + m_{\rm gas}+m_{\rm BH}$. The individual mass components ($m_{\rm DM}$, $m_{\rm *}$ $m_{\rm gas}$, and $m_{\rm BH}$) as well as the total mass $m_{\rm tot}$ are computed as the sum of all bound particles associated with the structure as returned by the \hbth algorithm.} While the masses are derived directly from the \hbth output, we utilize the \textsc{SOAP} (Spherical Overdensity and Aperture Processor) catalogue \citep{McGibbon2025} to obtain size-related structural properties.

\added{To quantitatively trace the efficiency of tidal stripping, we introduce the remaining bound mass fractions for each component, defined relative to their respective historical peak values. Specifically, the total remaining mass fraction of a satellite is expressed as:$f_{\rm tot}(t) \equiv m_{\rm tot}/m_{\rm tot, peak}$, where $m_{\rm tot, peak}$ is the maximum total bound mass achieved by the satellite's progenitor throughout its evolution history, and $m_{\rm tot}$ is its evolved mass at a given time of interest. By definition, $f_{\rm tot} = 1$ indicates that the satellite has not yet experienced any mass loss due to tidal stripping after reaching its peak mass, while $f_{\rm tot} \rightarrow 0$ corresponds to the heavily stripped regime. Analogously, we define the individual remaining mass fractions for the dark matter, stellar, and gas components as $f_{\rm DM} \equiv m_{\rm DM}(t)/m_{\rm DM, peak}$, $f_{*} \equiv m_{*}(t)/m_{*, \rm peak}$, and $f_{\rm gas} \equiv m_{\rm gas}(t)/m_{\rm gas, peak}$, respectively.}

Throughout this paper, we adopt an operational definition for the subhalo population based on their status at $z=0$. We classify subhaloes that retain at least 20 bound particles as \textit{resolved}. \added{Here, the particle count includes both dark matter and stellar particles combined; this inclusive tracking is consistent with the \hbth algorithm, which identifies and traces bound structures using both particle types as tracers.} Conversely, systems that fall below this resolution threshold prior to $z=0$ are collectively classified as \textit{lost}. 
\added{We note that the resolution criteria of \citet{vandenBosch2018} recommend a minimum of $\sim 250$ particles for reliable tidal evolution tracking. We have verified that satellites in the poorly-resolved regime ($20 \leq N_{\rm bound} < 250$) occupy only the transition valley between the resolved and lost populations in the mass-loss rate distribution, and that varying the tracking threshold between 20 and 250 has a negligible impact on the population-wide median trends presented in this work.} 
It is important to clarify that in our terminology, the ``lost'' population is an umbrella category that encompasses two distinct scenarios: (1) \textit{withering}, where a subhalo physically loses mass down to the resolution limit while maintaining an unresolved structure, and (2) \textit{disruption}, where the system fragments or unbinds completely, due to either physical disintegration, merger with another subhalo, or numerical artifacts. While ``artificial disruption (numerical) is a subset of the latter, we follow the findings of \citetalias{He2025} that the lost population in our simulation is dominated by physical withering rather than numerical artifacts. Therefore, unless otherwise specified, we treat ``lost'' as a tracer for systems undergoing extreme tidal mass loss. We explicitly investigate the properties and fates of this population in Section~\ref{sec:fate}. \added{Note that as a consequence of our tracking definition, our resolved catalogue explicitly preserves a population of extremely tidally-stripped galaxies that have lost all or most of their dark matter envelopes ($N_{\rm DM} = 0$ to $\lesssim 20$) but retain significant stellar components. These systems represent the most extreme manifestations of tidal mass loss and are central to the DMDG population studied in this work.}

Mergers among subhaloes provide another major physical channel governing the fate of satellites. In the \hbt family, subhalo mergers are explicitly resolved according to their phasespace coalescence and referred to as sinking events. As shown in \citetalias{He2025}, sunken subhaloes only contribute a minor fraction to the total subhalo population. We exclude sunken subhaloes in this work. A dedicated and complementary analysis of the hierarchical mergers among subhaloes identified by \hbt will be presented in \cite{Jiang2026}.


Throughout cosmic history, \hbth identifies a total of 31,655,573 satellite galaxies that once formed stars. Among this vast population, 4,513,484 satellites remain resolved and survive as self-bound systems at $z=0$.

\section{satellites stellar mass evolution}\label{sec:satellites}

\subsection{Case study of satellite mass evolution}

\begin{figure*}
    \centering
    \begin{subfigure}[b]{\textwidth}
        \includegraphics[width=\textwidth]{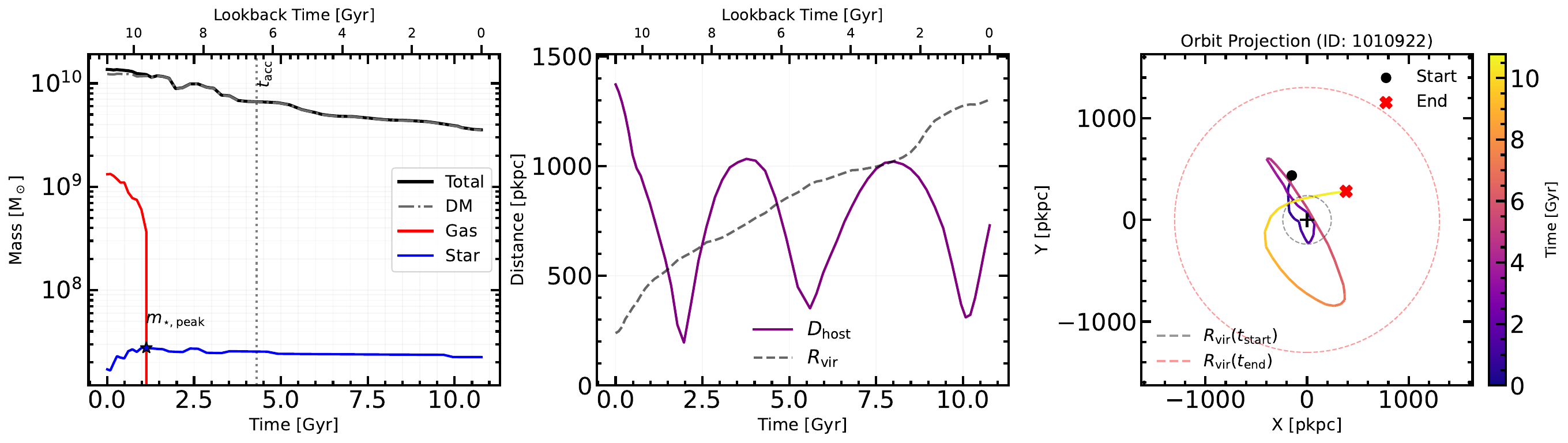}
    \end{subfigure}
    \hfill
    \begin{subfigure}[b]{\textwidth}
        \includegraphics[width=\textwidth]{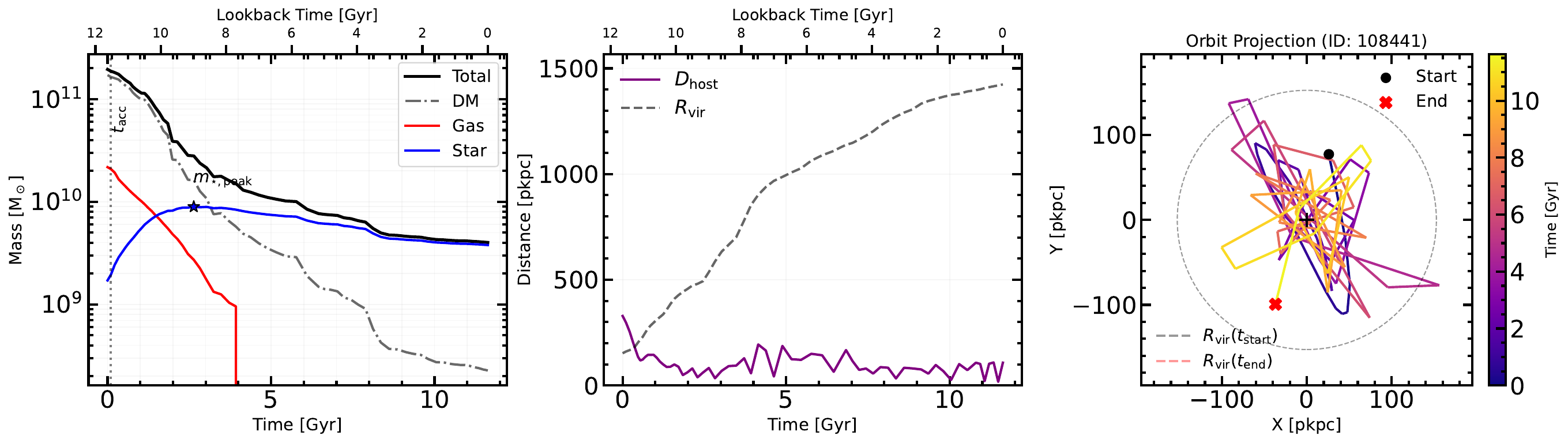}
    \end{subfigure}
    \hfill
    \begin{subfigure}[b]{\textwidth}
        \includegraphics[width=\textwidth]{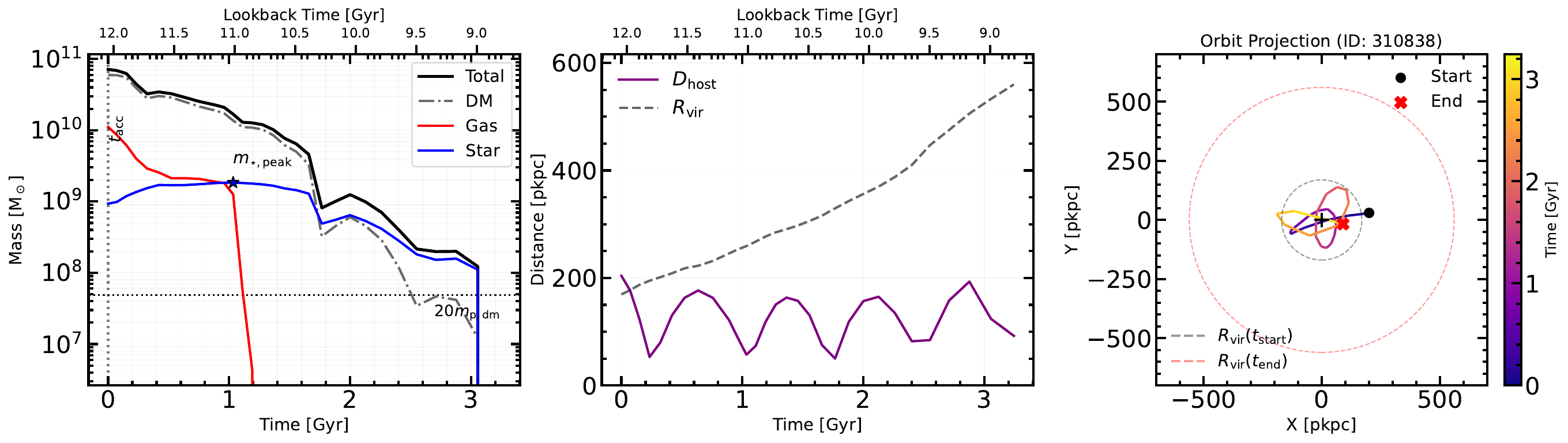}
    \end{subfigure}
    \hfill
    \begin{subfigure}[b]{\textwidth}
        \includegraphics[width=\textwidth]{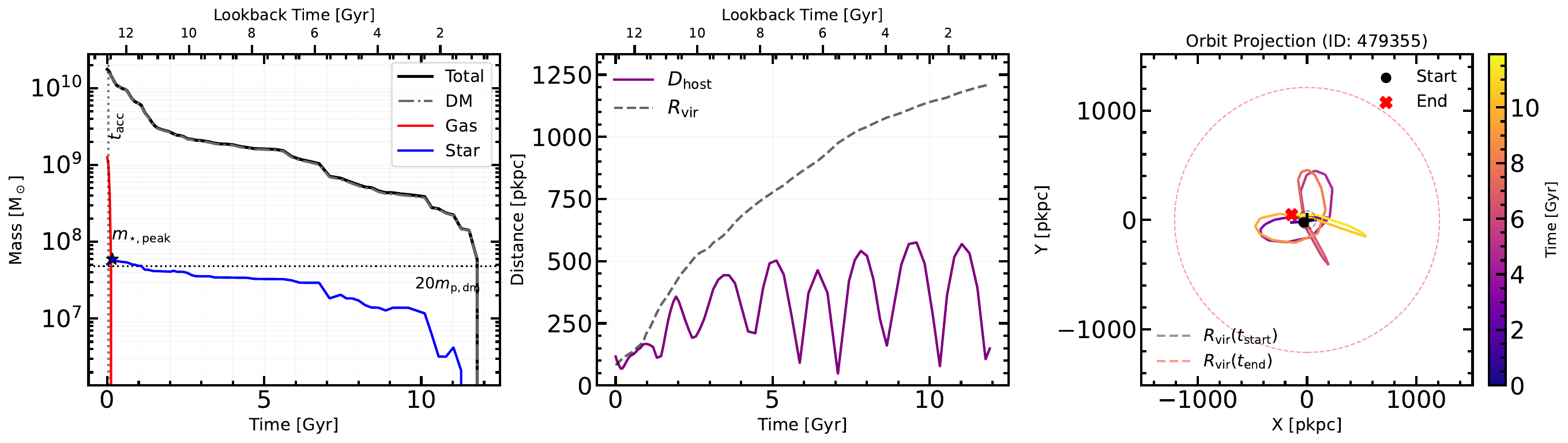}
    \end{subfigure}
    
    \caption{The mass and orbit evolution of example satellite subhaloes after reaching their peak mass $m_{\rm tot,peak}$. The left column shows the mass evolution. Different colours represent the mass evolution of different components, while the black line indicates the total bound mass of the subhalo. The grey vertical dotted line marks the accretion time $t_{\rm{acc}}$, defined as the time when the subhalo falls into the host halo. The upper two rows show resolved satellites that survive at $z=0$. The lower two rows show lost satellites; the DM mass resolution is indicated by the grey horizontal dashed line. The middle column shows the evolution of the distance to the host halo. The grey dashed line in this column shows the growth of the $R_{\rm vir}$ of the host halo for reference. The right column shows satellite orbit projection on the X-Y panel. The trajectory starts from the black dot and ends at the red cross, with the evolution time colour-coded. The centre of the host halo is presented by the black cross. The $R_{\rm vir}$ of the host halo at $t_{\rm start}$ ($t_{\rm{end}}$) is shown by a grey(red) dashed circle for reference.} 
    \label{fig:mass_evolution_example}
\end{figure*}

Before quantifying the statistical properties of the satellite population in \colibre, we illustrate the diversity of evolutionary pathways using four typical satellite systems from the L200m6 simulation. Fig.~\ref{fig:mass_evolution_example} presents the mass evolution, orbital distance, and projected trajectories for these cases. All selected examples are satellites of cluster-sized haloes with peak total masses exceeding $10^{10}\,\rm{M}_\odot$, yet they display distinct fates driven by their specific accretion histories and internal structures. 

While a system remains an isolated DM halo, it continuously accretes mass, including DM, gas, and stars. We identify the epoch at which the halo attains its maximum mass as $t_{\rm{peak}}$ (or $z_{\rm{peak}}$) and denote the mass at this epoch as $m_{\rm{peak}}$. Subsequently, the system enters the virial radius of a larger host halo; we define the accretion time, $t_{\rm{acc}}$, as the last snapshot where the subhalo is identified as an isolated system by \hbth. We show the orbit distance evolution in the middle panel and compare it with the $R_{\rm vir}$ evolution of the central galaxy, indicating the host growth.

After accretion, tidal effects from the host halo strip the outer layers of DM. Simultaneously, hydrodynamical processes such as ram-pressure stripping remove the gas component more rapidly than the DM \citep[e.g.,][]{Jiang2019}. As shown in the mass evolution panels of Fig.~\ref{fig:mass_evolution_example}, gas is typically completely removed within a few Gyr of the first pericentric passage. This gas loss quenches star formation; consequently, the stellar mass typically peaks ($m_{\rm{*,peak}}$) shortly after infall. While the stellar component is generally more concentrated and resilient to stripping than the extended DM halo, it eventually declines as tidal forces penetrate deeper into the potential well.

The top two rows of Fig.~\ref{fig:mass_evolution_example} display satellites that survive as resolved structures to $z=0$.
\begin{itemize}
    \item \textbf{TrackId: 1010922}:This system illustrates a ``backsplash'' orbit. The satellite initially enters the host halo, experiences a pericentric passage that strips its gas and quenches star formation (aligning $m_{\rm{*,peak}}$ with the first infall), but then exits the virial radius. Due to the growth of the host halo, it is re-accreted at a later time, resulting in a significant delay between $t_{\rm{peak}}$ and the final $t_{\rm{acc}}$. Despite these interactions, the satellite experiences only modest tidal stripping, maintaining a nearly constant stellar mass for over 10 Gyr.
    \item \textbf{TrackId: 108441}: In contrast, the second example represents an extreme survivor. Accreted at high redshift when the host halo was significantly smaller, this satellite orbits deep within the potential well. Its apocentre is roughly 100 kpc, while its closest host-centric distance dives to $\sim10$ kpc (the simulation outputs may not capture the pericentre), appearing almost centred in the orbital projection relative to the $z=0$ virial radius. Remarkably, while the DM component is stripped by a factor of $1000$ (reaching $f_{\rm DM} < 10^{-3}$), the satellite retains a substantial, long-lived stellar component ($m_* \approx 4 \times 10^9\,\rm{M}_\odot$ at $z=0$) with a half-mass radius of $\sim0.8$ kpc. This system serves as a DMDG formed via tidal stripping.
\end{itemize}

The bottom two rows show satellites that fall below the resolution limit (``lost'') prior to $z=0$.
\begin{itemize}
    \item \textbf{TrackId: 310838}: This satellite reveals that the DMDG phase can be a transient stage preceding total dissolution. Initially, the DM is stripped faster than the stars, and by $\sim 2$ Gyr after infall, the system becomes stellar mass dominated. However, unlike the survivor in the second row, this galaxy undergoes rapid subsequent stellar stripping and is eventually lost from the simulation.
    \item \textbf{TrackId: 479355}: Finally, we show a low-mass system with a very low initial stellar fraction ($m_{*}/m_{\rm{tot}} < 10^{-2}$, represented by only $\sim30$ star particles). Lacking a dense stellar core to provide resistance, the stellar mass loss tracks the subhalo mass loss almost synchronously after the subhalo is stripped down to 1/10 of its initial mass. This indicates that for systems where baryons do not dominate, the galaxy is as susceptible to tidal effects as its DM halo.
\end{itemize}

\subsection{Stellar tidal tracks}\label{sec:tidaltrack}

We aim to describe the tidal evolution of the stellar component of satellites: as the subhalo gradually loses DM and evolves under the influence of its host’s potential, what is the ultimate fate of the embedded galaxy? Will it eventually dissolve, or can it survive in a compact, tightly bound form? Addressing this question provides a crucial pathway to linking the structural evolution of subhaloes and their associated galaxies.

\citet{Penarrubia2008} designed a series of idealized $N$-body experiments in which subhaloes were placed on specified orbits within a fixed host potential. In these simulations, the bound particles near the potential centre of each subhalo were used as tracers of the stellar component, allowing them to study the structural evolution of satellite galaxies under tidal forces. Based on these controlled experiments, they found that the structural evolution of satellites depends primarily on their initial internal structure rather than on the details of their orbits. They defined the evolution of structural parameters as a function of subhalo mass loss as the tidal track. In this context, our main focus is the relationship between the fractional remaining of  \deleted{subhalo}\added{total bound} mass $f_{\rm tot}$ and the corresponding fractional remaining of stellar mass $f_{*}$.

\citetalias{Smith2016} proposed another method by analysing the tidal evolution of satellite galaxies in a cosmological hydrodynamical simulation of galaxy clusters. They proposed a simple empirical relation describing how the stellar mass fraction of surviving satellites at $z=0$ correlates with their DM mass loss:
\begin{equation}
f_{*} = 1 - \exp(-a_{\rm strip} f_{\rm DM}),
\label{eq:smith16}
\end{equation}
This model provides a useful quantitative link between the two components in a realistic cosmological context.

The question of the ultimate fate of satellite galaxies requires considering not only those that survive to the present day, but also those that have been fully disrupted. In this work, we investigate the relationship between subhalo mass loss and stellar mass loss for satellites in the \colibre L200m6 simulation. Due to the advantages of the \hbth algorithm, we have access to trace the full mass evolution history of each subhalo across cosmic time, including those that got lost before $z=0$.

\begin{figure}
    \centering
    \includegraphics[width=\linewidth]{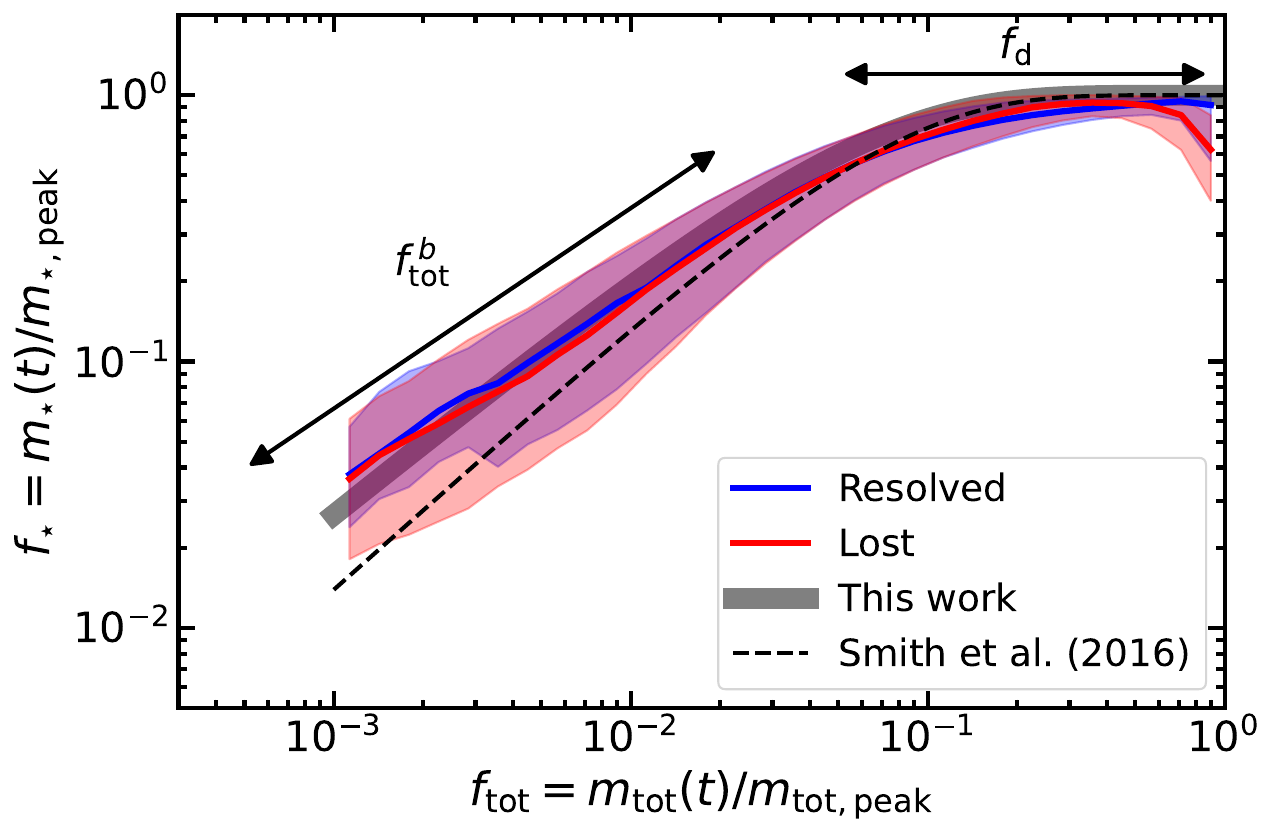}
    \caption{The stellar evolutionary (from right to left) tidal tracks of satellite galaxies in clusters with $M_{\rm host}>10^{14}\rm{M}_{\odot}$ from \colibre L200m6. The solid line indicates the median distribution of tidal tracks, with the shaded region representing the 16th–84th percentile distribution. Blue and red colours correspond to resolved and lost subhalo populations, respectively. The black dashed line shows the tidal track given by \citetalias{Smith2016}. 
    }
    \label{fig:tidaltrack}
\end{figure}

Fig.~\ref{fig:tidaltrack} presents the median tidal tracks, defined as the relation between the fraction remaining of \deleted{subhalo}\added{total bound} mass and that of stellar mass throughout the satellites’ evolutionary history. To compute these population-wide statistics and avoid biasing the results towards subhaloes that are sampled by more snapshots, we employ a two-step binning procedure. First, the evolutionary track of each individual subhalo is binned into logarithmic $f_{\rm tot}$ bins, and the median $f_{*}$ is calculated within each bin for that specific subhalo. This ensures that a single subhalo contributes at most one representative data point to any given $f_{\rm tot}$ bin, regardless of the number of snapshots it spends in that mass loss phase. Second, we aggregate these subhalo-specific medians across the entire population. For each global $f_{\rm tot}$ bin, we then calculate the 16th, 50th (median), and 84th percentiles from all subhaloes that pass through that bin. The subhalo population is divided into the resolved and lost groups. Overall, the $f_{*}$–$f_{\rm tot}$ distributions of both populations largely overlap and follow a consistent trend that can be described by a unified empirical relation. \added{Notably, the scatter in $f_*$ noticeably increases toward lower $f_{\rm tot}$ values. This widening distribution physically reflects the intrinsic dispersion in the late-phase stripping efficiency among individual satellites, which is regulated by our model's slope parameter $b$.} Building upon the \citetalias{Smith2016} model, we adopt a modified and more flexible functional form to describe the stellar mass evolution of satellites across their entire history:
\begin{equation}
    f_{*} \equiv  m_{*} /m_{*,\rm{peak}}= 1 - \exp[-( f_{\rm tot} / f_{\rm d})^b ],
    \label{eq:He26}
\end{equation}
This parametrisation differs from the \citetalias{Smith2016} prescription (Eq.~\eqref{eq:smith16}) in three important respects. First, we use the subhalo bound mass fraction $f_{\rm tot}$ instead of $f_{\rm DM}$ as the independent variable, which remains well-defined even when the dark component is almost entirely removed.

Second, the parameter $f_{\rm d}$ replaces $a_{\rm strip}$ used in Eq.~\eqref{eq:smith16} via $f_{\rm d} \equiv 1/a_{\rm strip}$. Physically, $f_{\rm d}$ sets the characteristic transition point between two distinct evolutionary phases: an early phase ($f_{\rm tot} \gtrsim f_{\rm d}$) where satellites predominantly lose DM while their stellar mass remains nearly constant ($f_{*}\sim 1$), and a second phase ($f_{\rm tot} \lesssim f_{\rm d}$) where the stellar component begins to experience significant mass loss.

Third, we introduce the exponent $b$ to regulate the relative efficiency of stellar stripping compared to DM stripping in this second phase. In the heavy-stripping regime ($f_{\rm tot} \ll f_{\rm d}$), Eq.~\eqref{eq:He26} asymptotically follows $f_{*} \propto f_{\rm tot}^{b}$, permitting flexible scalings. If $b = 1$, the fractional loss of stellar mass tracks that of the DM exactly, implying similar resilience between the two components. For $b > 1$, the stellar component is stripped more rapidly than the DM, reflecting a less bound or more extended stellar distribution. In extreme cases, a large $b$ may lead to the complete removal of the stellar mass before the DM subhalo is fully disrupted. Conversely, when $b < 1$, the stellar component is more bound and therefore more resistant to external tidal forces, consistent with compact, dense stellar systems that can survive even severe DM loss. This flexible form is readily comparable to the energy-space tidal-track models of \cite{Errani2022}; we provide a detailed comparison in Appendix~\ref{EN22}.

We note that the dashed line in Fig.~\ref{fig:tidaltrack} shows the empirical relation obtained by \citetalias{Smith2016}, who found an average value of $a_{\rm strip} = 14.2$ (equivalent to $f_{\rm d} \approx 0.07$). This value closely matches the transition point in our \colibre results, indicating that both studies identify a similar threshold for the onset of stellar stripping. In the \citetalias{Smith2016} model, the subsequent evolution strictly follows $b = 1$. While this linear scaling broadly aligns with the findings of \cite{Wang2022}, who showed that the stellar and DM components of severely stripped satellites share similar inner density profiles, we find that a fixed $b=1$ slightly overestimates the stellar mass loss in \colibre (lying roughly within the boundary of the $1\sigma$ scatter in Fig.~\ref{fig:tidaltrack}). Instead, our median trend in the heavy-stripping regime favors a slightly shallower relation, approximately $f_{*} \propto f_{\rm tot}^{0.9}$. This $b \approx 0.9$ scaling supports the general picture of similar inner profiles but indicates that the stellar component in \colibre satellites is slightly more centrally concentrated and tightly bound, making it slightly more resistant to tides than a pure $b=1$ model would predict.

\begin{figure}
    \centering
    \includegraphics[width=\linewidth]{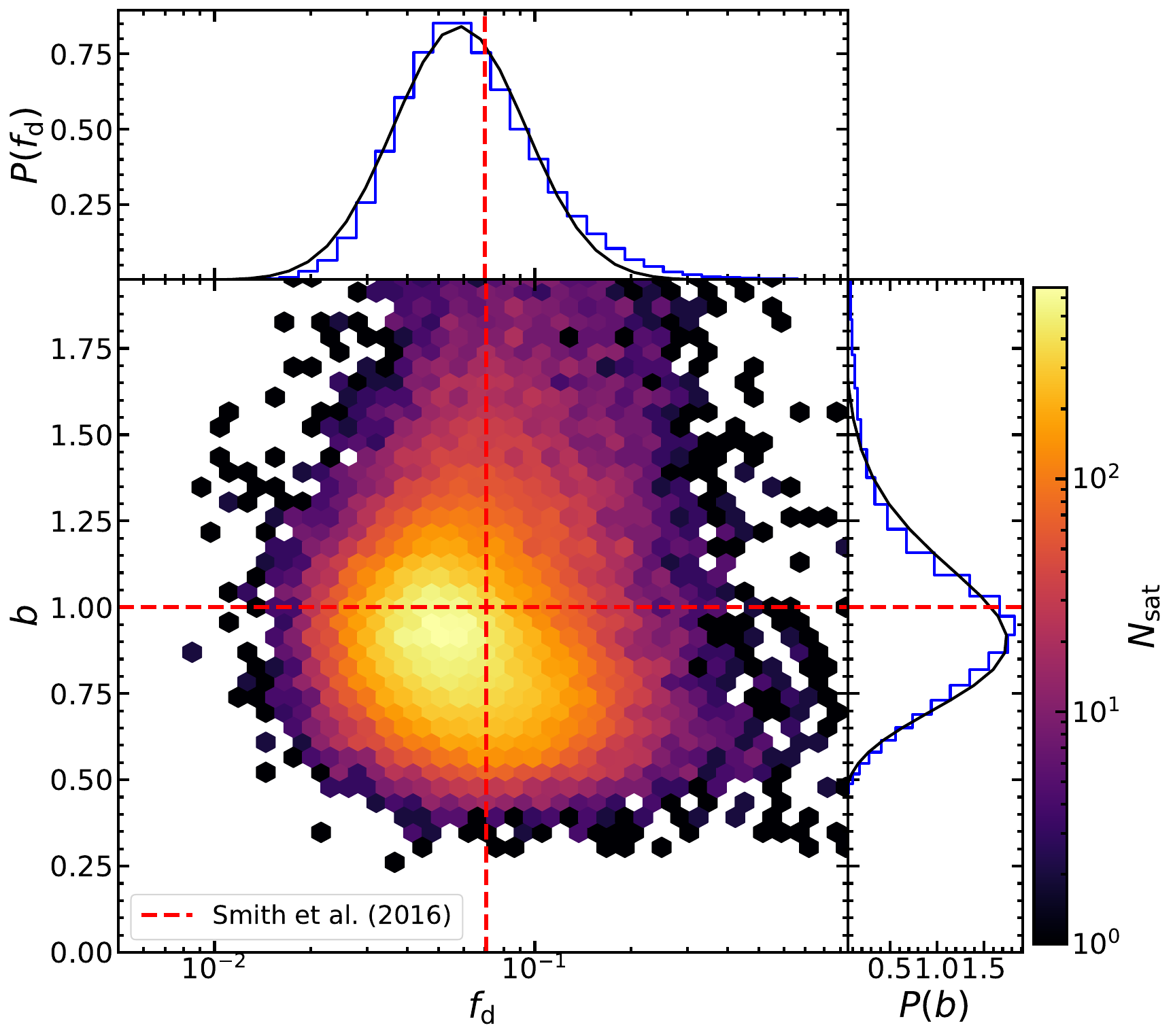}
    \caption{Joint and marginal distributions of the tidal track parameters $f_{\rm d}$ and $b$ for satellites in the L200m6 simulation. The sample is restricted to systems with a peak stellar mass $m_{\rm *,peak} > 10^{8}\,\rm{M}_{\odot}$ and a minimum remaining stellar mass fraction $f_{*, \rm min} < 0.5$. In the top and right sub-panels, the blue stepped histograms represent the probability density functions of $f_{\rm d}$ and $b$, respectively. The solid black curves denote the corresponding best-fitting log-normal distributions. Physically, a lower $f_{\rm d}$ value indicates a more delayed onset of stellar stripping relative to the DM. A lower $b$ value reflects a lower stellar stripping efficiency, characteristic of a more tightly bound and resilient stellar structure. The red dashed vertical and horizontal lines show the parameters found by \citetalias{Smith2016} for reference.} 
    \label{fig:ab_distribution}
\end{figure}


We fit Eq.~\eqref{eq:He26} to the tidal tracks of individual satellites that satisfy the following criteria:
\begin{enumerate}
    \item the satellite has a peak stellar mass of $m_{\ast,\rm{peak}} > 10^{8}\rm{M}_{\odot}$, corresponding to $\gtrsim 50$ stellar particles at its maximum stellar mass with L200m6 resolution;
    \item the satellite is resolved down to a minimum remaining stellar mass fraction of $f_{\rm{*,min}} < 0.5$;
    \item the satellite is resolved in at least four snapshots throughout its evolution.
\end{enumerate}

Fig.~\ref{fig:ab_distribution} shows the joint distribution of the best-fitting parameters $f_{\rm d}$ and $b$ obtained from all satellites that meet these criteria. Overall, the two parameters appear largely independent. The marginalized distribution of each parameter follows a lognormal distribution. We list the parameters of the distributions in Table~\ref{tab:track_parameters}.

\begin{table}
	\centering
	\caption{The best-fitting parameters for the distribution of tidal track parameters $f_{\rm d}$ and $b$ (Eq.~\eqref{eq:He26}). The distributions for both parameters are described by a log-normal probability density function: $P(\ln x) = \mathcal{N}(\ln x; \ln \mu, \sigma)$.} 
	\label{tab:track_parameters}
	\begin{tabular}{llccc} 
		\hline
		Parameter & $\ln \mu$ (Location) & $\sigma$ (Scale) & Median Value ($\mu$) \\
		\hline
		$f_{\rm d}$  & $-2.87$ & $0.5$ & $0.057$ & \\
		$b$  & $-0.1$ & $0.21$ & $0.9$ & \\
		\hline
	\end{tabular}
\end{table}

\section{Fates of satellites}\label{sec:fate}

\subsection{Bimodal distribution of satellite mass evolution}\label{sec:bimodal}



\begin{figure*}
    \centering
    \includegraphics[width=0.95\linewidth]{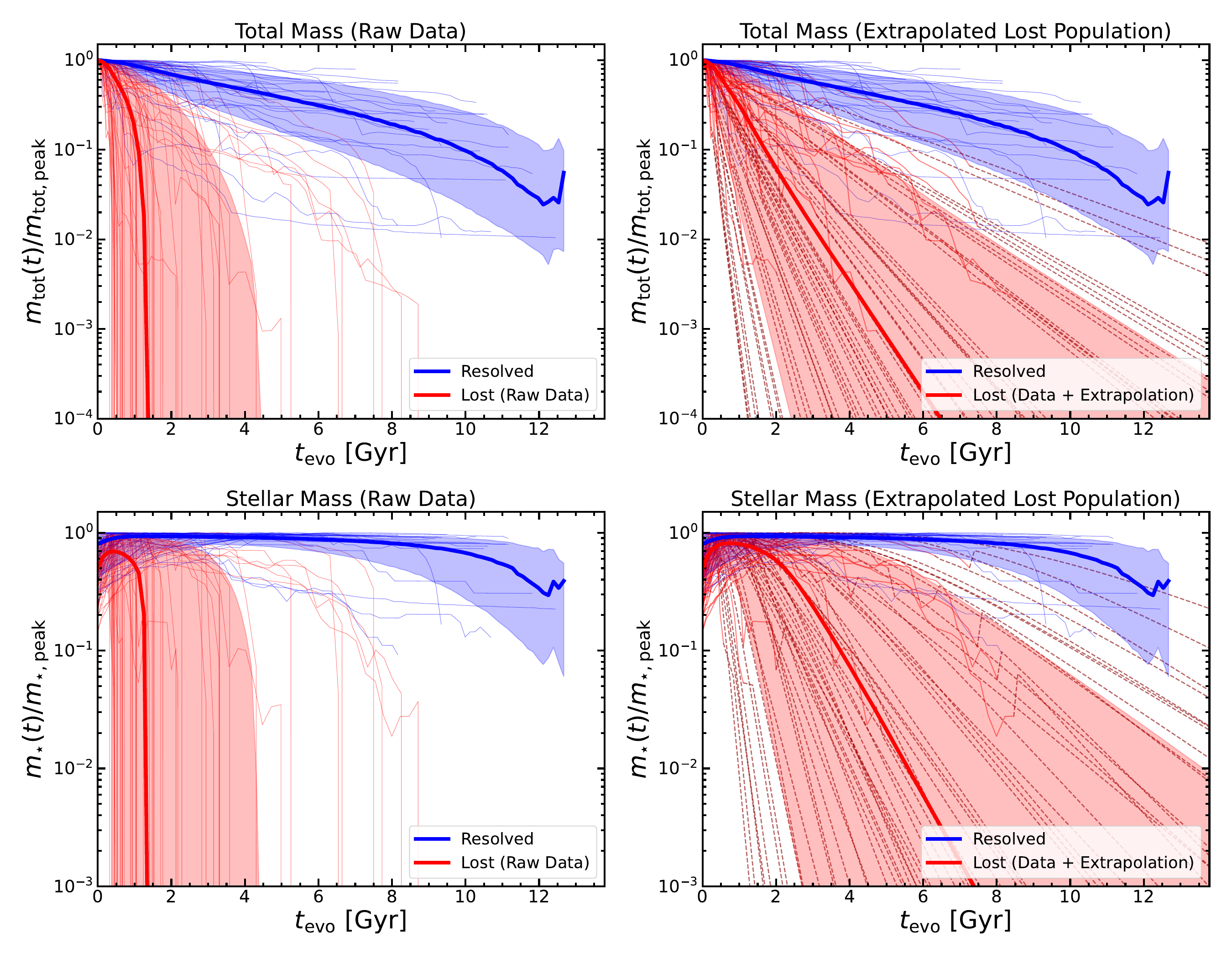}
    \caption{Comparison between the raw simulation data and the extrapolated tidal stripping model for satellite galaxies in cluster-scale haloes ($M_{\rm{host}} > 10^{14}\,\rm{M}_{\odot}$). All selected satellites have $m_{\rm{tot,peak}} > 10^{10}\,\rm{M}_{\odot}$. In all panels, thick solid lines and shaded regions represent the median and 16th–84th percentile ranges, while thin lines show the trajectories of 100 randomly selected individual systems. Blue and red colours denote the resolved and lost populations, respectively. \textit{Top-left}: Subhalo mass fraction ($f_{\rm tot}$) measured directly from the simulation. \textit{Top-right}: The same subhalo sample, but with the mass loss of the lost population reconstructed using our exponential extrapolation model (Eq.~\eqref{eq:masslossrate} $\&$ \ref{eq:extra}). \added{For the lost population, the solid thin lines represent the directly simulated evolution, while the dashed lines indicate the extrapolated mass loss beyond the last resolved snapshot.} \textit{Bottom-left}: Stellar mass fraction ($f_*$) measured directly from the simulation. Many lost satellites still possess significant stellar components at the moment of being lost. \textit{Bottom-right}: Predicted stellar mass evolution using the universal tidal track (Eq.~\eqref{eq:He26}) applied to the extrapolated subhalo mass.} 
    \label{fig:massloss}
\end{figure*}

By studying the evolution of subhaloes in the high-resolution Aquarius simulation~\citep{Springel2008}, \citetalias{He2025} found that the acquisition of subhaloes can be broadly divided into two channels corresponding to two distinct fates: 
\begin{enumerate}
    \item \textbf{Hierarchical Assembly Channel}: subhaloes first merge with other haloes (referred to as parent haloes), becoming 1st-order subhaloes, and subsequently accrete onto the final host halo as higher-order subhaloes.
    \item \textbf{Direct Accretion Channel}: subhaloes are accreted directly onto the host halo as 1st-order subhaloes.  
\end{enumerate}

\citetalias{He2025} found that subhaloes following the hierarchical assembly pathway are typically accreted at high redshift (larger $z_{\rm{acc}}$)~\citep[see][for an analytical model]{Jiang2025, Jiang2026}. These systems experience higher merger mass ratios with their parent haloes, stronger dynamical friction, and significant orbital decay due to host halo growth, causing them to shrink rapidly toward the central regions of the parent halo. The high-density environment facilitates tidal stripping, leading to extreme mass loss. \added{This enhanced susceptibility of pre-processed satellites to tidal disruption and environmental effects has been noted in earlier studies \citep[e.g.][]{Penarrubia2012, Bahe2019}.} As a result, such subhaloes often become lost below the resolution limit of simulations within a short timescale, contributing significantly to the ``lost subhaloes'' commonly observed in DMO simulations.

In contrast, subhaloes accreted directly typically have lower $z_{\rm{acc}}$ and smaller merger mass ratios with their parent haloes. Their orbits remain more stable after accretion, and they reside in gentler tidal fields, resulting in a relatively mild mass loss rate. The survival of subhaloes in this channel is also in agreement with theoretical studies using idealized simulations under similar orbital conditions~\citep{Ogiya2019, Errani2021}.

It should be noted that Aquarius is a DMO simulation. 
In hydrodynamic simulations, subhaloes consist of three main components: DM, stars, and gas. Their structural and orbital evolution can differ from the DMO case and are considerably more challenging to model. 
To illustrate how these two channels manifest in our hydrodynamical cosmological simulation, we present the mass evolution trajectories of 100 randomly selected cluster satellites in the top-left panel of Fig.~\ref{fig:massloss}. The sample is divided into the \textit{resolved} population (blue) and the \textit{lost} population (red). While the resolved systems show smooth, gradual mass loss, the lost systems exhibit precipitous declines before vanishing from the simulation.

However, a direct statistical analysis of these trajectories is hindered by a ``survivorship bias'': the simulation can only track the lost population down to the resolution limit, censoring their subsequent evolution. To correct for this effect and recover the full mass loss distribution, we introduce an extrapolation scheme based on the average mass loss rate, $\Gamma$. We define $\Gamma$ as:
\begin{equation}
    \Gamma \equiv - \frac{ \log_{10} (m_{\rm{final}} / m_{\rm{peak}})} { \Delta t},
    \label{eq:masslossrate}
\end{equation}
where $m_{\rm{final}}$ is the subhalo mass at the last resolved snapshot, and $\Delta t$ is the duration since peak mass. It is worth noting that while \citetalias{He2025} utilized the sophisticated semi-analytic model {\tt\string SatGen} to predict mass loss in DMO simulations, applying such models to hydrodynamical simulations is non-trivial. The presence of baryons introduces complex torques, adiabatic contraction/expansion, and gas dynamics that alter the orbital and structural evolution compared to the DMO case. Therefore, for simplicity and robustness, motivated by \citet{vanDenBosch2005, Jiang2016}, who showed that the orbit-averaged subhalo mass loss follows an approximately exponential decay with time, we adopt this empirical average mass loss rate $\Gamma$ as a first-order approximation to characterize the stripping efficiency. \added{We stress that this constant-rate extrapolation is not intended to accurately reproduce the detailed mass-loss history of individual subhaloes. Rather, $\Gamma$ serves as a simplified, population-level diagnostic introduced primarily for illustration: it allows us to bypass survivorship bias and reconstruct the overall distribution of mass-loss trajectories for the lost population.}


Using this rate, we extrapolate the mass evolution of lost satellites beyond their last resolved snapshot $t_{\rm lost}$:
\begin{equation}
    f_{\rm tot}(t) = f_{\rm tot,final} \cdot 10^{\left[-\Gamma (t - t_{\rm lost})\right]}, \quad \text{for } t > t_{\rm lost},
    \label{eq:extra}
\end{equation}
as indicated by the red dashed thin lines in the top-right panel of Fig.~\ref{fig:massloss}. This procedure reconstructs the ``hidden'' portion of the statistic. The shaded blue and red regions in the both top panel of Fig.~\ref{fig:massloss} represent the 16th–84th percentile ranges of the mass loss for the resolved and lost populations, respectively. While the top-left panel provides an intuitive view of the complete, physically extrapolated distribution of subhalo mass loss trajectories over time.

\begin{figure}
    \centering
    \begin{subfigure}[b]{0.45\textwidth}
        \includegraphics[width=\textwidth]{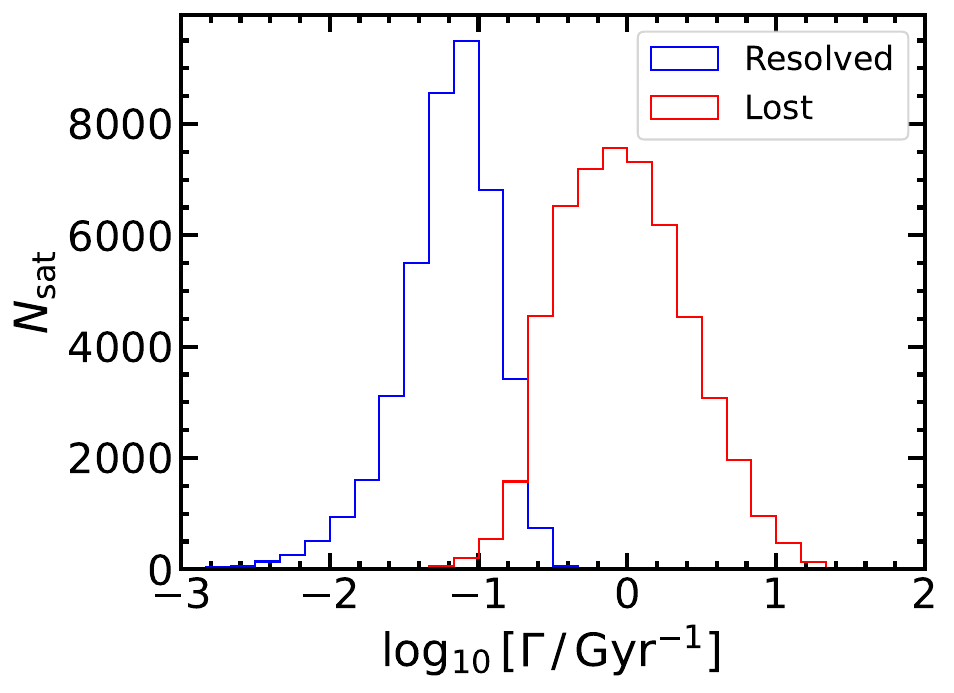}
        \caption{Distribution of the average mass loss rate $\Gamma$ for subhaloes with $m_{\mathrm{tot, peak}} > 10^{10} \rm{M}_{\odot}$ in cluster haloes with $M_{\rm host}>10^{14}\rm{M}_\odot$. Blue and red lines represent the resolved and lost subhalo populations, respectively.}
    \end{subfigure}
    \hfill
    \begin{subfigure}[b]{0.45\textwidth}
        \includegraphics[width=\textwidth]{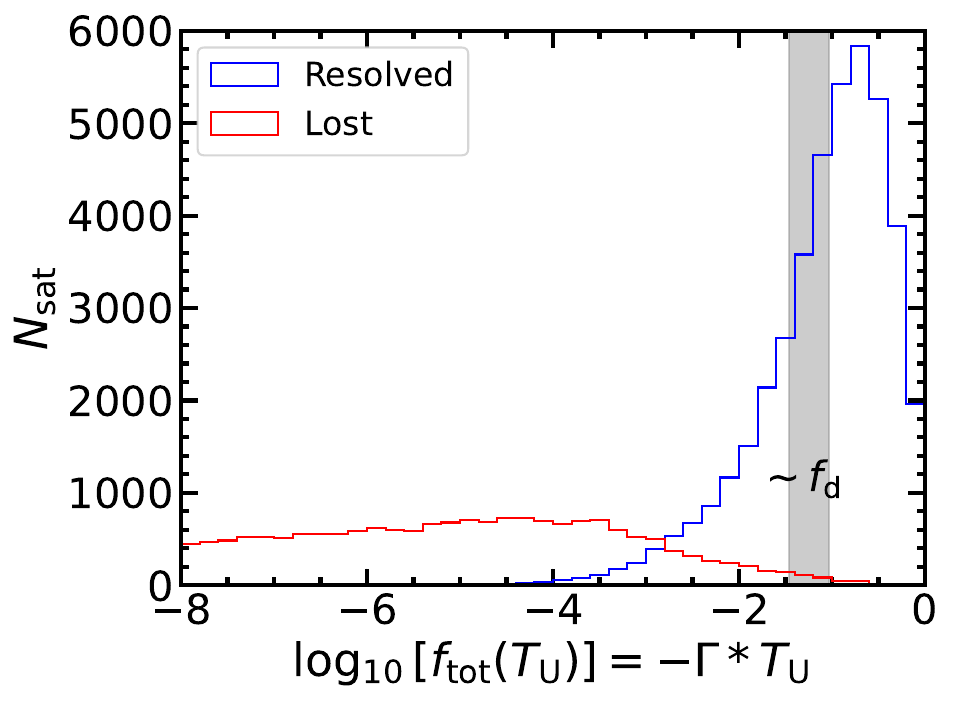}
        \caption{Predicted mass loss distribution of the same subhalo sample, assuming they evolve at the measured rate $\Gamma$ over the age of the Universe $T_{\rm U}$. The vertical shaded region indicates the $\pm 1\sigma$ range of the stellar tidal track turn-around points ($f_{\rm d}$), Note that the scatter of $f_{\rm d}$ is quite narrow relative to the dynamical range of the $f_{\rm tot}$ distribution. }
    \end{subfigure}
    \hfill

    \begin{subfigure}[b]{0.45\textwidth}
        \includegraphics[width=\textwidth]{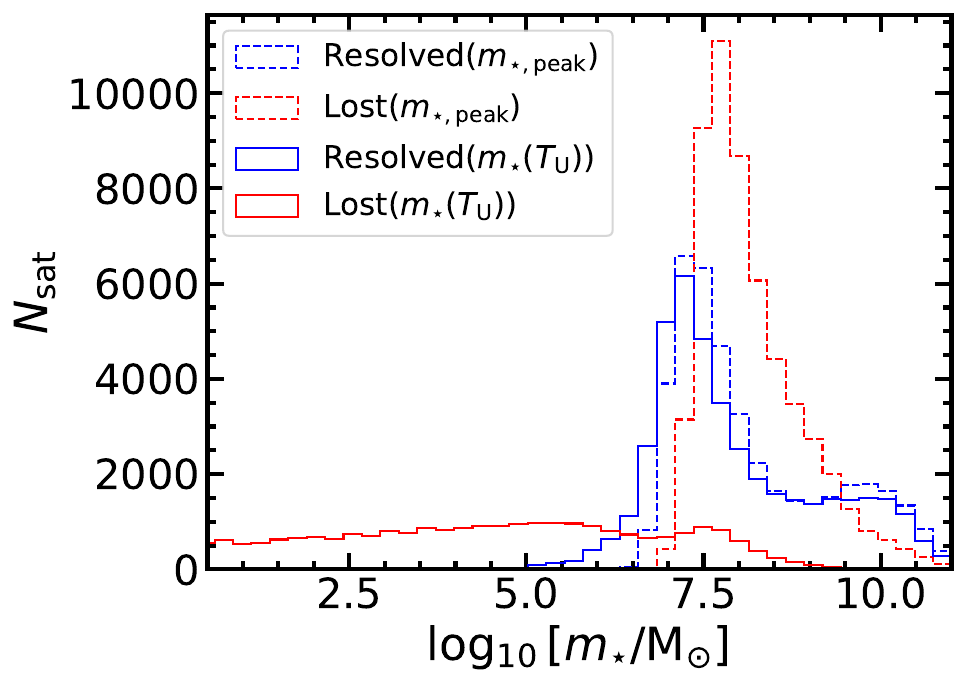}
        \caption{Ultimate satellite stellar mass function (solid lines) after evolving for the age of the Universe. The dashed lines show the satellite unevolved stellar mass function for the resolved and lost population, respectively. }
    \end{subfigure}

    \caption{The distribution of subhalo mass loss and rates.}
    \label{fig:str_massloss}
\end{figure}

The distribution of this mass loss rate, $\Gamma$, is shown in the top panel of Fig.~\ref{fig:str_massloss}. Remarkably, despite the simplicity of our definition and the complexity of hydrodynamical effects, the distribution is clearly bimodal. This finding strongly corroborates the results of \citetalias{He2025}: the satellite population is indeed composed of two distinct families. Typical resolved satellites (the ``slow'' population) lose only $\sim 15\%$ of their mass per Gyr, whereas the lost population (the ``fast'' population) undergoes extreme stripping, losing up to $90\%$ of their mass per Gyr. Such satellites usually fall below the resolution limit of current state-of-the-art cosmological simulations within a few Gyr.

Next, we assume that all these satellites evolve under this mass loss rate up to the age of the Universe $T_{\rm U}$ to obtain the final mass loss statistic. This ultimate mass loss distribution is shown in the middle panel of Fig.~\ref{fig:str_massloss}. Satellites that experience moderate mass loss exhibit a log-normal distribution of their mass fraction $f_{\rm tot}$, typically above $10^{-4}$. This result is consistent with the findings of \citetalias{Han2016b} in the Aquarius simulations, which also demonstrated convergence with respect to mass resolution. In contrast, satellites that undergo extreme mass loss display a much more extended distribution, with $f_{\rm tot}$ falling below $10^{-8}$ over a cosmological timescale.

Based on the results of the previous section and Fig.~\ref{fig:tidaltrack}, the relation between satellite stellar mass loss and subhalo mass loss follows a two-phase tidal track described by Eq.~\eqref{eq:He26} very well. Its turnover point of the two-phase, where the stellar mass of a satellite remains effectively frozen, corresponds to a relatively fixed range of $f_{\rm d} \sim 10^{-1.5}$–$10^{-1}$ ($1\sigma$ scatter), as indicated by the grey column in the middle panel of Fig.~\ref{fig:str_massloss}. 
Once the subhalo mass loss falls below this range, the stellar components of satellites are consistently stripped at a rate comparable to the overall subhalo mass loss rate.
Most resolved satellites that undergo modest mass loss rates lie at or above this turnover range. In other words, these satellites are still able to bind most of their formed stars, thereby maintaining their stellar mass constant after reaching their peak. Below this range, those satellites located at the tail of the modest mass loss distribution will enter the second phase of stellar mass loss if they evolve for a sufficiently long time. By contrast, satellites experiencing extreme mass loss rates almost entirely fall below the lower bound of the $f_{\rm d}$ range. Consequently, the stars they host will inevitably be stripped and ultimately disrupted by tidal forces at rates comparable to those of their host subhaloes. These results also provide a natural explanation for the lack of ``orphan'' galaxies in some previous hydrodynamical simulations, as the galaxies of the disrupted subhaloes tend to be disrupted.

Based on the tidal track fitting results in Section~\ref{sec:satellites}, we randomly generate $f_{\rm d}$ and $b$ for each satellite in the selected cluster according to the log-normal distribution shown in Table.~\ref{tab:track_parameters}. Substituting the extrapolated $f_{\rm tot}^{\rm{ext}}$ 
into Eq.~\eqref{eq:He26} then allows us to predict the long-term stellar mass evolution of each satellite. 
The extrapolated stellar mass loss for lost satellites is illustrated by the red dashed curve in the bottom-right panel of Fig.~\ref{fig:massloss}. For resolved satellites, stellar mass loss curves are directly extracted from simulation data, as shown by the blue curves in the same panel. This analysis clearly demonstrates that the universal tidal track translates the bimodal distribution of subhalo mass loss into a corresponding bimodality in stellar mass loss. 
Most surviving satellites maintain $ f_* \simeq 1 $ even after evolving for 10 Gyr, whereas the disruption of the subhalo systematically leads to the disruption of the embedded galaxy. 

In the bottom-left panel of Fig.~\ref{fig:massloss}, we present the raw stellar mass loss curves (i.e. without our extrapolation approach) measured directly from the simulation for comparison. These trajectories indicate that many lost satellites disappear from the simulation abruptly, even while still hosting a substantial stellar component. In contrast, our extrapolated model demonstrates that these systems should physically undergo extensive stellar stripping, withering away gradually rather than undergoing the sudden dissolution seen in the raw tracking data. This discrepancy underscores the impact of artificial disruption, where satellites are numerically lost before their physical withering is complete. We examine this influence in the next subsection.


We multiply the stellar mass loss fraction at $T_{\rm U}$ by the satellite peak (unevolved) mass. In this way, we can obtain the final evolved stellar mass distribution. The stellar mass functions for evolving $T_{\rm{U}}$ are presented in the bottom panel of Fig.~\ref{fig:str_massloss}. The dashed line shows the unevolved stellar mass function associated with subhaloes of $M_{\rm tot,peak} > 10^{10}\rm M_{\odot}$. The blue solid line represents the predicted final stellar mass function of resolved subhaloes, which exhibits only a slight shift relative to their unevolved stellar mass function. In contrast, subhaloes that undergo extreme mass loss rates also experience drastic stripping of their stellar components. Their final stellar mass function extends downward by several orders of stellar mass, which can in some sense be regarded as representing disrupted satellites (these satellites may either merge with the host halo or with their parent satellites, depending on their orbits).

It is worth noting that here we only use a simple mass loss rate to estimate the general trend of the final stellar mass evolution of satellites. This approach cannot rigorously predict their precise behavior at extremely low masses (e.g.,  a few dozen solar masses, or below the scale of globular clusters). Nevertheless, what can be concluded is that, due to the extreme mass loss induced by the tidal environment, these satellites cannot maintain their stellar peak mass. 

The dynamic range of subhalo bound mass fractions, $f_{\rm tot}$, is large, spanning from unity down to values far below $10^{-4}$ for the lost population. In stark contrast, the transition threshold for the tidal track, $f_{\rm d}$, occupies a relatively narrow range of $10^{-1.5}\sim10^{-1}$. Because the distribution of $f_{\rm tot}$ for the extreme mass loss population plunges far below this narrow threshold, these systems are statistically destined to cross the transition point. The vast majority will inevitably fall below the current resolution limit of large-volume numerical simulations ($\sim 10^{6}\rm{M}_{\odot}$)—which is traditionally regarded as the disruption of a satellite galaxy. Consequently, the universal tidal track maps the bimodal distribution of subhalo mass loss directly onto the stellar component: just as subhaloes are segregated into two populations, the disruption of a subhalo leads to the disruption of its embedded galaxy.

\subsection{Assessing the impact of artificial disruption}\label{sec:ad}

\begin{figure}
    \centering
    \includegraphics[width=\linewidth]{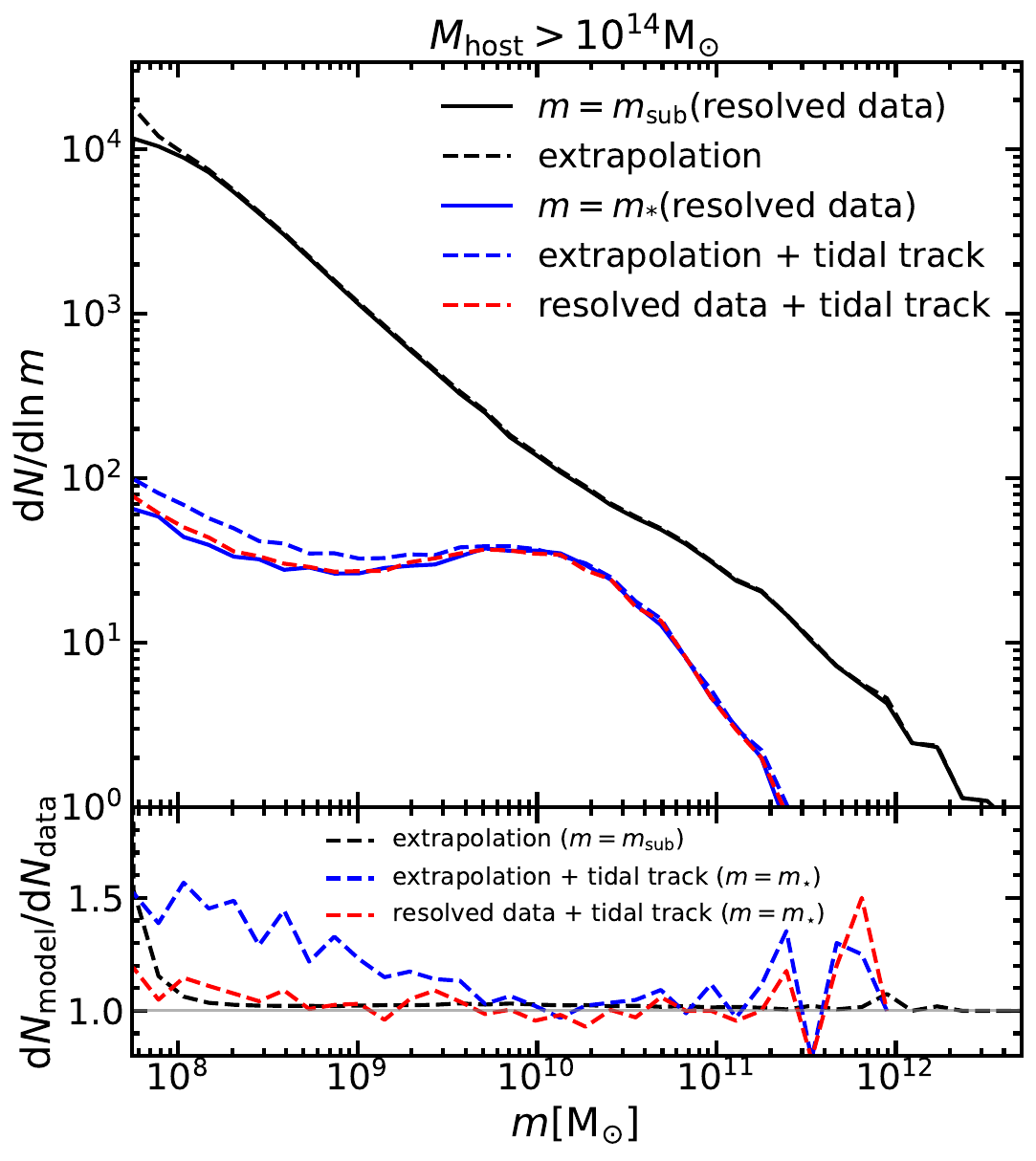}
    \caption{The subhalo mass function (black lines) and satellite stellar mass function (blue lines) averaged over host haloes with virial masses $M_{\rm host} > 10^{14}\, \mathrm{M}_\odot$. \textit{Top Panel}: Solid lines represent the resolved population directly measured from the simulation data. Dashed lines include both resolved satellites and the lost population, recovered via our extrapolation model: the black dashed line represents the subhalo mass function including lost satellites; the blue dashed line represents the stellar mass function including lost satellites, whose stellar masses are derived by applying our tidal track model Eq.~\eqref{eq:He26} to the extrapolated subhalo mass; red shaded line serves as a validation test where the tidal track model is applied to the resolved population. It is in close agreement with the direct stellar mass measurement (blue solid line), demonstrating the accuracy of the tidal track model.
    \textit{Bottom Panel}: The ratio of the model-predicted mass functions to the direct simulation results. The deviation of the black dashed line is negligible for $m_{\rm tot} > 10^{8}\rm{M}_\odot$. In contrast, the blue dashed line rises above unity for $m_* \lesssim 10^{10}\, \mathrm{M}_\odot$. This excess highlights the impact of numerical resolution on the faint end of the satellite stellar mass function.}
    \label{fig:ad}
\end{figure}

Based on the extrapolated subhalo mass loss and the tidal track model, we can further estimate the effects of artificial disruption on the stellar mass function. In Fig.~\ref{fig:ad}, the black and blue solid lines represent the subhalo and stellar mass functions, respectively, for cluster satellites at $ z = 0 $. The corresponding extrapolated subhalo and stellar mass functions at $ z = 0 $, derived from our model, are shown as black and blue dashed lines. The red dashed line is obtained by applying randomly generated tidal tracks to the resolved satellite population in the simulation, using their measured $ f_{\mathrm{tot}} $ values as input. This result demonstrates that our tidal track model successfully reproduces the stellar mass function and its evolution as seen in the simulation, serving as a reference. \added{To further verify the validity of our mass-loss prescription in the highly stripped regime, we perform an internal validation test in Appendix~\ref{sec:ftotbin} by splitting the resolved population into subsets based on their remaining bound mass fractions. The model consistently matches the simulation data even down to the lowest resolved $f_{\rm tot}$ bin, justifying its extrapolation to the lost population.}

The footprint of artificial disruption is also evident in the distribution of stellar mass loss curves. In the right panels of Fig.~\ref{fig:massloss}, we contrast the raw stellar mass loss curves from the simulation with our modelled trajectories, which apply the tidal track to the extrapolated subhalo mass. By extrapolating the subhalo evolution, we reveal that many ``lost'' satellites should survive above stellar mass fractions of $f_* \gtrsim 10^{-3}$. In the absence of this correction, these systems appear to dissolve as an artifact of the premature numerical disruption of their host subhaloes.

The difference between the dashed and solid lines in Fig.~\ref{fig:ad} represents this missing satellite population, which is expected to be resolved above the current simulation mass limit but is lost due to numerical effects. For the subhalo mass function, the effect is negligible at $m_{\mathrm{tot}} > 10^8 \, \mathrm{M}_{\odot}$. This result aligns with the findings of \citetalias{He2025}, who concluded that artificial disruption is not a significant concern for state-of-the-art simulations. We note that while \citetalias{He2025} analyzed the DMO Aquarius simulation (level-2), the effective resolution of the \colibre  L200m6 run used here is comparable. Aquarius-A2 employs a particle mass resolution of $m_{\rm p} \sim 10^4 \, \mathrm{M}_{\odot}$ to resolve subhaloes within a Milky Way-mass host ($M_{\mathrm{host}} \sim 10^{12} \, \mathrm{M}_{\odot}$), yielding a dynamic range of $m_{\rm p}/M_{\mathrm{host}} \sim 10^{-8}$. Similarly, \colibre L200m6 resolves cluster-scale environments ($M_{\mathrm{host}} \sim 10^{14} \, \mathrm{M}_{\odot}$) with $m_{\rm p} \sim 10^6 \, \mathrm{M}_{\odot}$, maintaining the same relative mass resolution of $\sim 10^{-8}$. Consequently, we confirm that even with the inclusion of hydrodynamics, the subhalo mass function remains robust against numerical disruption at this resolution level.

In contrast, the extrapolated stellar mass function begins to deviate from the raw simulation data at $m_* < 10^{10} \, \mathrm{M}_{\odot}$. Specifically, within the mass range of $10^9\text{--}10^{9.5} \, \mathrm{M}_{\odot}$, the simulation underpredicts the expected stellar mass function by approximately 20 per cent, even for satellites consisting of hundreds of star particles. This discrepancy grows toward lower masses, the ratio of extrapolated to simulated counts reaches 1.5 at $m_* \sim 10^8 \, \mathrm{M}_{\odot}$, which corresponds to the resolution limit of approximately 50 particles in the L200m6.

This discrepancy arises from two main factors that uniquely affect the evolved mass functions at $z=0$, whereas the unevolved (peak) mass functions share identical convergence properties. 

First, stellar mass loss is significantly delayed relative to subhalo mass loss because stars are more centrally concentrated than the extended DM envelope. As illustrated by our tidal track model, a subhalo may be heavily stripped down to its resolution limit (e.g., retaining $1/10$ of its peak mass) while its embedded galaxy remains almost entirely intact ($f_* \simeq 1$). If the simulation artificially disrupts this subhalo due to numerical resolution effects, it prematurely and abruptly removes a massive, surviving stellar component. This creates a hidden population of "numerical orphans": galaxies that are artificially deleted despite physically surviving. Second, the mapping between peak subhalo mass and stellar mass is non-linear, which causes the faint-end slope of the satellite stellar mass function to be significantly shallower than that of the subhalo mass function. Due to this shallower slope, the numerical loss of a given absolute number of low-mass systems translates into a much larger fractional deficit in the stellar mass function than in the subhalo mass function. 

The combination of these effects acts as \deleted{a} \added{an} amplifier for numerical limits: while artificial disruption accounts for only a few per cent deficit in the subhalo mass function, it manifests itself as a $\sim 20\% \ (50\%)$ artificial suppression in the satellite stellar mass function at $m_{*}\sim10^{9}\ (10^{8})\, \rm{M}_{\odot}$. To accurately model satellite statistics, it is crucial to distinguish these "numerical orphans" from physically withered systems whose stellar components have genuinely been stripped. We will rigorously formalize the analytical mechanics of this differential sensitivity and the corresponding orphan modeling in a follow-up paper (He et al. in prep).

Such a deviation could be considered marginal when $m_{*}\gtrsim 10^{9}\,\rm{M}_\odot$. However, recent resolution convergence studies suggest a more complex picture. Comparisons within the IllustrisTNG suite \citep{Engler2021, Lovell2025} reveal that the suppression of the satellite stellar mass function at lower resolution is primarily driven by a systematic change in the stellar-to-halo mass relation, rather than solely by an enhanced rate of subhalo disruption. Specifically, lower-resolution simulations tend to produce fainter galaxies for a given halo mass, owing to the resolution dependence of subgrid star formation and feedback models. Consequently, while artificial disruption imposes a distinct survival bias on the satellite population, this effect may be secondary to the intrinsic lack of convergence in galaxy formation physics across different resolution levels.

Finally, we identify this missing population that is numerically disrupted but physically should retain a bound stellar component, as the hydrodynamic equivalent of ``orphan galaxies''. Historically, orphans were introduced in semi-analytic models to address the over-merging problem. Our results demonstrate that this is not a concern for DMO simulations; hydrodynamic simulations are more subject to artificial disruption that selectively deletes satellite galaxies. Consequently, to accurately recover the faint end of the luminosity function and clustering statistics at small scale, even hydrodynamic simulations require explicit ``orphan'' prescriptions or advanced tracking techniques to compensate for numerical disruption.



\subsection{Stellar-to-subhalo mass ratio evolution}\label{sec:xi_evo}

\begin{figure*}
\centering
    \begin{subfigure}[b]{0.49\textwidth}
        \includegraphics[width=\textwidth]{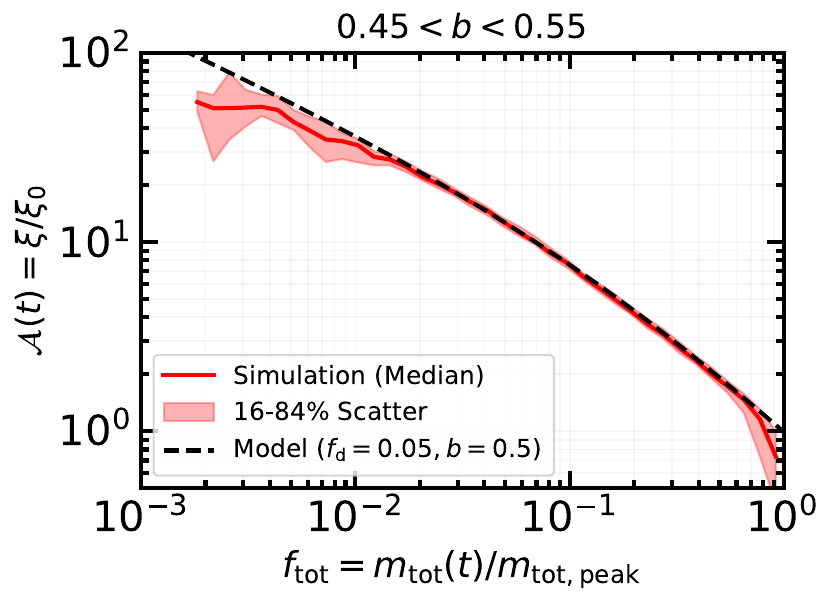}
    \end{subfigure}
    \hfill
    \begin{subfigure}[b]{0.49\textwidth}
        \includegraphics[width=\textwidth]{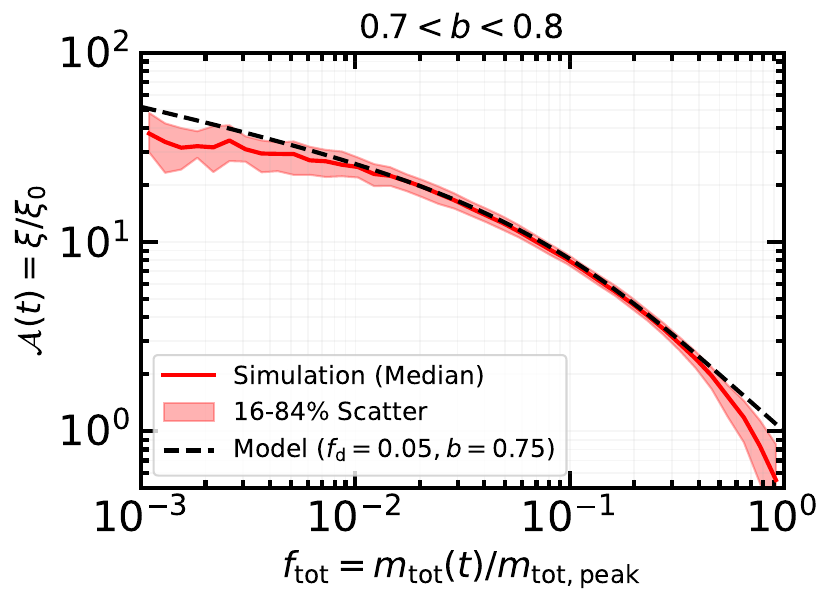}
    \end{subfigure}  
    
    \vspace{-0.1cm} 
    
    \begin{subfigure}[b]{0.49\textwidth}
        \includegraphics[width=\textwidth]{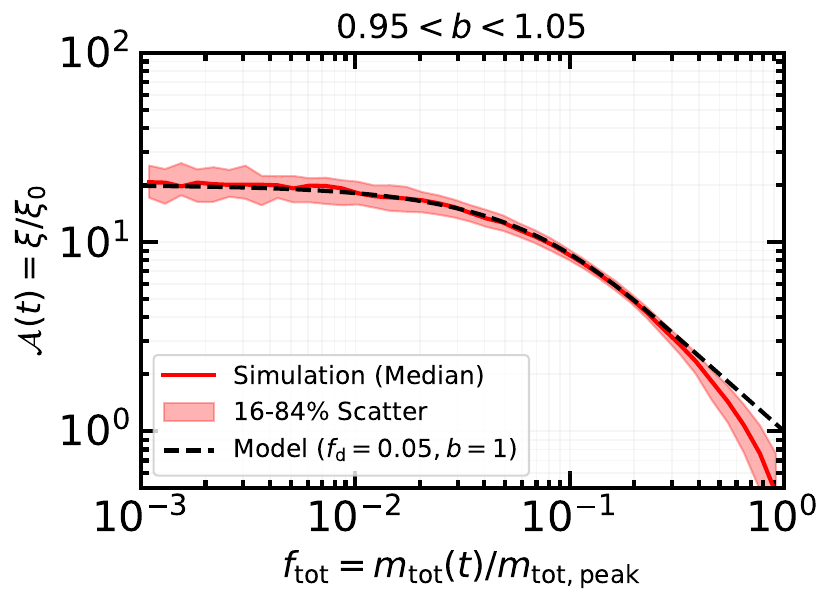}
    \end{subfigure}
    \hfill
    \begin{subfigure}[b]{0.49\textwidth}
        \includegraphics[width=\textwidth]{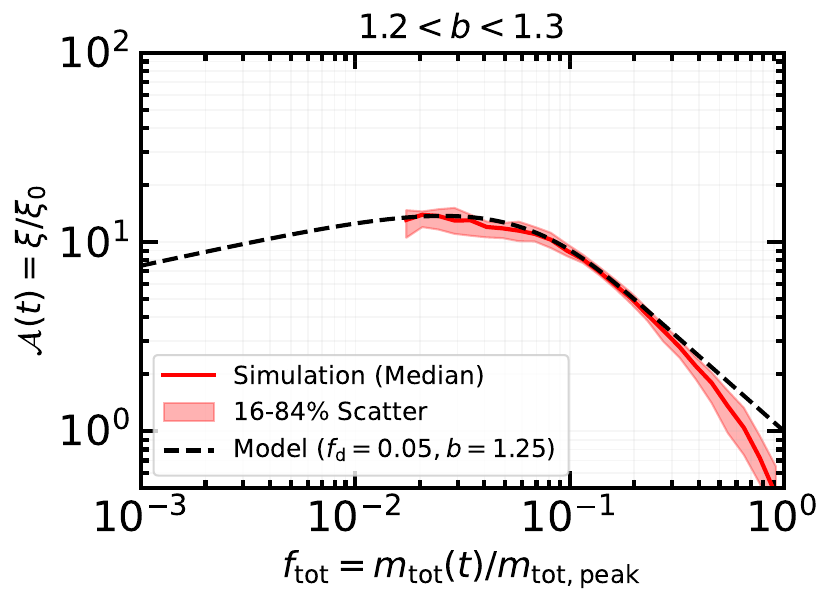}
    \end{subfigure}
    
    \caption{Evolution of the normalized stellar-to-subhalo mass ratio (the tidal  ``amplification factor'' $\mathcal{A}$), as a function of the $f_{\rm tot}$. The four panels display satellite tracks binned by their best-fit tidal track parameter $b$ (centered at $b \approx 0.5, 0.75, 1.0, 1.25$). The red solid lines and shaded regions represent the median and the $16{\rm th}$--$84{\rm th}$ percentile scatter of the simulation data, respectively. The black dashed lines show the theoretical prediction derived from Eq.~\eqref{eq:boost}, assuming a fixed structural parameter $f_{\rm d}=0.05$. The model accurately captures the diversity of evolutionary tracks.}
    \label{fig:xi_boost}
\end{figure*}

Tidal stripping preferentially removes the extended DM halo before affecting the centrally concentrated stellar component. This differential mass loss leads to an increase in the stellar-to-subhalo mass ratio, $\xi \equiv m_{*}/m_{\rm{tot}}$. We define the tidal ``amplification'' factor $\mathcal{A}$, as $\xi$ normalized by \added{$\xi_0=m_{*,\rm peak} / m_{\rm tot, peak}$, which is the peak stellar-to-subhalo ratio (note that they are measured at their respective peak values)}:
\begin{equation}
\mathcal{A}(f_{\rm tot}) \equiv \frac{\xi(t)}{\xi_0} = \frac{f_{*}}{f_{\rm tot}} = \frac{1-\exp[-(f_{\rm tot}/f_{\rm d})^b]}{f_{\rm tot}},
\label{eq:boost}
\end{equation}
where we used the Eq.~\eqref{eq:He26} in the last step. This equation describes how efficiently a satellite can ``concentrate'' its baryons relative to its DM as it undergoes stripping. 

To evaluate the predictive power of the tidal track model across diverse stripping scenarios, we present the evolution of the tidal amplification factor, $\mathcal{A}(t)$ in Fig.~\ref{fig:xi_boost}, for a representative subset of satellites within the L200m6 \colibre simulation. To isolate the impact of the stripping efficiency parameter $b$, we restrict the sample to satellites with stripping parameters in the range $f_{\rm d} \in [0.04, 0.06]$, ensuring that the transition between the two evolutionary phases occurs at a comparable $f_{\rm tot}$. The satellites are subsequently selected in four bins of their best-fit $b$ values, centred at approximately $0.5$, $0.75$, $1$, and $1.25$ with a bin width of $0.1$.
This systematic breakdown illustrates how the stellar-to-subhalo mass ratio responds to varying degrees of stellar resilience. The median tracks from the simulation (red solid lines) demonstrate a remarkable agreement with the analytical predictions derived from Eq.~\eqref{eq:boost} (black dashed lines), confirming that our model accurately captures the diversity of individual satellite mass evolution. The systematic deviations at the initial states ($f_{\rm tot} \sim 1$), come from the fact that we do not model post-infall star-formation. 

For satellites with $b < 1$ (top panels of Fig.~\ref{fig:xi_boost}), the stellar component is highly resilient. In the second phase ($f_{\rm tot} \ll f_{\rm d}$), the stellar mass fraction scales as $f_* \propto f_{\rm tot}^b$, causing the amplification factor to diverge as a power law: $\mathcal{A} \propto f_{\rm tot}^{b-1}$. Since $b-1 < 0$, $\mathcal{A}$ increases monotonically as the subhalo loses its mass. This regime represents the ``classic'' stripping scenario where the DM halo is whittled away, leaving behind a naked, baryon-dominated system.
However, we note that a physical upper limit applies: $\xi$ cannot exceed unity (since $m_{\rm tot} \ge m_*$). Thus, the amplification factor is bounded by $\mathcal{A}_{\rm max} = 1/\xi_0$. Once this limit is reached, the system becomes a pure stellar clump, and subsequent evolution must follow $f_* \propto f_{\rm tot}$ (i.e., $\mathcal{A} = \text{const}$), rendering Eq.~\eqref{eq:boost} formally invalid beyond this point. This is the reason why we see so little deviation of the red solid line from the black dashed line at a very low $f_{\rm tot}$.

When $b \approx 1$ (bottom-left panel), \deleted{the stripping of stars and DM becomes coupled,}\added{the spatial distributions of stars and dark matter are highly aligned, causing their relative mass-loss rates to become directly proportional in the late stages of stripping.} Consequently, the amplification factor initially rises but asymptotically approaches a saturation value of $\mathcal{A}_{\rm lim} \approx 1/f_{\rm d}=a_{\rm strip}$. In this scenario, the satellite can become more baryon-dominated than it was at infall, but the enhancement is capped. The DM halo is never fully removed relative to the stars because both components are stripped at comparable rates in the late stages.

For $b > 1$ (bottom-right panel), the evolution of $\mathcal{A}$ becomes non-monotonic. Initially, $\mathcal{A}$ rises as outer DM is stripped. However, once the tidal radius penetrates the stellar core, the loosely bound stellar distribution (implied by high $b$) is stripped faster than the remaining DM cusp ($f_* \propto f_{\rm tot}^b$ with $b>1$). Consequently, $\mathcal{A}$ reaches a peak and then declines ($\mathcal{A} \propto f_{\rm tot}^{b-1}$ where $b-1 > 0$). These systems represent ``failed'' candidates for baryon dominance; although they lose mass, they fail to retain their stars efficiently enough to become DM deficient.

\subsection{The emergence of dark matter deficient galaxies} \label{sec:dmdg}

\begin{figure}
    \centering
    \includegraphics[width=\linewidth]{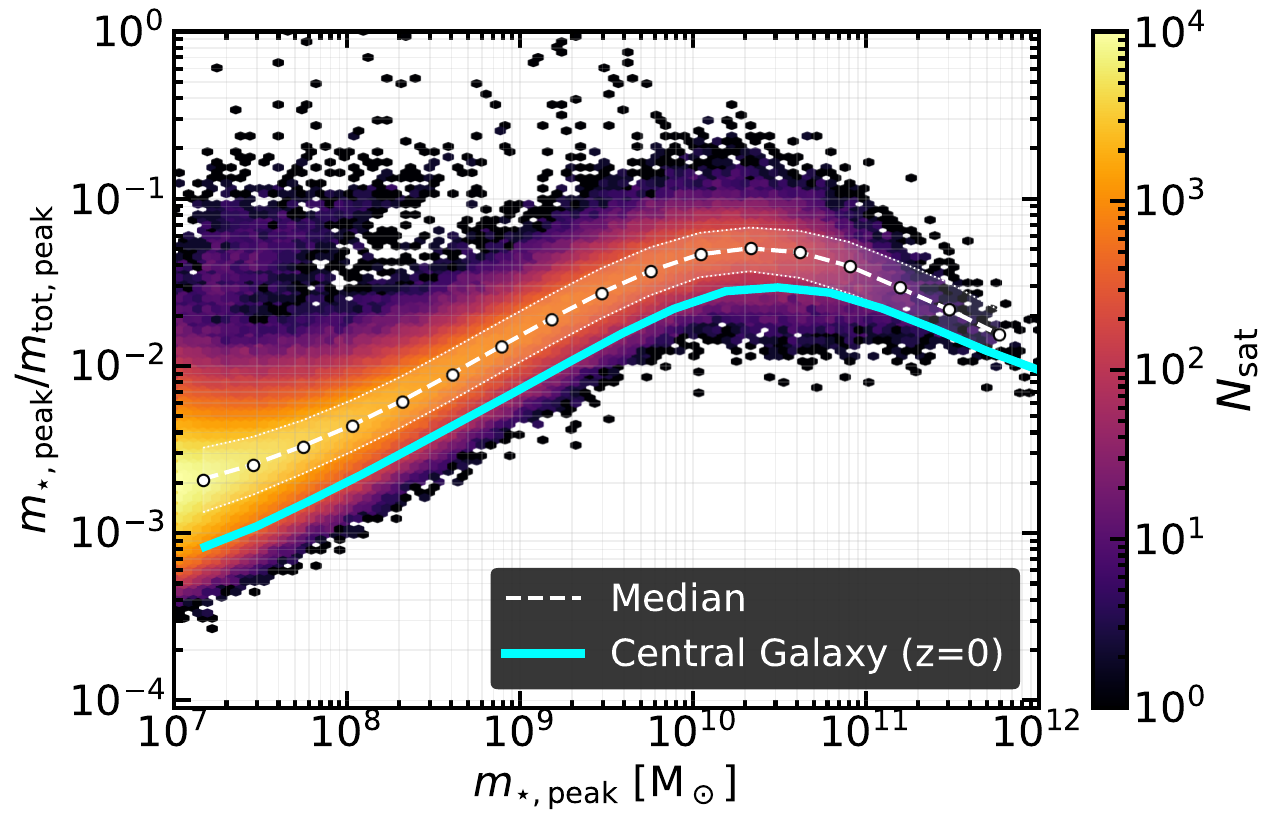}
    \caption{The peak stellar-to-subhalo mass ratio $m_{*, \rm peak}/m_{\rm tot, \rm peak}$, as a function of stellar peak mass. The colour map represents the number of satellites in each bin ($N_{\rm sat}$). The white dashed line and circles trace the median relation, while the dotted lines indicate the $16{\rm th}$–$84{\rm th}$ percentile scatter. The thick cyan line represents the median of stellar-to-halo mass relation for central galaxies at $z=0$ for reference.
}
    \label{fig:SHMR}
\end{figure}

\begin{figure*}
    \centering
    \includegraphics[width=\linewidth]{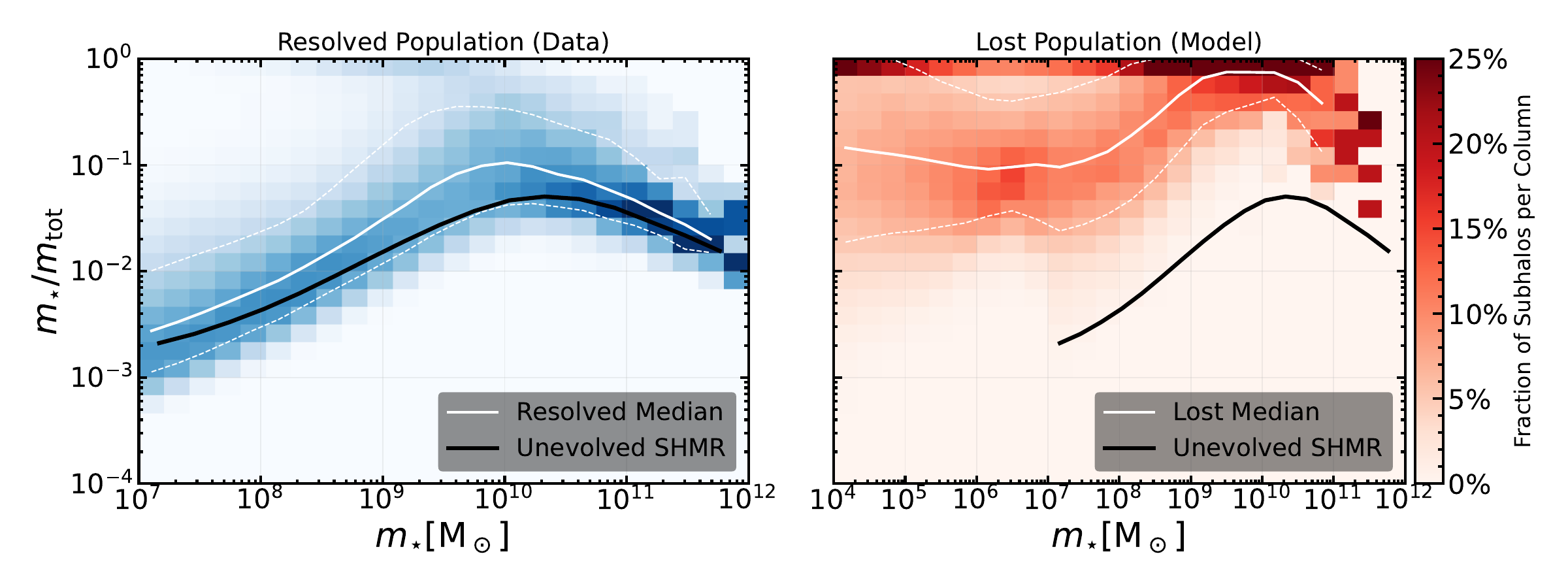}
    \caption{The stellar-to-subhalo-mass ratio evolved to the present day ($z=0$) for the resolved (left, blue) and extrapolated lost (right, red) satellite populations. The stellar masses of the resolved population are obtained directly from the simulation data, while the masses of the lost populations are inferred using the tidal track model. The y-axis shows the ratio $\xi=m_*/m_{\rm tot}$ at $z=0$. The colour map indicates a relative fraction in each $m_{*}$ bin (along the column), rather than the absolute number of satellites. The white solid lines show the median relation, while the dashed lines indicate the $16{\rm th}$–$84{\rm th}$ percentile scatter. The distributions show a significant increase in $\xi$ (i.e. moving upward) compared to the unevolved SHMR in Fig.~\ref{fig:SHMR}, as shown by the black solid lines in two panels, illustrating the ``amplification'' effect caused by tidal stripping.} 
    \label{fig:SHMR_evo}
\end{figure*}

\begin{figure}
    \centering
    \includegraphics[width=\linewidth]{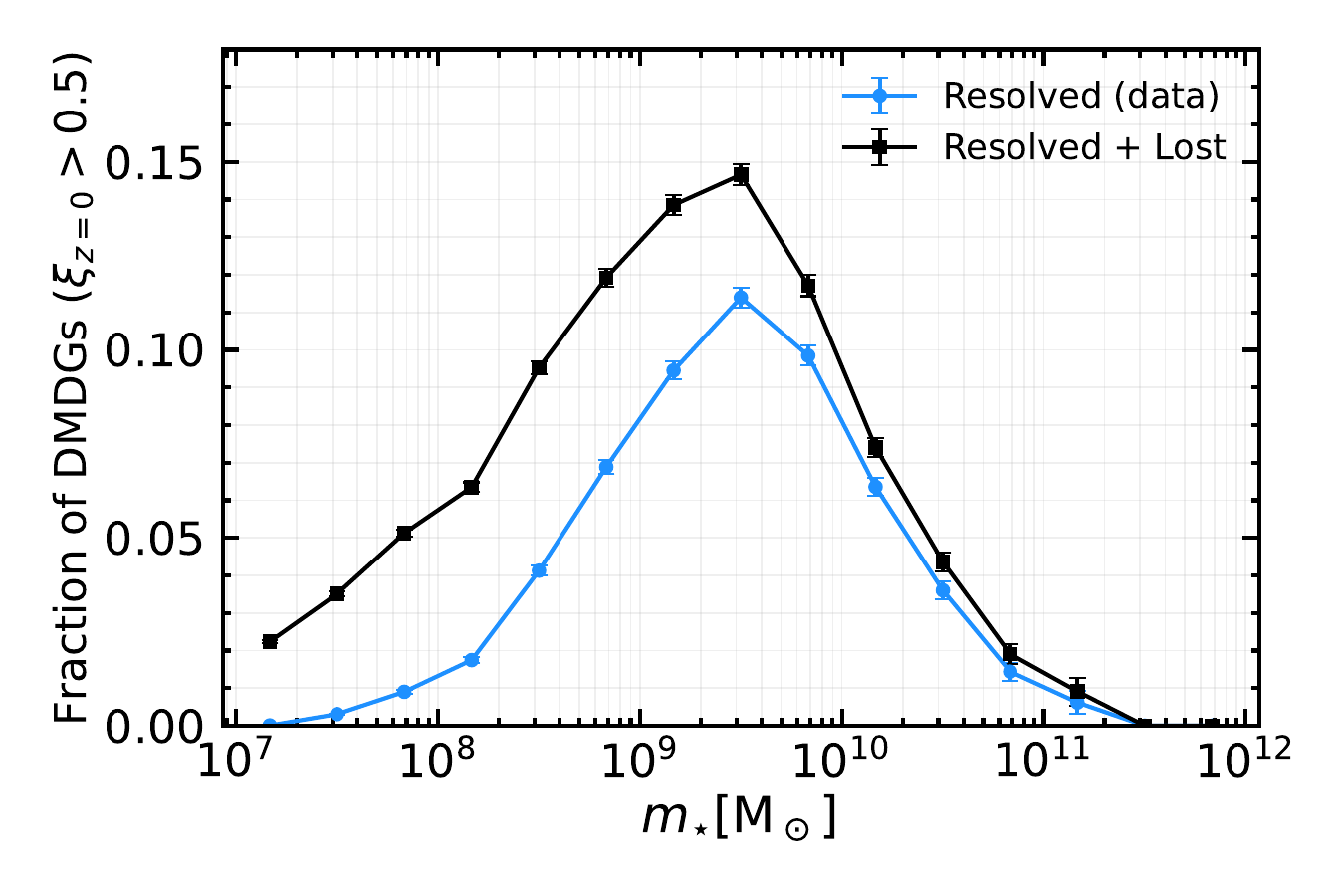}
    \caption{The fraction of DMDGs (satellites with a stellar-to-subhalo mass ratio $\xi(z=0) > 0.5$), as a function of their evolved stellar mass at $z=0$. The blue circles represent the fraction measured solely from the resolved satellite population found in the simulation. The black squares represent the total population, which combines resolved satellites with the lost population (satellites recovered via extrapolation and tidal track model). Error bars represent Poisson uncertainties. }
    \label{fig:fdmdg}
\end{figure}

Based on the context discussed in the previous subsection, the present-day stellar-to-subhalo mass ratio $\xi(z=0)$ can be expressed as a multiplicative process:
\begin{equation}
\xi(z=0) = \xi_{0} \times \mathcal{A}(f_{\rm tot}, f_{\rm d}, b),
\label{eq:xi_mult}
\end{equation}
here $\mathcal{A}$ is the tidal amplification factor determined by the stripping history and tidal track.

In Fig.~\ref{fig:SHMR}, we visualize the first ingredient of Eq.~\eqref{eq:xi_mult}: the distribution of $\xi_0$ for all satellites with $m_{*,\rm peak}>10^{7}\,\rm{M}_{\odot}$. The distribution peaks at $\xi_0 \sim 0.05$ for galaxies near $m_{\rm *, peak} \sim 10^{10.2} \,\rm{M}_{\odot}$ (corresponding to $m_{\rm tot, peak} \sim 10^{12} \,\rm{M}_{\odot}$) and drops rapidly to $\xi_0 \lesssim 10^{-3}$ for faint dwarf galaxies. This unevolved relation closely resembles the central-galaxy stellar-to-halo mass relation (SHMR) at $z=0$ (shown by the thick cyan line in Fig.~\ref{fig:SHMR}), except for a modest upward offset reflecting post-infall star formation, \added{which aligns with the difference between the red and black curves at $f_{\rm tot} \sim 1$ in Fig.~\ref{fig:xi_boost}}. We note the presence of a scattered ``cloud'' of outliers in the top-left region of the distribution, representing low-mass systems with anomalously high peak stellar fractions ($\xi_0 \gtrsim 10^{-2}$). Upon investigation, we find that these outliers originate from two rare, transient populations. The first case primarily occurs at low redshift and appears to be an algorithmic artifact of our halo finder, as these objects typically exist for only a single snapshot before disappearing. The second case occurs at high redshift and has a physical origin driven by complex mergers. During such events, a low-mass galaxy may rapidly accrete another system's baryonic components; this causes both its stellar and gas fractions to spike, producing an anomalously large stellar-to-subhalo mass ratio right before the galaxy is merged into by a larger system. Because these objects are rare and short-lived, they do not affect the overall relations of the satellite population.

Using the extrapolation procedure and the tidal-track model introduced in sections~\ref{sec:bimodal} and~\ref{sec:ad}, we model the evolved $\xi$ at $z=0$ for both the lost satellites, as shown in the right panel of Fig.~\ref{fig:SHMR_evo}. To factor out the changing number density with stellar mass, the panels show a column-normalised density (relative fraction within each $m_{*}$ bin) rather than absolute counts.

The left panel (blue) shows the directly measured results of the resolved population from simulation data, which exhibits a modest amplification; the median is moved upward about 0.5 dex relative to that in Fig.~\ref{fig:SHMR}. With the amplification of tides, we can naturally obtain a ``tail'' of the distribution reaching high $\xi\sim 1$, indicating the emergence of a population of resolved DMDGs. We simply define the population with $\xi > 0.5$ as DMDGs.

On the other hand, the right panel (red) reveals that the lost population that have fallen below the simulation resolution limit, is dominated by systems with extreme amplification factors. A large fraction of these lost satellites reach $\xi \sim 1$ (the theoretical maximum where $m_{\rm tot} \approx m_*$) just prior to or coincident with their numerical loss. This confirms that most DMDGs are often those undergoing the most severe stripping. 

We quantify the abundance of these systems in Fig.~\ref{fig:fdmdg}, which shows the fraction of satellites that are DMDGs as a function of stellar mass. \added{Here, the DMDG fraction at a given stellar mass is defined as the number of DMDGs divided by the total number of satellite galaxies within the same stellar mass bin, i.e., $f_{\rm DMDG}(m_{*}) = N_{\rm DMDG}(m_{*}) / N_{\rm sat}(m_{*})$.} As illustrated in the figure, the distribution is notably peaked, with the highest fraction of DMDGs occurring in the stellar mass range $10^{9} < m_{*} / \rm{M}_{\odot} < 10^{10} $. This mass dependence is a natural consequence of the requirement derived from Eq.~\eqref{eq:xi_mult}. 
As we discuss below for dwarf galaxies:
\begin{itemize}
    \item For massive dwarfs ($m_{*} \sim 10^{9.5} \rm{M}_{\odot}$): The initial baryon fraction is maximized near the peak of the SHMR ($\xi_0 \sim 0.05$). Consequently, a moderate amplification of $\mathcal{A} \sim 10$ is sufficient to push $\xi$ above 0.5. As shown in the previous section, such an amplification is achievable for typical tidal tracks ($b \lesssim 1$) once the subhalo loses $\sim 90\%$ of its mass.
    \item For faint dwarfs ($m_{*} < 10^{8} \rm{M}_{\odot}$): The peak stellar fraction is low ($\xi_0 < 10^{-3}$). Achieving DMDG status would require an extreme amplification of $\mathcal{A} > 500$. Most subhaloes fall below the resolution limit or become numerically disrupted long before they can achieve such high contrast between stellar and DM loss.
\end{itemize}

The black line in Fig.~\ref{fig:fdmdg} represents the total population (resolved + lost), and is systematically higher than the blue line (resolved only). This indicates that a significant number of DMDGs are transient precursors to the lost state. Because these systems reside in the extreme survival tail of tidal evolution, the DMDG population is uniquely sensitive to numerical disruption. Consequently, the resolution and the associated disruption thresholds in a simulation will undoubtedly influence the predicted DMDG abundance. We will explore this numerical sensitivity and the impact of resolution on DMDG abundance in detail in a dedicated follow-up work (He et al. in prep).

We identify a total of 8,499 resolved DMDGs in the \colibre L200m6 volume at $z=0$. \added{For cluster haloes with $M_{\rm host} >10^{14}\rm{M}_{\odot}$, we find an average of 33 DMDGs per host galaxy cluster.} \added{This abundance is substantially higher than reported in previous-generation hydrodynamical simulations, although a fully like-for-like comparison is complicated by differences in the adopted DMDG definition, host environment, and reported metric. Comparing number densities, the EAGLE simulation \citep{eagle2015} yields only $73$ such systems \citep[see also][]{Jing2019, Sifon2024}; even after accounting for the factor of eight difference in volume, \colibre predicts a DMDG number density roughly two orders of magnitude higher than EAGLE. A comparable contrast appears on a per-host basis: within The Three Hundred project, \citet{Contreras2024} identified 302 DMDGs across 324 cluster regions ($\sim1$ per cluster), against our 33 per cluster over a similar host-mass regime. Studies reporting satellite fractions show a similarly broad range: \citet{MDA2024} found that DMDGs comprise  $1-2\%$ of satellites at $z=0$ in TNG simulations, whereas \citet{Jackson2021}, adopting the less restrictive criterion $m_{\rm halo}/m_{*} < 10$ in the higher-resolution NewHorizon simulation, reported a much larger fraction $\sim 30\%$ among satellite dwarfs.} 

While the differing definitions preclude a strict one-to-one comparison, the broad trend—larger DMDG abundances in higher-resolution simulations—is itself suggestive. We interpret the high abundance in \colibre as a robust physical result, driven by both numerical and physical advances in the model: the enhanced DM mass resolution suppresses the spurious numerical disruption, while \colibre's advanced treatment of the cold ISM further promotes their formation and survival. By moving star formation away from an artificially pressurized ISM (such as the one used in EAGLE) and restricting it to the gravitationally unstable regions of the cold, molecular phase, the simulation naturally produces much more tightly self-bound stellar distributions. These highly resilient stellar cores are fundamentally better equipped to survive extreme DM stripping without undergoing complete disruption. It is expected that these DMDGs must undergo extreme tidal stripping to reach such high baryon fractions, and understanding the specific formation conditions and orbital histories that enable this state remains a key objective for future research.

\added{We note that the amplification factor $\mathcal{A}$ as defined in this work is based on the total bound mass of satellites. Observational estimates of DMDGs such as NGC-1052-DF2 and DF4 instead compare the stellar mass to the dynamical mass measured within a few effective radii, where both dark matter and stellar mass loss are expected to be smaller than in the outer regions. Bridging this gap requires dynamically modelling the structural profile and size evolution of satellite galaxies under strong tidal fields, which goes beyond the scope of the present work. DF2 and DF4 are remarkable not only for being dark matter-deficient but also for their ultra-diffuse nature.  Our results suggest an interesting tension with this observation. As shown by the tidal track model, achieving a large stellar-to-halo mass ratio requires a small $f_{\rm d}$, corresponding to a highly delayed onset of stellar stripping. In Section~\ref{sec:discussion}, we demonstrate that $f_{\rm d}$ correlates with the compactness of the stellar component relative to its host subhalo ($l_{\rm *,half}/l_{\rm vir}$): more compact satellites tend to have smaller $f_{\rm d}$ values and are therefore more likely to evolve into DMDGs. This is, however, precisely the opposite of the ultra-diffuse morphology observed in DF2 and DF4, suggesting that the formation of dark-matter-deficient ultra-diffuse galaxies through tidal stripping alone may be particularly challenging or intrinsically rare. A dedicated follow-up study within the \colibre simulation framework should systematically investigate the DF2 and DF4 analogues.}


\section{Discussion}\label{sec:discussion}

\subsection{Tidal track parameters}


\citetalias{Smith2016} found in their simulation results that galaxies of different sizes exhibit distinct efficiencies of stellar stripping. They quantified this dependence using the ratio $l_{\rm eff} / l_{\rm vir}$, where $l_{\rm eff}$ is the effective stellar radius of the satellite galaxy and $l_{\rm vir}$ is the virial radius of its host subhalo. When $l_{\rm eff}$ is relatively large, the stellar component is more extended and, like the outer DM, becomes more susceptible to tidal stripping. This corresponds to a larger value of $f_{\rm d}$ (i.e., an earlier onset of stellar loss). Conversely, when $l_{\rm eff} $ is small, the stars are concentrated near the centre of the potential well and are therefore more resistant to tidal effects. In such cases, significant stellar mass loss occurs only after substantial DM stripping, corresponding to a lower $f_{\rm d}$. By grouping galaxies according to their size and fitting the tidal tracks of each subsample, \citetalias{Smith2016} reported that $f_{\rm d} = 0.042$ for compact systems with $l_{\rm eff} / l_{\rm vir} < 0.025$; $f_{\rm d} = 0.084$ for intermediate systems with $0.025 < l_{\rm eff} / l_{\rm vir} < 0.04$; and $f_{\rm d} = 0.116$ for the most extended galaxies with $l_{\rm eff} / l_{\rm vir} > 0.04$.

In the \colibre simulation, we similarly explore the correlation between the fitted parameters and the structural properties of satellites. For clarity and rigour, we adopt the stellar half-mass radius, $l_{\rm *, half}$ normalized by the subhalo's virial radius, $l_{\rm vir}$ to characterize the stellar-to-suhbalo size ratio. We note a minor methodological difference: we measure $l_{\rm *, half} / l_{\rm vir}$ at the epoch of peak stellar mass, meaning our values may be systematically larger than those measured for purely isolated field galaxies. 

\begin{figure}
    \centering
    \begin{subfigure}[b]{0.5\textwidth}
        \includegraphics[width=\linewidth]{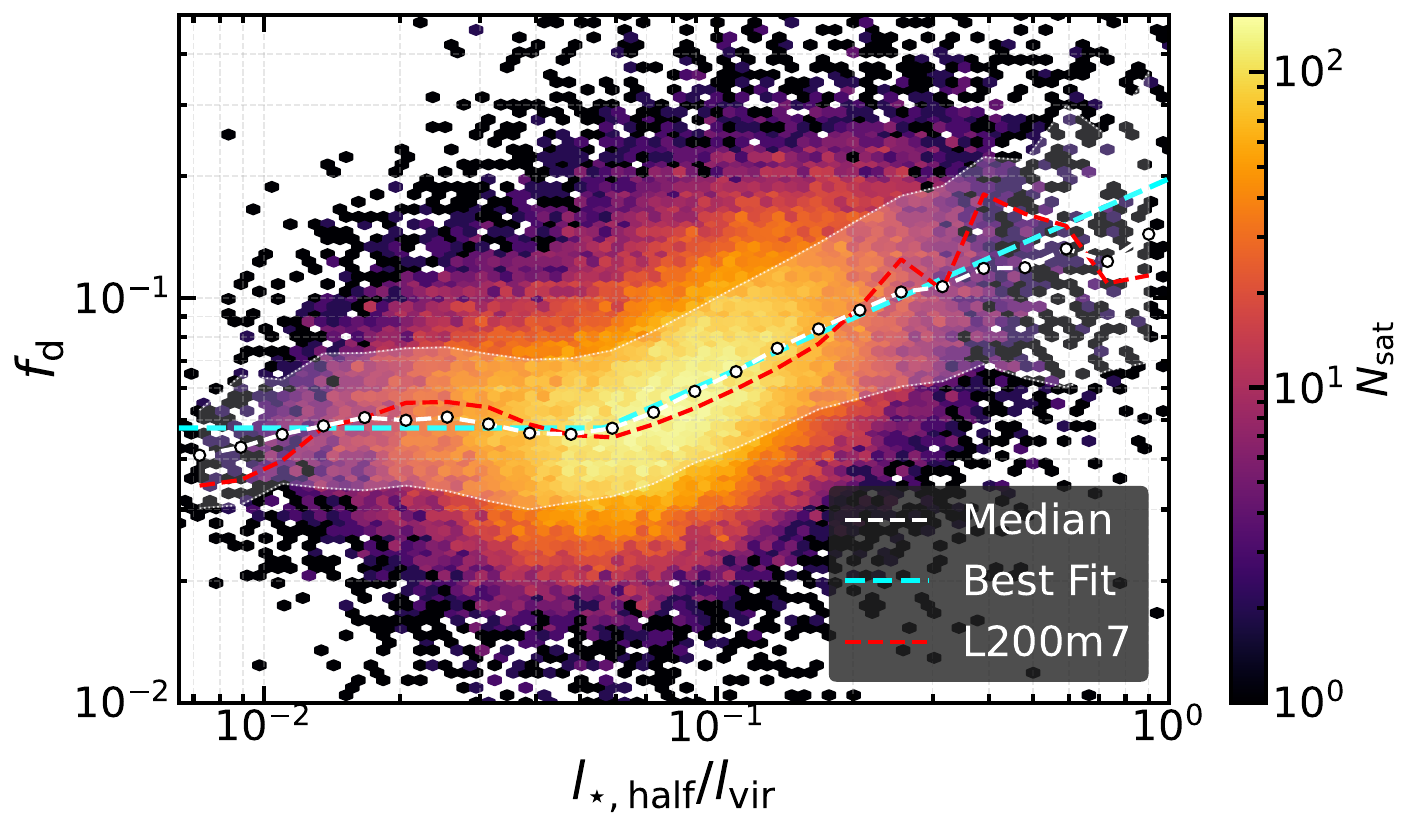}
    \end{subfigure}
    \vspace{0.5cm}
    \begin{subfigure}[b]{0.5\textwidth}
        \includegraphics[width=\linewidth]{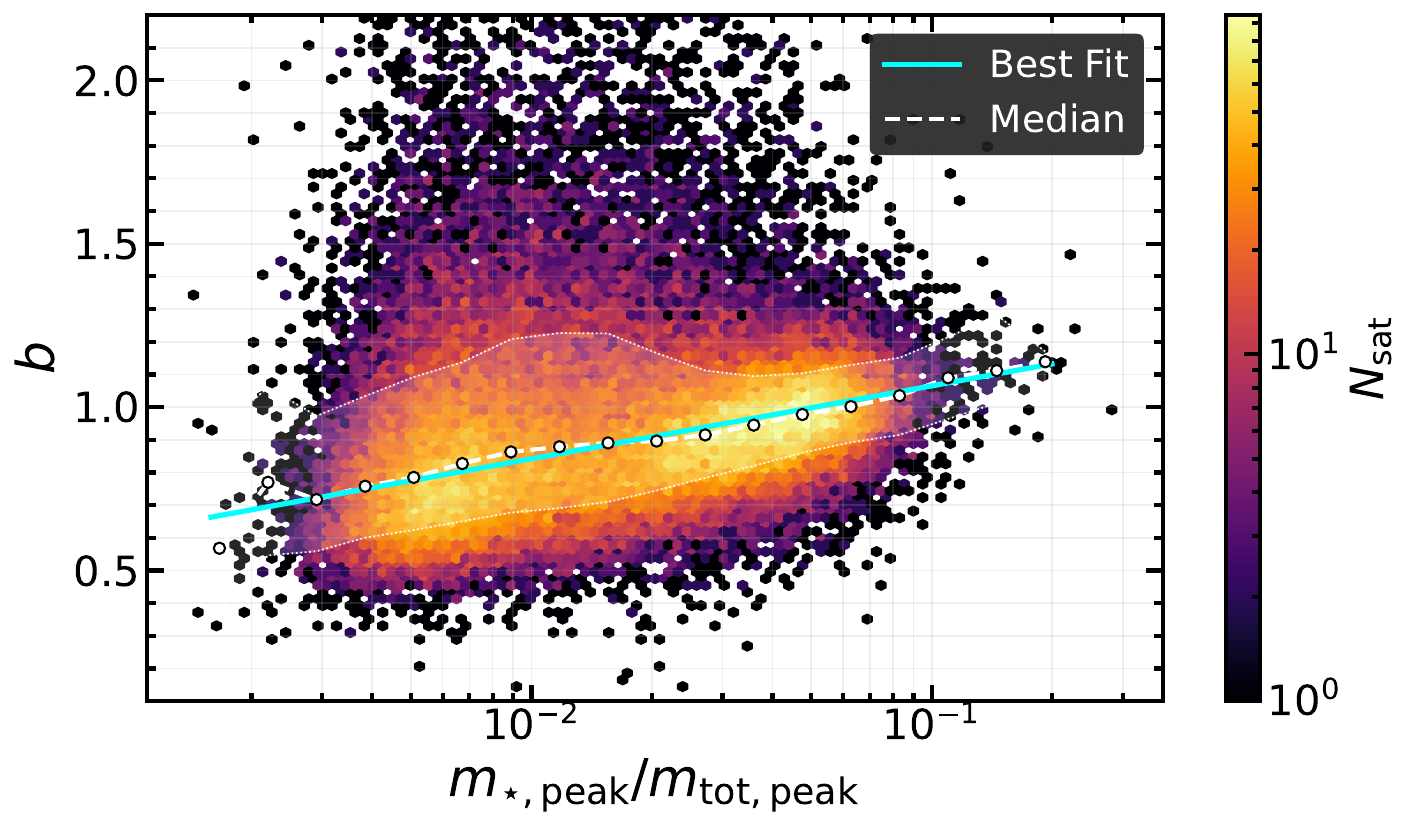}
    \end{subfigure}
    \caption{\textit{Top Panel}: The relationship of parameter $f_{\rm{d}}$ in Eq.~\eqref{eq:He26} to the stellar half-mass radius of satellites, scaled by the virial radius of their host subhaloes. The white dashed line represents the median of the distribution, and the shaded region represents the 16th-84th percentile. \added{The red dashed line shows the median relation from the lower-resolution L200m7 run.} \textit{Bottom Panel}: The relationship of parameter $b$ in Eq.~\eqref{eq:He26} to the stellar peak mass of satellites, scaled by the subhalo peak mass. The white dashed line represents the median of the distribution and the dashed region represents the 16th-84th percentile. The cyan curves are the best fittings of the median.}
    \label{fig:xi_relation}
\end{figure}

As shown in the top panel of Fig.~\ref{fig:xi_relation}, the stripping threshold $f_{\rm d} = 1/a_{\rm strip}$ exhibits a positive, albeit highly scattered, broad trend with the ratio $l_{\rm *, half} / l_{\rm vir}$. More extended galaxies tend to have larger $f_{\rm d}$ values, meaning their stars are stripped earlier, consistent with the physical picture of \citetalias{Smith2016}. The median relation can be approximated by a piecewise function:
\begin{equation}f_{\rm d} = \max \left[ 0.19\left( \frac{l_{\rm *, half}}{l_{\rm vir}} \right)^{0.48}, 0.049 \right].\end{equation}
Interestingly, the median relation flattens into a ``floor'' at $f_{\rm d} \approx 0.049$ for the most compact systems ($l_{\rm *, half} / l_{\rm vir} \lesssim 0.05$). This floor is likely a consequence of the finite spatial resolution (gravitational softening) in the simulation. For low-mass dwarfs, the softening length artificially limits the maximum central density and the depth of the potential well, preventing the stellar component from being infinitely resilient. Consequently, the simulation imposes a hard lower limit on how much subhalo mass loss is required before stellar stripping inevitably begins. 

\added{To test this interpretation, we overlay the median $f_{\rm d}-l_{*,\rm half}/l_{\rm vir}$ relation from the lower-resolution L200m7 run (red dashed line), where the sample is restricted to $m_{*,\rm peak}>8\times 10^{8}\rm{M}_{\odot}$ to ensure sufficient stellar particle resolution (see Section~\ref{sec:convergence}). For well-resolved systems at intermediate and large $l_{*,\rm half}/l_{\rm vir}$, the L200m7 median is broadly consistent with the L200m6 result. At the compact end, the relation in L200m7 develops a non-monotonic upturn at the smallest $l_{*,\rm half}/l_{\rm vir}$, indicating an unphyscial behaviour. More broadly, cosmological simulations are inherently limited in their ability to resolve the most compact satellite galaxies. The resulting sample incompleteness in this regime, combined with the artificial inflation of central stellar distributions, means that this relation for the most compact systems should be interpreted with caution and will benefit from calibration using higher-resolution simulations or controlled idealised experiments.}

For the shape parameter $b$, we do not find a strong dependence on satellite size. Instead, a weak but discernible anti-correlation emerges between $b$ and the peak stellar-to-subhalo mass ratio, $\xi_0 = m_{\ast,\rm peak}/m_{\rm tot,peak}$, as shown in the bottom panel of Fig.~\ref{fig:xi_relation}. The median of this relation can be roughly described by:
\begin{equation}b = 1.2 + 0.165 \log_{10}\left( \frac{m_{\ast,\rm peak}}{m_{\rm tot,peak}} \right).\end{equation}

\added{Physically, the parameter $b$ should be governed by the inner structural profiles of both the stellar and dark matter components. Because directly resolving inner density slopes can be limited by numerical resolution in cosmological simulations, we employ the stellar-to-subhalo mass ratio as a proxy. This ratio effectively captures how baryonic processes modify the inner structure of the satellite, thereby scaling the parameter $b$ accordingly.}

Overall, while the median trends physically align with theoretical expectations, we emphasize that both $f_{\rm d}$ and $b$ display massive scatter against these structural indicators, resulting in a relatively weak overall correlation. We note that environmental factors, such as orbital eccentricity and host potential depth, can also influence tidal stripping efficiencies, particularly for massive satellites falling into dense cluster environments \citep{HeJ2024}. However, according to idealized DMO experiments \citep{Errani2021, Errani2022}, the underlying tidal track is largely independent of orbital ellipticity. Therefore, variations in orbit and environment likely act as second-order effects that contribute to the broad scatter observed in Fig.~\ref{fig:xi_relation}, rather than driving the fundamental structural trends. In addition to environmental properties, we also examined the potential stabilizing effect of supermassive black holes (BHs) within the satellites. We find that the BH mass shows no discernible correlation with the stripping threshold $f_{\rm d}$, and exhibits only a weak correlation with the shape parameter $b$, indicating that satellites with more massive BHs tend to undergo slightly slower stellar stripping relative to dark matter in the late evolutionary phase. Generally, however, the overall impact of BHs on the tidal tracks is not significant. Establishing a tighter, quantitative relation that fully disentangles structural, environmental, and numerical effects remains challenging in a full cosmological context; we therefore propose to calibrate a more robust parameter set using a large suite of controlled satellite simulations in future work.

A complementary route to constraining satellite tidal evolution is to use observations to directly infer how stellar mass is removed as subhaloes are stripped. \citet{Wan2025} exploit JWST measurements of faint satellites of clusters at $1\lesssim z\lesssim3.5$ \citep{Suess2023} to place empirical bounds on the tidal-track parameter that controls how rapidly stellar mass is lost relative to DM. They adopt the \citetalias{Smith2016} functional form for the stellar bound fraction, but use the converted parameter $f_{1/2} = -\ln(1/2)/a_{\rm{strip}}$ in their formula. To measure $f_{1/2}$, \citet{Wan2025} employ a flexible galaxy disruption model that predicts the evolution of stellar mass as a function of the remaining DM bound fraction. By varying $f_{1/2}$ across a wide range—from highly fragile to extremely durable scenarios—they generate a suite of predicted satellite abundances. \citet{Wan2025} forward–model the observed satellite stellar mass functions using this parametrisation, marginalizing over host selection, projection/interloper effects, and an orphan-subhalo correction, and thereby derive ranges of $f_{1/2}$ that are consistent with the JWST data.

They permit a direct comparison between observationally inferred $f_{1/2}$ and the $a_{\rm strip}$ values measured in hydrodynamical simulations. \citet{Wan2025} found that the observations admit $f_{1/2}$ roughly in the range $0.015 \sim 0.09$ depending on the assumed stellar profile. Thus, the JWST-based constraints overlap the range inferred from hydrodynamic tidal tracks and span values that include the \citetalias{Smith2016} fiducial value ($f_{1/2}$=0.0488).

In this work, we find that the best fitting stripping parameters $f_{\rm d}$ for individual satellites typically lie in the $0.035 - 0.094$ range (within $1\sigma$ scatter) and that the population median is close to the \citetalias{Smith2016} value. Converting this to $f_{1/2}$ yields $f_{1/2}\sim 0.024-0.065$, which lies fully within the observationally allowed interval reported by \citet{Wan2025}.  In other words, the JWST constraints are broadly consistent with the degree of stellar resilience to stripping predicted by \colibre hydrodynamical simulation. This concordance supports the interpretation that, at least for the massive, inner satellites probed by JWST, the tidal evolution in modern hydrodynamical models is compatible with the data.

The comparison yields useful implications. First, JWST satellite counts already provide meaningful, independent support for the picture that stars are more resilient than halo outskirts (i.e. $f_{1/2} \ll 1$), and they are consistent with $f_{\rm d}$ values of the same order of magnitude as those found in \colibre. Second, the remaining degeneracies (galaxy structural assumptions, numerical systematics, and sample variance) imply that precise discrimination between tidal-track models will require (i) larger JWST samples and improved photometric redshifts, (ii) careful forward modelling of galaxy profiles and selection effects. \citet{Wan2025} already pointed the way forward by demonstrating how varying $f_{1/2}$ alters predicted satellite mass functions; combining such forward models with higher-quality JWST samples will soon tighten the allowed $f_{\rm d}(f_{1/2}) -  b$ parameter space and thus directly inform our physical understanding of satellite survivability.

\subsection{Convergence study}\label{sec:convergence}

\begin{figure}
    \centering
    \begin{subfigure}[b]{0.45\textwidth}
        \includegraphics[width=\linewidth]{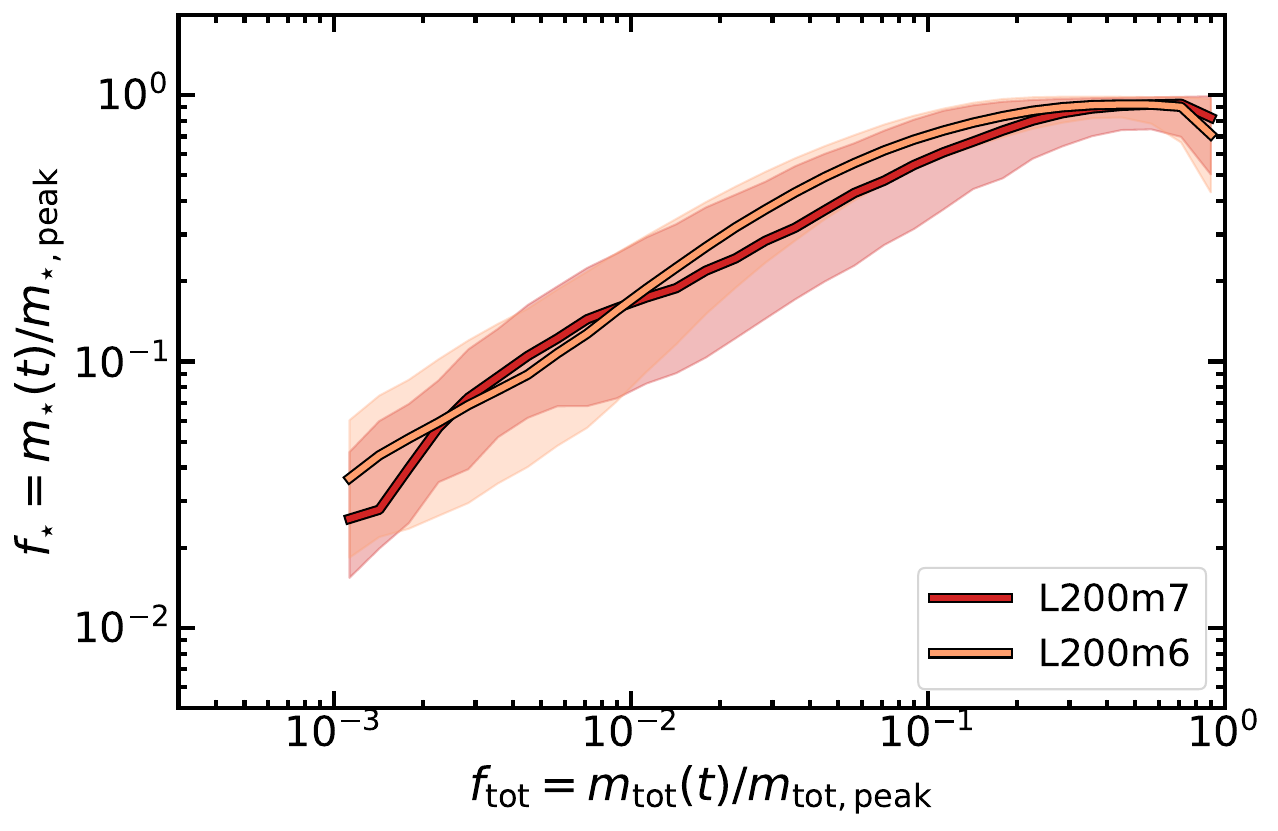}
    \end{subfigure}
    \vspace{0.cm}
    \begin{subfigure}[b]{0.45\textwidth}
        \includegraphics[width=\linewidth]{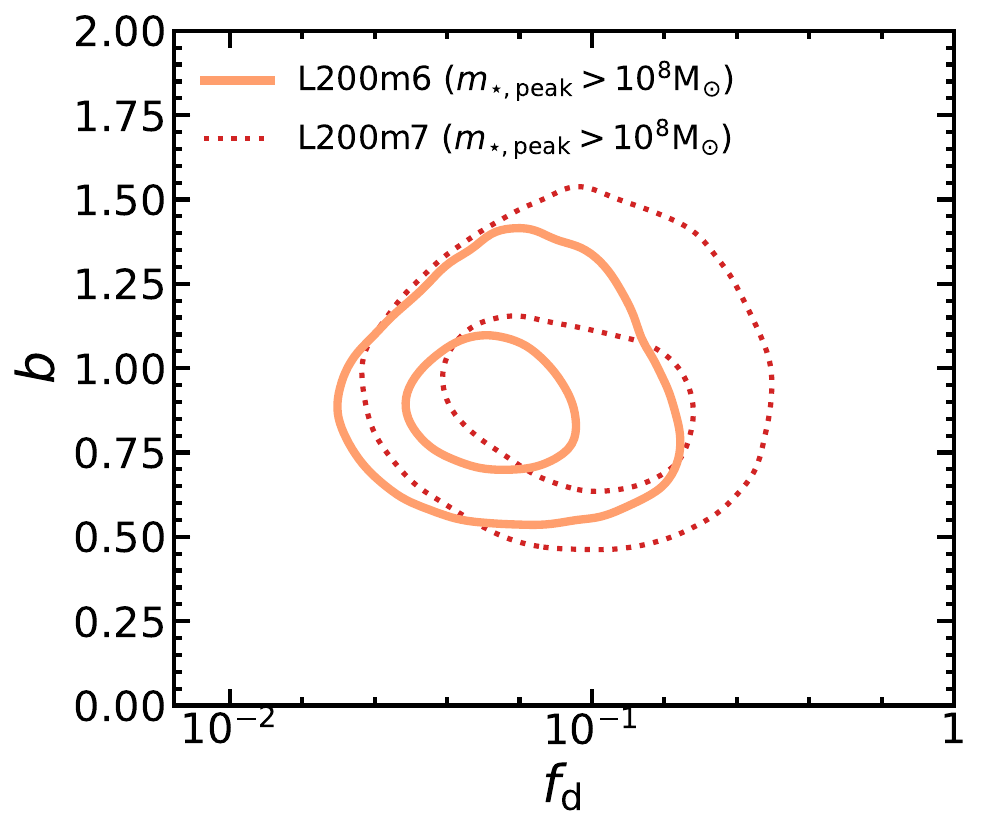}
    \end{subfigure}
    \vspace{0.5cm}
    \begin{subfigure}[b]{0.45\textwidth}
        \includegraphics[width=\linewidth]{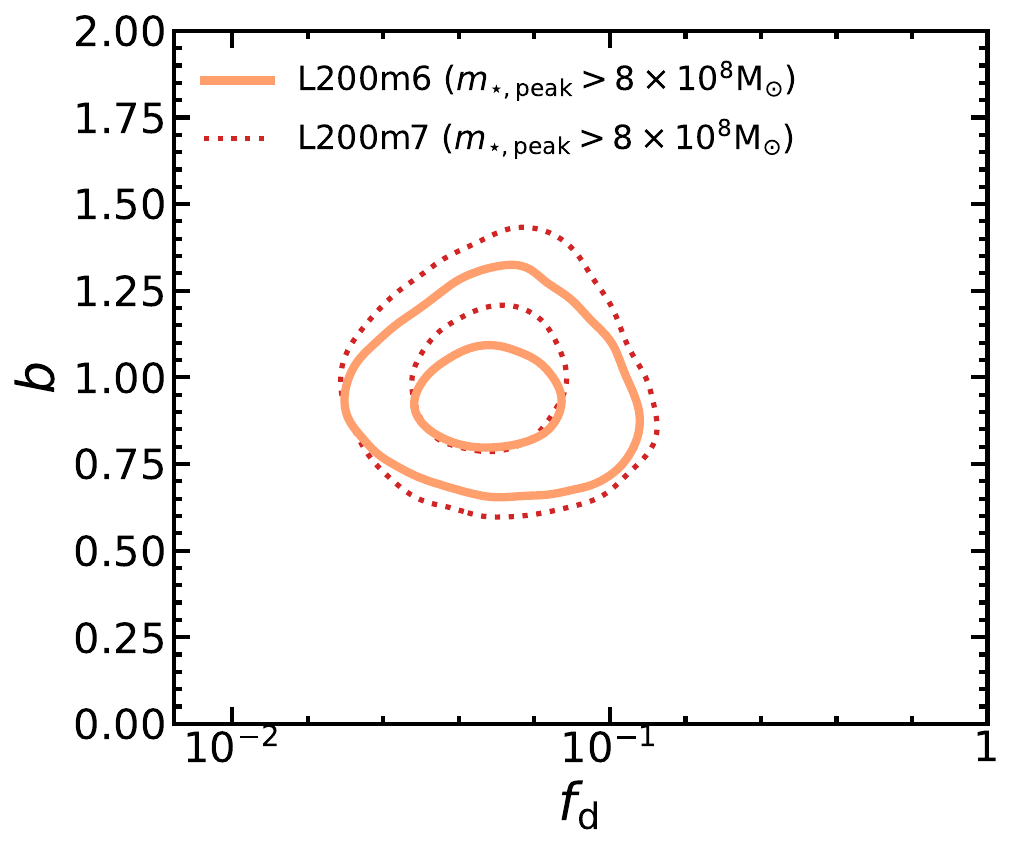}
    \end{subfigure}    
    \caption{Resolution convergence test for satellite galaxies in the L200 volume. \textit{Top Panel}: The median stellar mass fraction $f_*$ as a function of the subhalo mass fraction $f_{\rm tot}$ for the L200m6 (orange) and L200m7 (red) runs. The shaded areas correspond to the 16th–84th percentile scatter. \textit{Middle Panel}: Comparison of the joint probability distributions of the best-fitting tidal track parameters $f_{\rm d}$ and $b$ for the L200m6 (solid orange contours) and L200m7 (dotted red contours) simulations, using an identical mass selection threshold ($m_{\rm *,peak} > 10^8 \, \rm{M}_{\odot}$). The contours represent the $1\sigma$ and $2\sigma$ levels of the distributions. \textit{Bottom Panel}: Same as the middle panel, but the minimum stellar peak mass threshold run has been increased by a factor of eight (to $m_{\rm *,peak} > 8 \times 10^8 \, \rm{M}_{\odot}$) to ensure selected galaxies in L200m7 have sufficient particle number.}
    \label{fig:L200_convergence}
\end{figure}

\begin{figure}
    \centering
    \begin{subfigure}[b]{0.45\textwidth}
        \includegraphics[width=\linewidth]{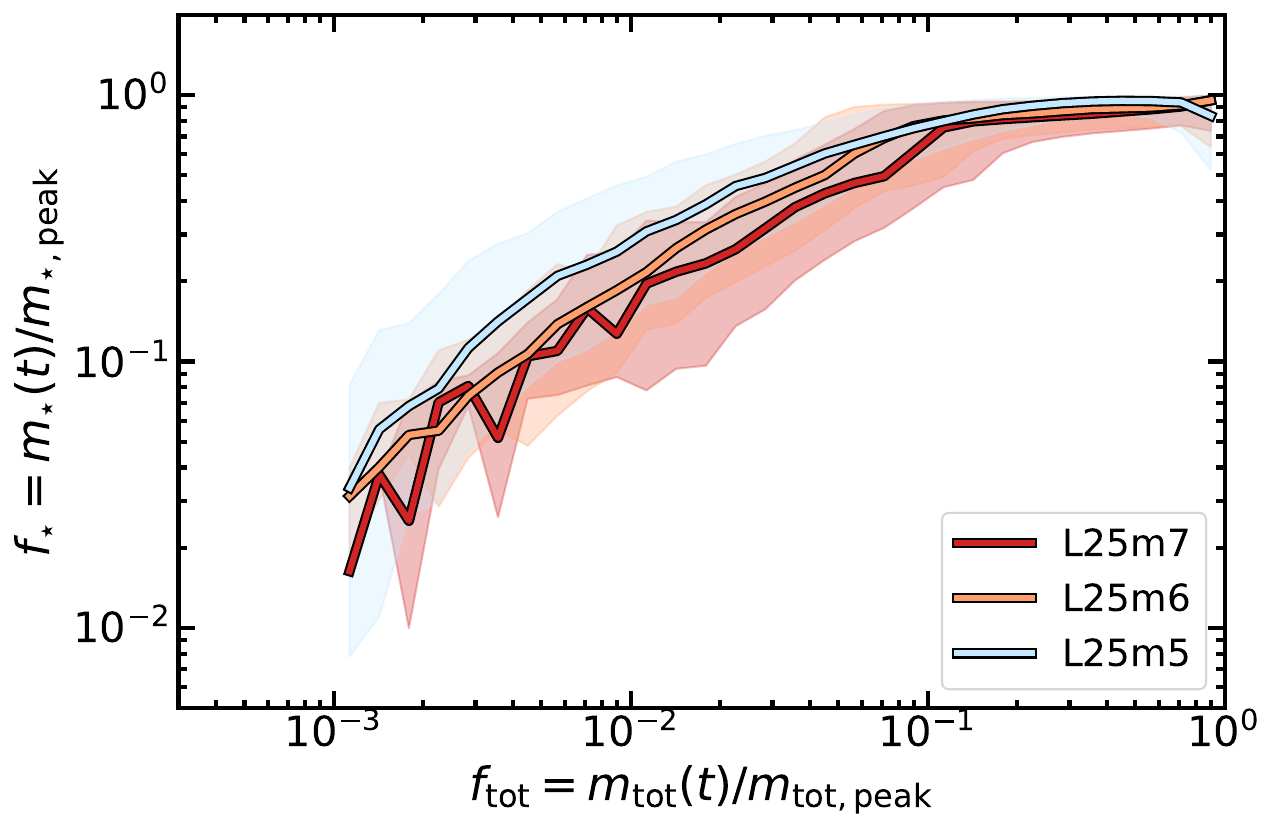}
    \end{subfigure}
    \vspace{0.5cm}
    \begin{subfigure}[b]{0.45\textwidth}
        \includegraphics[width=\linewidth]{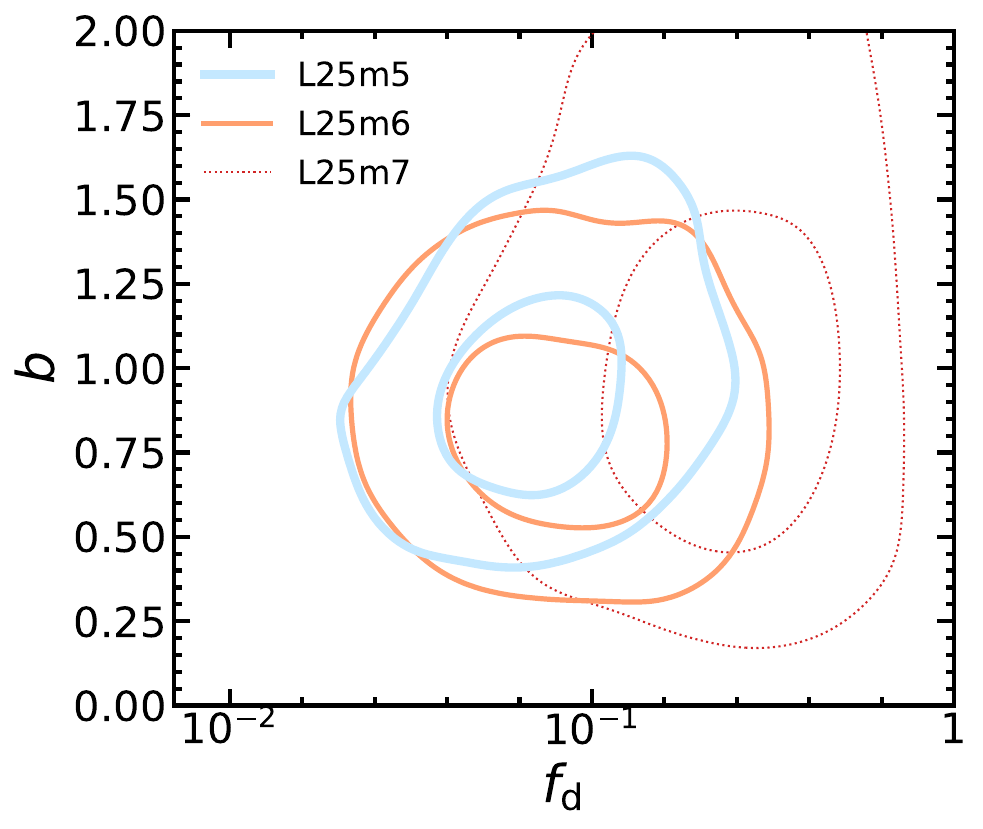}
    \end{subfigure}
    \caption{Same as Fig.~\ref{fig:L200_convergence}, but for the L25 volume, comparing the L25m5 (blue), L25m6 (orange), and L25m7 (red) runs. \textit{Top Panel}: The median stellar evolutionary tracks for satellite galaxies across the three resolution levels. The shaded regions show the $16{\text{th}}$–$84{\text{th}}$ percentiles. \textit{Bottom Panel}: The joint distributions of the tidal track parameters $b$ and $f_{\mathrm{d}}$. The solid blue, solid orange, and dotted red contours correspond to the L25m5, L25m6, and L25m7 runs, respectively, indicating the $1\sigma$ and $2\sigma$ levels of the distributions.}
    \label{fig:L25_convergence}
\end{figure}

To ensure the robustness of our results, it is essential to perform a resolution convergence study. However, comparing the tidal evolution of satellite galaxies across different resolution levels is non-trivial. In cosmological hydrodynamical simulations, the stochastic nature of star formation and feedback \citep{Borrow2023} and stellar feedback implies that the exact formation epoch, internal structural evolution, and subsequent orbital trajectory of a specific galaxy can diverge significantly across varying resolutions, even when the phases of the initial conditions are the same \citep{Lovell2025}. Consequently, performing a strict one-to-one matching of individual satellites is highly challenging and often misleading. Instead, we evaluate numerical convergence by comparing the overall statistical distributions of the tidal track parameters for satellite populations selected using identical physical criteria.

We first compare our fiducial L200m6 run with its lower-resolution counterpart, L200m7. Applying the identical selection criteria used in Section~\ref{sec:satellites} (including the stellar peak mass threshold $m_{\rm *,peak} > 10^8 \, \rm{M}_{\odot}$), we present the resolution comparison in Fig.~\ref{fig:L200_convergence}. The top panel displays the median stellar tidal tracks of m6 and m7 (the m6 results have been shown in Fig.~\ref{fig:tidaltrack}). Overall, the tracks are broadly consistent between the two resolutions. To quantitatively assess the differences, the \deleted{bottom} \added{middle} panel compares the joint probability distributions of the best-fitting track parameters, $f_{\rm d}$ and $b$. Unlike the 2D density histogram used previously in Fig.~\ref{fig:ab_distribution}, here we display the $1\sigma$ and $2\sigma$ contour levels to facilitate a clearer cross-resolution comparison. The contours reveal a slight shift: the $f_{\rm d}$ distribution for the L200m7 run is broader and shifted toward higher values compared to L200m6. Physically, a higher $f_{\rm d}$ indicates that the stellar component begins to be stripped earlier (i.e., requiring less subhalo mass loss). This behavior is expected in lower-resolution simulations, where larger gravitational softening lengths and lower particle counts artificially ``puff up'' the central stellar distribution, making it less tightly bound and more susceptible to tidal stripping. \added{To test this interpretation, we perform a supplementary convergence test in the bottom panel of Fig.~\ref{fig:L200_convergence}. Here, we raise the minimum stellar peak mass threshold in both runs to $m_{\rm *,peak} > 8 \times 10^8 \, \rm{M}_{\odot}$, ensuring that all selected galaxies are well above the resolution limit in each simulation. With this stricter selection, the $f_{\rm d}$ and $b$ distributions for L200m6 and L200m7 are in significantly better agreement, with overlapping $1\sigma$ contours. We have also verified that an intermediate test—raising only the L200m7 threshold to $m_{\rm *,peak} > 8 \times 10^8 \, \rm{M}_{\odot}$ to match the minimum star particle count of our fiducial L200m6 selection—yields a comparable degree of convergence. Together, these tests confirm that the tidal track parameters are robust and converge well when galaxies are resolved with a sufficient number of particles, and that the residual discrepancy under our fiducial selection is driven by the poorly-resolved tail of the L200m7 sample.} 

To further test convergence at higher resolutions, we turn to the smaller L25 ($25 \, \rm{cMpc}$) volume, which includes runs at the m5, m6, and m7 resolution levels. Because the L25 volume contains far fewer massive satellites, we relax the minimum stellar peak mass threshold to $m_{\rm *,peak} > 10^7 \, \rm{M}_{\odot}$ to ensure a sufficiently large statistical sample for this comparison. We note that matching the particle counts by scaling the mass thresholds, as done in the L200 bottom panel, is not feasible here because applying a higher mass cut in the L25m7 run would leave too few satellites to construct robust statistical contours. The results for the matched mass threshold are shown in Fig.~\ref{fig:L25_convergence}. As shown in the top panel, the median tracks for the three resolution levels remain largely comparable. The comparison clearly demonstrates that the m5 (blue) and m6 (orange) runs are highly consistent, with their $f_{\rm d}$ and $b$ contours largely overlapping. In contrast, the m7 (red) distribution is significantly offset to the right. This divergence is more pronounced in the L25 sample than in the L200 sample because the relaxed mass cut introduces more poorly-resolved dwarf galaxies into the m7 bin, amplifying the aforementioned resolution effects. 

Overall, while absolute convergence is notoriously difficult to achieve in galaxy formation models due to the resolution dependence of subgrid physics, these tests demonstrate that the dependence of the tidal track parameters on resolution is relatively mild once a sufficient resolution threshold is reached. The good convergence between the m5 and m6 tracks gives us confidence that the stellar stripping efficiencies derived from our fiducial L200m6 simulation provide a fairly reliable representation of satellite tidal evolution.

\subsection{Gas evolution track}

\begin{figure}
    \centering
    \includegraphics[width=\linewidth]{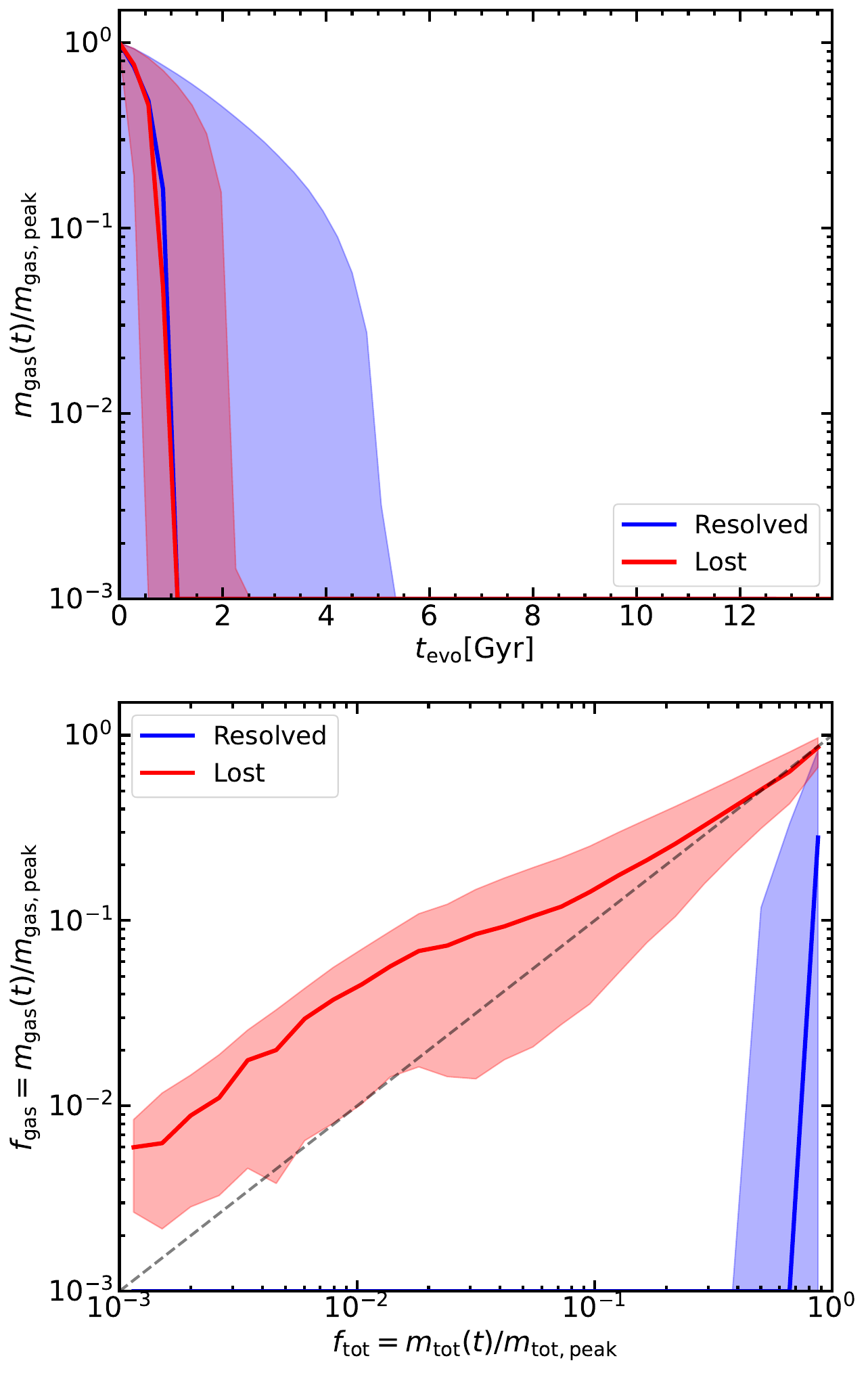}
    \caption{The evolution of gas mass for the same sample of satellites shown in Fig.~\ref{fig:tidaltrack}. Blue and red colours correspond to resolved and lost populations, respectively. Shaded regions indicate the $16{\rm th}$–$84{\rm th}$ percentile scatter. Top panel: The remaining gas mass fraction, $m_{\rm gas}(t)/m_{\rm gas, peak}$, as a function of evolution time $t_{\rm{evo}}$ since $t_{\rm{tot,peak}}$. Gas removal is rapid for both populations, with median tracks showing complete depletion within $\sim 1$ Gyr, significantly faster than the stripping of DM or stars.
    Bottom panel: The gas evolution track, showing the fractional gas mass loss versus the fractional subhalo mass loss. The dashed diagonal line represents a scenario where gas and DM are stripped at identical rates. While resolved satellites (blue) lose their gas long before experiencing significant subhalo mass loss, the lost population (red) follows the diagonal more closely, indicating that for these systems, gas and DM removal are driven primarily by the same strong tidal forces.}
    \label{fig:gasloss}
\end{figure}

Analogous to the stellar tidal tracks discussed in Section~\ref{sec:tidaltrack}, we examine here the evolution of the gas mass of satellites as a function of time and as a function of the remaining bound mass. In this context, the gas mass is defined as the total mass of all self-bound gas associated with the subhalo, including all phases (ionized, atomic, and molecular). Fig.~\ref{fig:gasloss} shows these trends for the same satellite sample used in Fig.~\ref{fig:tidaltrack}. As already seen in the case study of Fig.~\ref{fig:mass_evolution_example}, gas is typically removed very rapidly after infall: most satellites lose the bulk of their gas within a few gigayears, at a rate generally faster than that of either the DM or the stars.

In contrast to the bimodal behaviour observed for subhalo and stellar mass loss, the gas-loss histories of the resolved and lost satellite populations are remarkably similar in their median trends. Both populations typically lose all of their gas within $\sim1$ Gyr after infall. The main difference lies in the scatter: resolved satellites show a broader range of gas-survival times, with the 84th percentile reaching $\sim5.5$ Gyr, compared with $\sim2.5$ Gyr for lost satellites.

This combination of a bimodal pattern in subhalo stripping but a uniform pattern in gas removal naturally leads to divergent gas evolution tracks for resolved and lost satellites. This behaviour contrasts with the near-universal stellar tidal track found in Section~\ref{sec:tidaltrack}. As shown in the bottom row of Fig.~\ref{fig:gasloss}, the resolved satellites typically lose all of their gas even when their total subhalo mass has barely been stripped. In other words, gas is preferentially and efficiently removed early, before any substantial DM mass loss occurs.

Conversely, the gas evolution tracks of the lost satellites (red lines) lie close to the diagonal $m_{\rm gas}/m_{\rm gas,peak} \approx m_{\rm tot}/m_{\rm tot,peak}$. This indicates that the gas mass decreases at a rate comparable to the DM mass loss caused by tides. In other words, gas removal for these heavily stripped satellites is dominated by tidal forces rather than by stellar feedback or ram pressure stripping. 

One potential source of confusion is that many resolved (i.e. surviving) satellites lose their gas early and typically before any significant DM stripping, by processes that are not primarily tidal in origin. Early gas removal reduces the central potential and lowers the inner DM density, rendering the subhalo structure more diffuse and, in principle, more susceptible to tidal forces. However, our results indicate that this weakened central binding does not generally lead to immediate or inevitable subhalo loss. 

By contrast, the lost subhalo population appears to be dominated by strong, orbit-driven tidal events (deep pericentric passages, pre-processing in more massive progenitor hosts, or rapid growth of the host potential) that remove mass from the system on a short timescale and lead to wholesale withering. 
This interpretation is consistent with previous comparisons between hydrodynamical and DMO runs:
hydrodynamical satellites do lose mass faster and show higher disruption (lost) rates than their DMO counterparts (the baryonic processes change internal structure), but the most extreme cases of stripping are still associated with strong tidal encounters \citep[e.g.][]{Bahe2019, Barry2023, Lovell2025}. Quantifying the relative contributions of gas removal, internal structural change, and orbital history to net disruption remains an important goal for future work.

\section{Conclusions and Summary}\label{sec:summary}



In this work, we have investigated the co-evolution of satellite galaxies and their host subhaloes under tidal forces using the state-of-the-art cosmological hydrodynamical simulation \colibre L200m6 run. Our primary focus was to establish a robust link between stellar mass loss and subhalo mass loss
and understand their ultimate fates. 

Our main findings are summarized as follows:

\begin{enumerate}


\item A Universal Two-Phase Tidal Track: we identify a universal relation that tightly couples the stellar mass fraction of satellites ($f_{\rm{*}}=m_{\rm{*}}(t)/m_{\rm{*,peak}}$) to the remaining bound mass fraction of their host subhaloes ($f_{\rm tot}=m_{\rm{tot}}(t)/m_{\rm{tot,peak}}$) (see Fig.~\ref{fig:tidaltrack}). This evolution proceeds in two distinct phases: an initial phase of preferential DM stripping during which the stellar component remains largely intact, followed by a coupled stripping phase triggered when the subhalo mass fraction falls below a characteristic threshold ($f_{\rm tot} \sim f_{\rm d}=1/a_{\rm strip}$). We accurately describe this behaviour using a refined empirical relation (Eq.~\eqref{eq:He26}) parameterized by the stripping threshold, $f_{\rm d}$, and a shape parameter, $b$, both of which follow independent log-normal distributions (see Table~\ref{tab:track_parameters} and Fig.~\ref{fig:ab_distribution}). 
This flexible formulation provides an improvement over the fixed-shape model of \cite{Smith2016} and is in good agreement with the analytical predictions of \cite{Errani2022}.

\item Bimodal Fate: the fates of subhaloes exhibit a bimodal distribution, which is linked to their accretion history. Subhaloes accreted early via hierarchical assembly experience extreme tidal mass loss and are often withered ( ``lost'' population). In contrast, those accreted later via direct accretion suffer more modest mass loss and typically survive as ``resolved'' systems (Fig.~\ref{fig:massloss}). The universal tidal track maps the bimodal distribution of subhalo mass loss into stellar mass loss. The disruption of subhaloes also leads to the disruption of embedded galaxies.
\item Artificial Disruption: using the extrapolation of the resolved subhalo mass loss rate as well as the tidal track model, we estimate the potential influence of artificial disruption on satellite subhalo and stellar mass function at $z=0$. Numerical disruption has a negligible effect on the amplitude of the subhalo mass function but a marginal effect on the stellar mass function ($m_{*} >10^{8}\rm{M}_{\odot}$) (Fig.~\ref{fig:ad}). 
\item Formation of Dark-Matter-Deficient-Galaxies (DMDGs): the tidal track model allows us to predict the evolution of the stellar-to-subhalo mass ratio $\xi \equiv m_{\rm{*}}/m_{\rm{tot}}$. We find that the emergence of DMDGs ($\xi > 1/2$) is not guaranteed and depends critically on the parameters $f_{\rm d}$ and $b$, as well as the peak stellar mass fraction $\xi_0=m_{\rm *,peak}/m_{\rm tot,peak}$ (Fig.~\ref{fig:SHMR_evo}). Only satellites with sufficiently high $\xi_0$ can evolve into the DMDG regime before potential disruption. The fraction of DMDGs peaks in the satellite stellar mass range $10^{9}-10^{10}\rm{M}_{\odot}$. We report 8499 DMDGs among the resolved satellites in the colibre (200 cMpc)$^{3}$ volume simulation at m6 resolution. This number is much higher than that in previous generations of hydrodynamical cosmological simulations. 
\item Structural Dependence: the tidal stripping parameter $f_{\rm d}$ shows a correlation with the satellite's compactness, specifically the ratio of the stellar effective radius to the subhalo's virial radius. More compact stellar distributions are more resilient to stripping, corresponding to lower values of $f_{\rm d}$.

\end{enumerate}

Our results provide a comprehensive and quantitative description of the evolution of satellite galaxies in tidal fields within the \colibre simulations. The established tidal track maps the bimodal fates of subhaloes onto the embedded satellites, demonstrating that the disruption of a subhalo ultimately leads to the disruption of its galaxy. The model and findings presented here offer critical insights for interpreting the satellite galaxy population, including their spatial distribution and luminosity function. The model can also be directly employed by semi-analytical models of galaxy evolution to produce more physical and accurate predictions for satellite galaxies. 

In future work, we will focus on incorporating these tidal evolution recipes into a complete model that spans the full range of galaxy structure parameters and further dissecting the interplay between baryonic feedback and tidal stripping in shaping the final fates of satellite galaxies.





\section*{Acknowledgements}


The authors warmly thank Julio Navarro for his insightful comments which improved the physical discussion of this work. This work is supported by NSFC (12595312), National Key R \& D Program of China (2023YFA1607800, 2023YFA1607801), China Manned Space Program (CMS-CSST-2025-A04), 111 project (No. B20019), and Office of Science and Technology, Shanghai Municipal Government (grant Nos. 24DX1400100, ZJ2023-ZD-001). Z.Z.L. acknowledges the NJU Double First Class funding and the Fundamental Research Funds for the Central Universities (KG202502). FH and EC acknowledge funding from the Netherlands Organization for Scientific Research (NWO) through research programme Athena 184.034.002. SP acknowledges support by the Austrian Science Fund (FWF) through grant-DOI: 10.55776/V982. JT acknowledges support of a STFC Early Stage Research and Development grant (ST/X004651/1). ABL acknowledges support by the Italian Ministry for Universities (MUR) program “Dipartimenti di Eccellenza 2023-2027” within the Centro Bicocca di Cosmologia Quantitativa (BiCoQ), and support by UNIMIB’s Fondo Di Ateneo Quota Competitiva (project 2024-ATEQC-0050). 

\section*{Data Availability}
The data underlying this article were produced as part of the \colibre project. The simulation data and halo catalogues will eventually be made publicly available as part of the collaboration's phased data-release programme. Until then, researchers wishing to use the simulations may contact the \colibre team. Further information on the project is available at \url{https://colibre.strw.leidenuniv.nl/}. The specific data supporting the plots and results presented in this work will be shared upon reasonable request to the corresponding author.
 



\bibliographystyle{mnras}
\bibliography{example} 




\appendix

\section{Comparison with Errani22 model}\label{EN22}


Based on idealized numerical simulations, \citet{Errani2022} (hereafter \citetalias{Errani2022}) derived a stellar-subhalo mass loss relationship by tracing stellar particles using massless tracers. In this appendix, we compare our results with those of \citetalias{Errani2022}.

\citetalias{Errani2022} demonstrated that the tidal mass stripping of subhaloes can be effectively characterized in energy space. By defining the normalized particle energy as:

\begin{equation}
    \mathcal{E}=1 - E/\Phi_0,
\end{equation}

Here, $E= E_{\rm{k}}+\Phi$ represents the binding energy, where $\Phi$ is the potential from the subhalo itself, and $\Phi_0$ denotes the minimum of the potential well. The lower the value of $\mathcal{E}$,  the more tightly bound the particle is. Once $\mathcal{E} > 1$, the particle will no longer be gravitationally bound to the subhalo.

They found that tidal forces preferentially strip particles in higher energy states. \citetalias{Errani2022} analyzed the energy distribution of initially bound particles that remained after different orbital periods of tidal stripping. As stripping progresses, the surviving particles become increasingly concentrated in the most tightly bound states. \citetalias{Errani2022} established that the energy distribution of remaining particles follows a specific relationship with mass loss:

\begin{equation}
    \mathrm{d}N/\mathrm{d}\mathcal{E}|_{\rm{i,t}}=\frac{\mathrm{d}N/\mathrm{d}\mathcal{E}|_{\rm{i}} }{1 + (p\mathcal{E}/\mathcal{E}_{\rm{mx,t}})^q},
    \label{eq:e22}
\end{equation}
Here, $\mathrm{d}N/\mathrm{d}\mathcal{E}|_{\rm{i}}$ represents the initial energy distribution of the bound particles. The values are as followed: $p=0.85$, $q=12$, $\mathcal{E}_{\rm{mx,t}}\approx0.77f_{\rm tot}^{0.43} $.

The stellar particles in subhaloes, similar to DM particles, orbit within the gravitational potential of the subhalo. Therefore, the relationship described in Eq.~\eqref{eq:e22} can also be applied to the evolution of stars of satellite galaxies. \citetalias{Errani2022} assumed that the stellar initial binding energy distribution of a dwarf satellite galaxy can be well fitted by the following function:

\begin{equation}
\frac{\mathrm{d}N_{*}}{\mathrm{d}\mathcal{E}}=\mathcal{E}^{\alpha}\exp(-(\mathcal{E}/\mathcal{E}_{\rm{s}})^{\beta}).
\label{eq:star_energy}
\end{equation}

This distribution exhibits a power-law slope of $\alpha$ toward the most bound states and is exponentially truncated at higher energies. The distribution peaks at $\mathcal{E}{*}=\mathcal{E}_{\rm{s}}(\alpha/\beta)^{1/\beta}$. For a dwarf galaxy with a spherical exponential density profile, it can be turned into that $\alpha=\beta=3$.

By applying the energy-space evolutionary relationship from Eq.~\eqref{eq:e22} to the initial stellar energy distribution, one can derive the distribution of still-bound stellar particles at different stages of subhalo mass loss. Naturally, when the subhalo experiences only mild stripping, tidal forces have little effect on the satellite galaxy because the stellar particle distribution is truncated at high energies. However, when the subhalo undergoes significant stripping, stellar and DM particles are sequentially removed from high to low energy states following the same rule.

Notably, for satellite galaxies with an exponential density profile, the stellar energy-space distribution has a steeper power-law index, resulting in fewer stars in the most bound states near the subhalo centre. Consequently, these stars are more susceptible to tidal stripping compared to DM particles, which dominate the central regions. This difference in distribution leads to distinct evolutionary pathways for the stellar and DM components under tidal forces.

\begin{figure*}
    \centering
    \begin{subfigure}[b]{0.45\textwidth}
        \includegraphics[width=\textwidth]{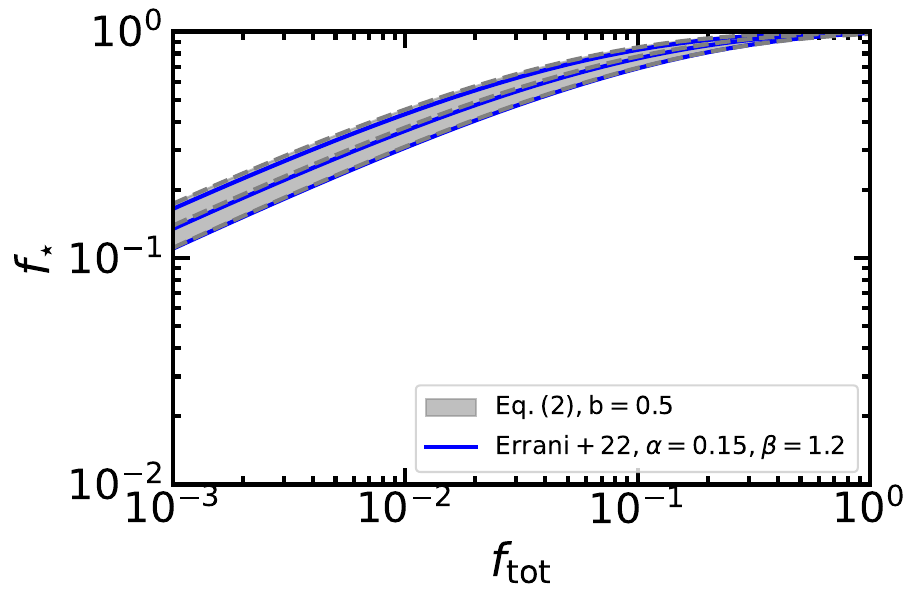}
    \end{subfigure}
    \hfill
    \begin{subfigure}[b]{0.45\textwidth}
        \includegraphics[width=\textwidth]{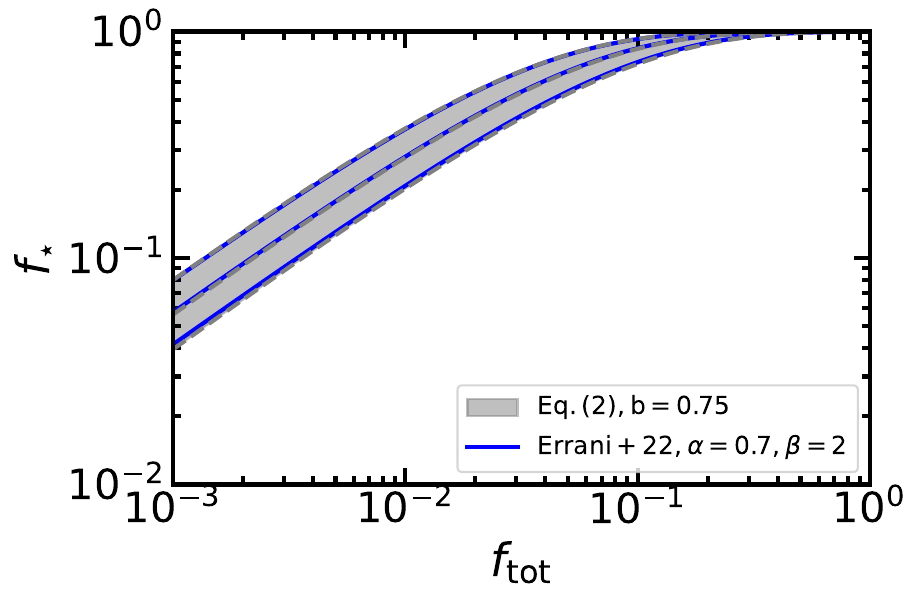}
    \end{subfigure}
    
    \vspace{0.5cm} 
    
    \begin{subfigure}[b]{0.45\textwidth}
        \includegraphics[width=\textwidth]{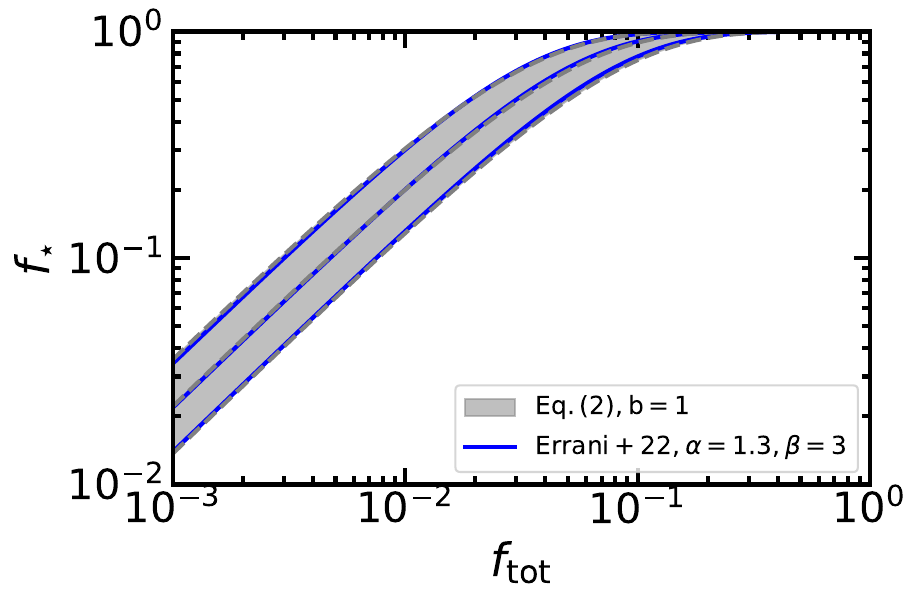}
    \end{subfigure}
    \hfill
    \begin{subfigure}[b]{0.45\textwidth}
        \includegraphics[width=\textwidth]{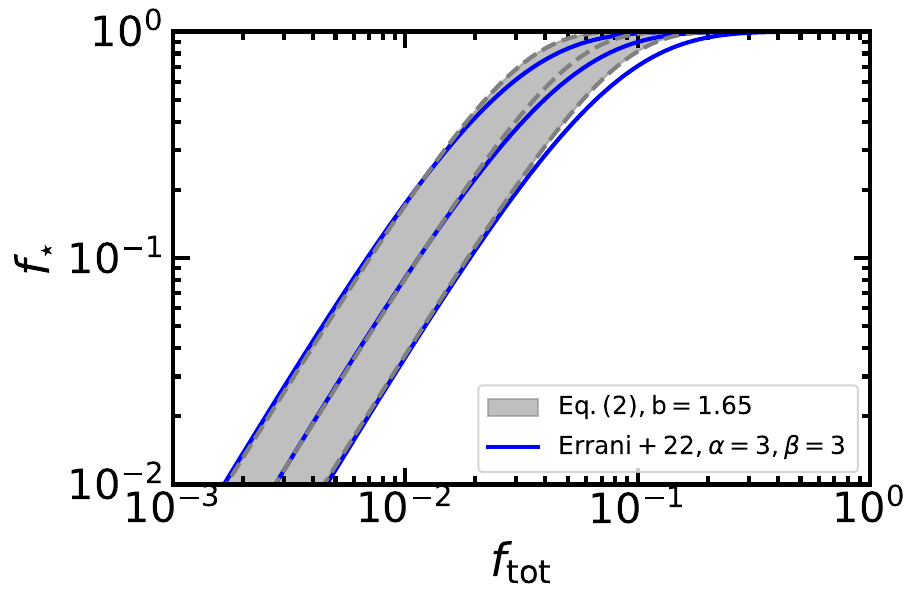}
    \end{subfigure}
    \caption{Comparison with the \citetalias{Errani2022} model. In each panel, grey dashed lines and the enclosed region show the results of Eq.~\eqref{eq:He26}. The blue lines are an analytical calculation of \citetalias{Errani2022} model with fixed parameters $\alpha$ and $\beta$ given in the legend, varying $\mathcal{E}_{\rm s}$ for three blue lines. The bottom-right panel shows the fiducial case in \citetalias{Errani2022}.}
    \label{fig:comp_e22}
\end{figure*}

In Fig.~\ref{fig:comp_e22}, we compare our results extracted from the cosmological simulation with the \citetalias{Errani2022} model. In each panel, we plot three grey dashed lines to show the median $f_{\rm d}$ along with the $\pm 1\sigma$ scatter, \added{(as listed in Table~\ref{tab:track_parameters})} . In the top left, top right and bottom left panels, the grey dashed lines are tidal tracks with parameter $b=0.5,0.75$ and $1$. If we choose the correct stellar energy profile, the tidal track predicted by \citetalias{Errani2022} model matches the tidal track we proposed in Eq.~\eqref{eq:He26}. Specifically, if $b$ in Eq.~\eqref{eq:He26} equals 0.5, it means that the stellar initial binding energy profile has a slope of $\alpha=0.15$ in the most-bound states. $b=0.75$ and $1$ in Eq.~\eqref{eq:He26} correspond to $\alpha=0.7$ and $1.3$ in Eq.~\eqref{eq:e22}, respectively. 

The slope of the binding energy profile is correlated with the inner slope of the density profile, which can be converted into each other through Eddington Inversion. That is to say, the cosmological hydrodynamical simulation prefers $\alpha \lesssim 1$ and the stellar inner slope $\lesssim - 1$. The stellar inner density should be as steep as or even steeper than that of DM. The slope of the stellar energy profile at the most-bound direction becomes nearly identical to that of DM. This consistency is supported by \citet{Wang2022}, who found that simulated Milky Way dwarf galaxies exhibit stellar and DM density profiles so similar as to be practically indistinguishable.

However, under their fiducial spherical exponential stellar profile case (i.e. $\alpha=\beta=3$ in Eq.~\eqref{eq:star_energy}), as shown in the bottom right panel of Fig.~\ref{fig:comp_e22}, we observe that stellar stripping in the second phase becomes faster than the DM mass loss rate, corresponding to $b=1.65$. 

We emphasise that the \citetalias{Errani2022} model primarily applies to dwarf galaxies with negligible stellar mass fractions. For brighter satellite galaxies where baryons dominate, the \citetalias{Errani2022} framework must account for baryonic gravitational potentials—particularly their influence on central potential wells. \added{This is also crucial for accurately modelling DMDGs, because the \citetalias{Errani2022} model framework represents stars as massless particles.} Our simulations and models suggest that in tidally stripped bright satellites, stellar particles \added{can} dominate the satellite's mass budget. Consequently, regardless of the stellar density profile, the stellar mass loss rate precisely traces the satellite's total mass loss rate at this stage. This regime warrants further exploration in subsequent studies.

\section{Goodness of fit}

\begin{figure}
    \centering
    \begin{subfigure}[b]{0.45\textwidth}
        \includegraphics[width=\linewidth]{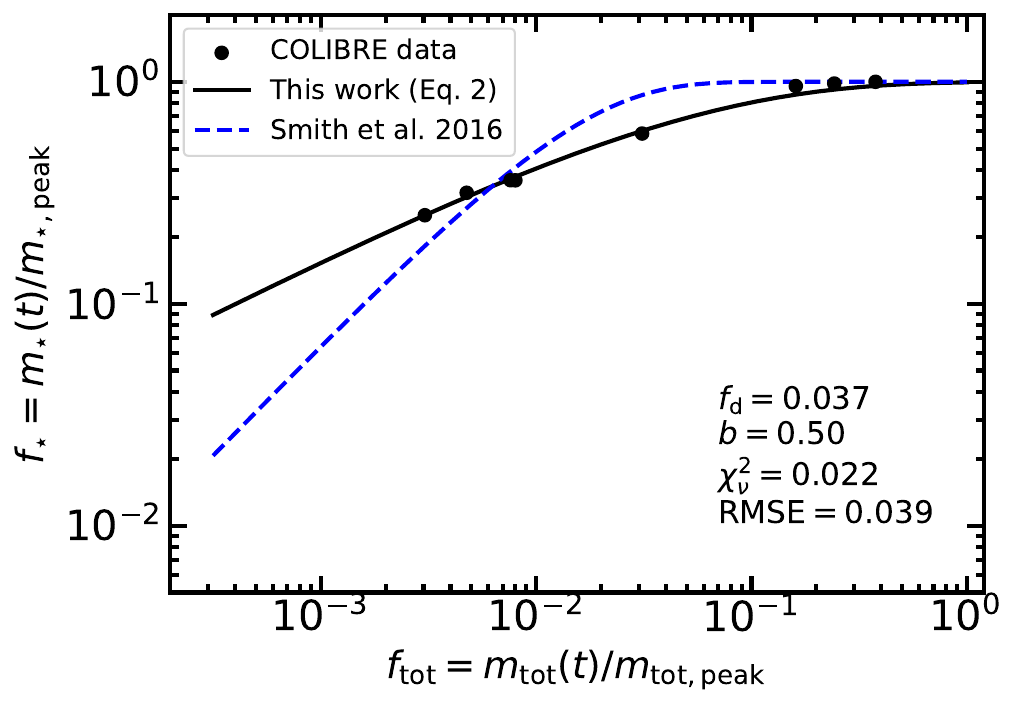}
    \end{subfigure}
    \vspace{0.1cm}
    \begin{subfigure}[b]{0.45\textwidth}
        \includegraphics[width=\linewidth]{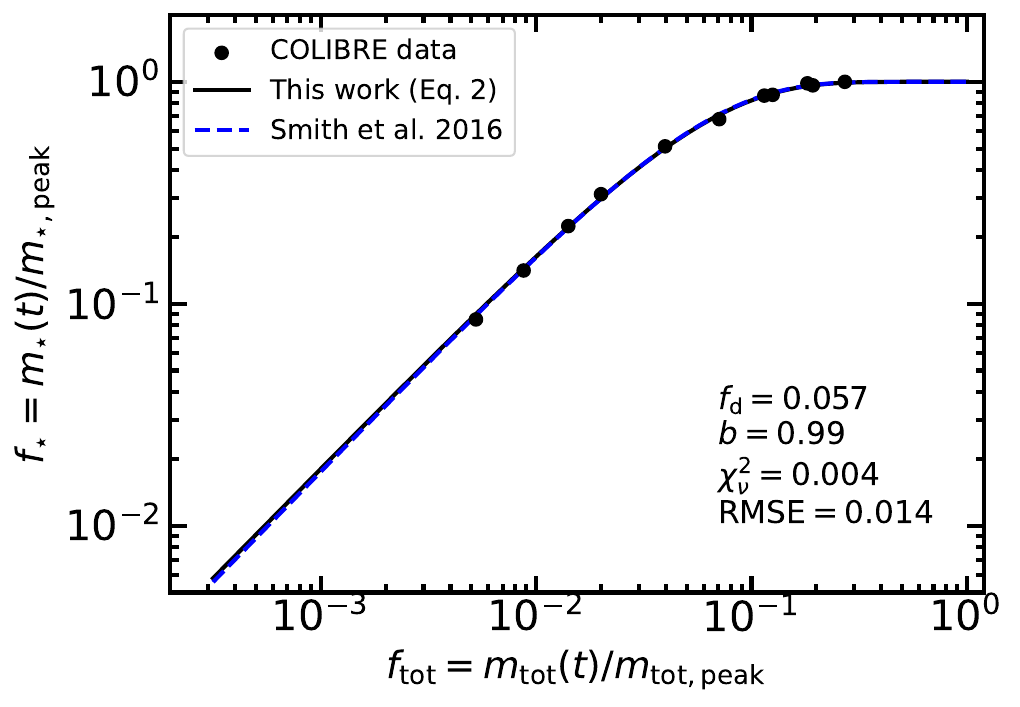}
    \end{subfigure}
    \vspace{0.1cm}
    \begin{subfigure}[b]{0.45\textwidth}
        \includegraphics[width=\linewidth]{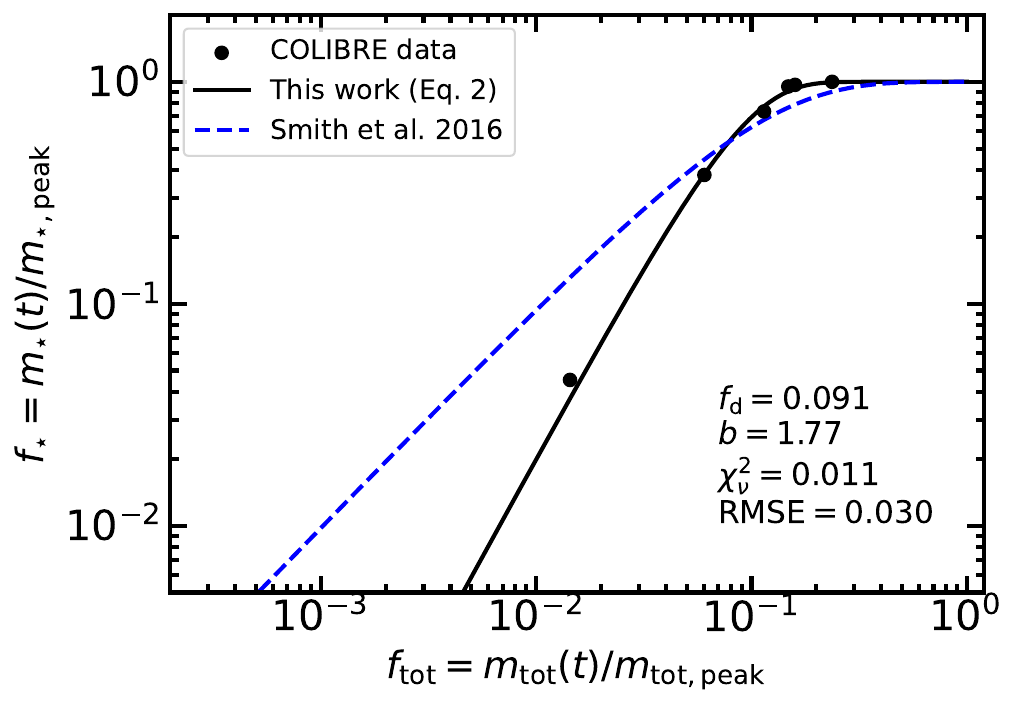}
    \end{subfigure}
    \caption{Representative examples of satellite galaxy stellar mass tidal evolution tracks from L200m6 simulations. The black circles represent the simulation data, showing the remaining stellar mass fraction ($f_{*}$) as a function of the remaining bound mass fraction ($f_{\rm tot}$). The solid black and dashed blue curves denote the best-fit results using our model (Eq.~\eqref{eq:He26}) and \citetalias{Smith2016} model (Eq.~\eqref{eq:smith16}), respectively. The best-fitting parameters ($f_{\rm d}$ and $b$), along with the weighted reduced chi-squared ($\chi_\nu^2$) and root-mean-square error (RMSE) for Eq.~\eqref{eq:He26}, are annotated in each panel.}
    \label{fig:fitting_example1}
\end{figure}

\begin{figure}
    \centering
    \includegraphics[width=\linewidth]{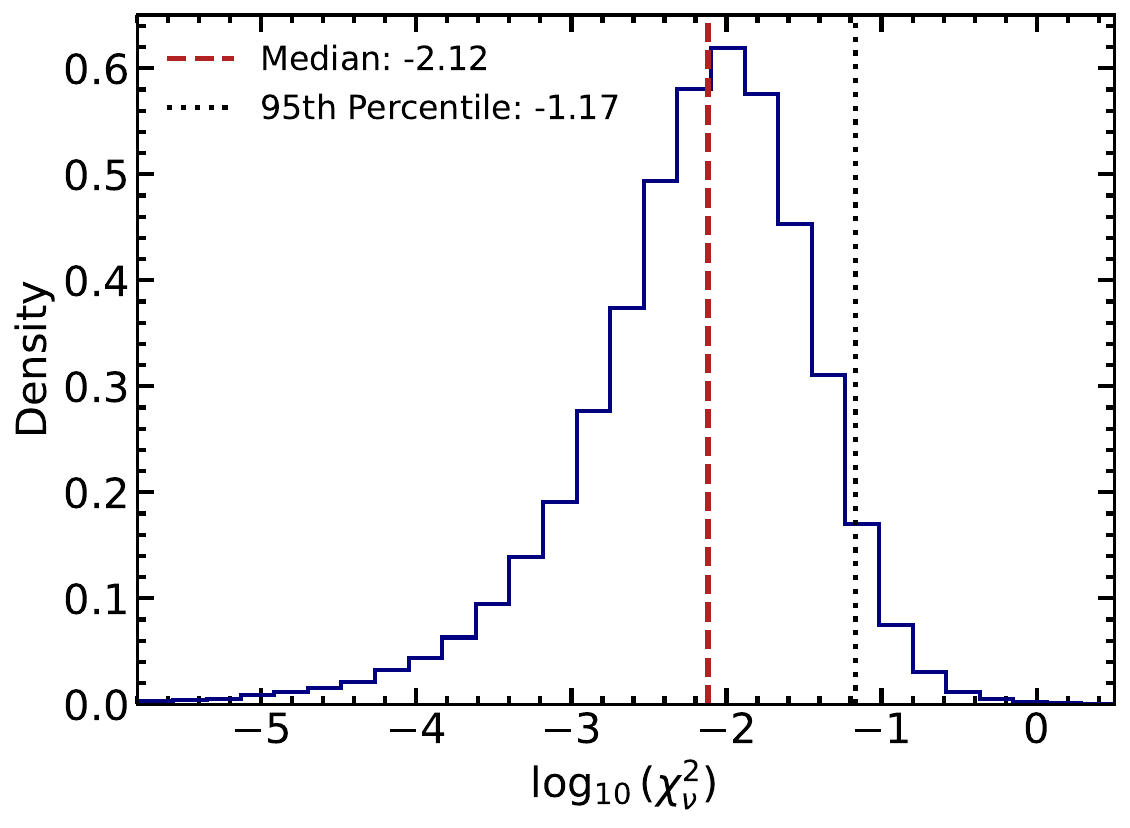}
    \caption{The probability density distribution of the weighted reduced chi-squared ($\log_{10}\chi_\nu^2$) for the fits of Eq.~\eqref{eq:He26} to the satellite population. To ensure the model accurately captures the late stages of tidal disruption, a weighting scheme of $w \propto 1/\sqrt{f_{\rm tot}}$ was applied during the fitting process. The vertical red dashed line indicates the median of the distribution at $\log_{10}(\chi_\nu^2) = -2.12$, showing that the vast majority of tracks are excellently described by the analytical formula. The vertical black dotted line marks the 95th percentile at $\log_{10}(\chi_\nu^2) = -1.17$.} 
    \label{fig:red_chi2}
\end{figure}

\begin{figure}
    \centering
    \begin{subfigure}[b]{0.45\textwidth}
        \includegraphics[width=\linewidth]{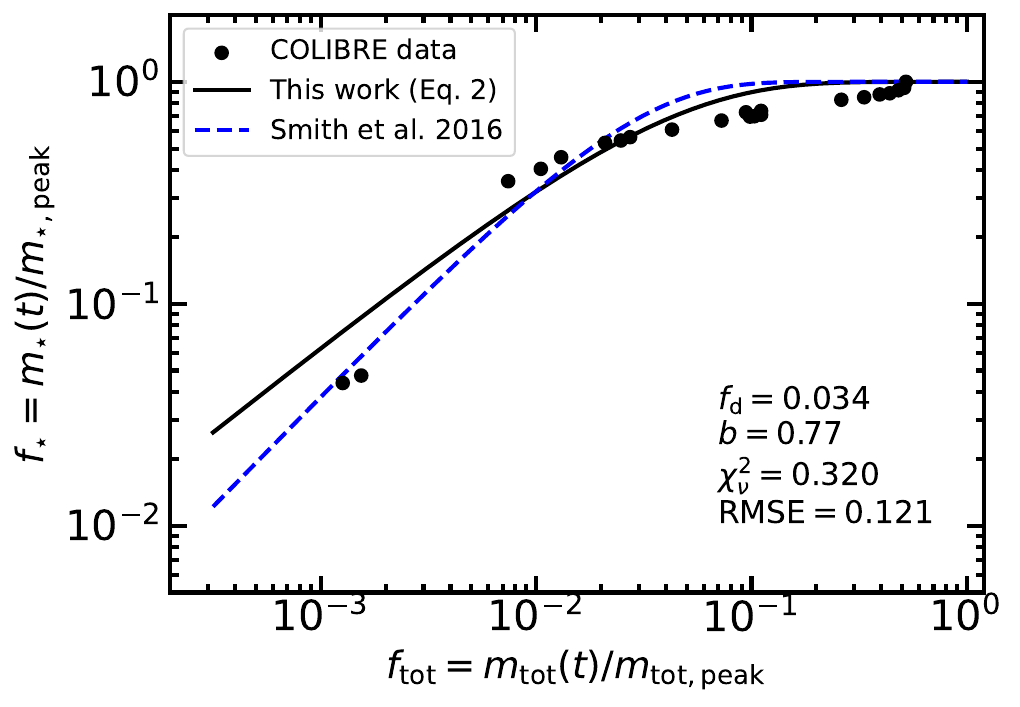}
    \end{subfigure}
    \vspace{0.1cm}
    \begin{subfigure}[b]{0.45\textwidth}
        \includegraphics[width=\linewidth]{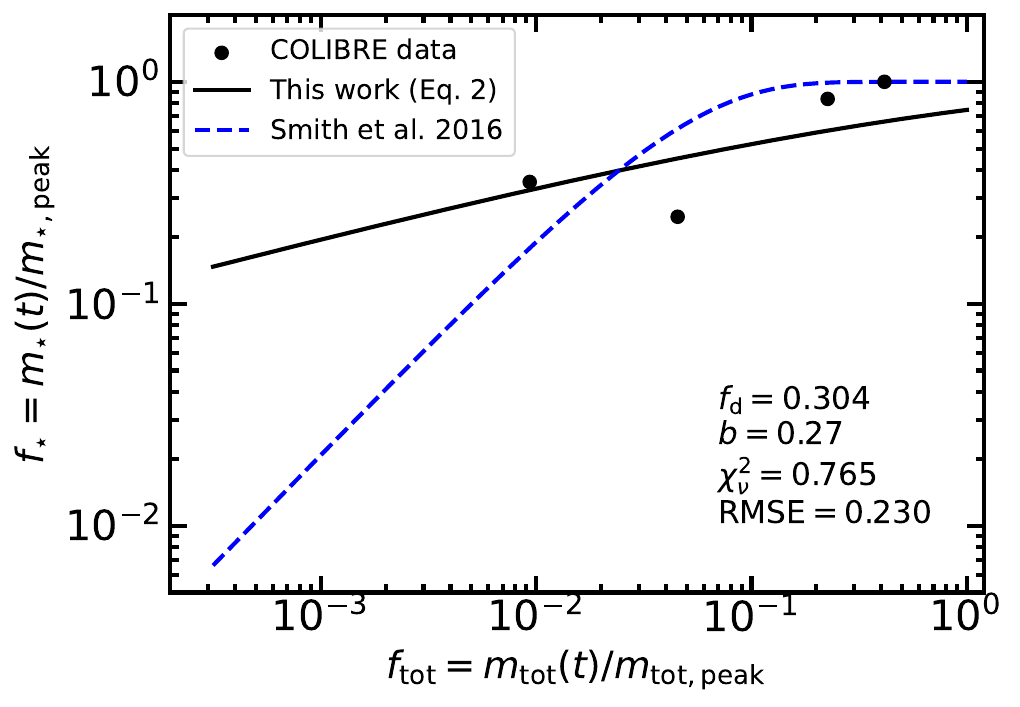}
    \end{subfigure}
    \caption{ Examples of outlier tidal tracks, presented in the same format as Fig.~\ref{fig:fitting_example1}. These subhaloes belong to the high-$\chi_\nu^2$ tail in Fig.~\ref{fig:red_chi2}. For these rare cases, both our model (Eq.~\eqref{eq:He26}) and Eq.~\eqref{eq:smith16} struggle to reproduce the simulated data precisely.} 
    \label{fig:fitting_example2}
\end{figure}

In this section, we provide further details regarding the fitting methodology applied to the satellite tidal tracks and evaluate the goodness-of-fit. 
To determine the best-fitting parameters $f_{\rm d}$ and $b$ for each subhalo, we utilize a non-linear least-squares minimization algorithm. The fitting is performed in the $f_{\rm tot}$ vs. $f_{*}$ plane, utilizing the available snapshot data as discussed in Section~\ref{sec:tidaltrack} and Table~\ref {tab:track_parameters}.

A notable aspect of tracking stellar mass evolution is accurately capturing the "heavy stripping" regime, where the subhalo has lost the vast majority of its bound mass ($f_{\rm tot} \ll 1$) and the stellar mass begins to decline sharply. Unweighted least-squares fitting tends to be overwhelmingly dominated by the early evolutionary phase (where both $f_{\rm tot}$ and $f_{*}$ are close to unity). To ensure our fits are sensitive to the heavily stripped tail, we apply a weighting scheme to the data points during the minimization process, assigning a weight $w_i \propto 1/\sqrt{f_{\rm tot, i}}$ to each snapshot point $i$. \added{This form is motivated by Poisson statistics: since $f_{\rm tot}$ traces the bound particle number, the associated counting uncertainty scales as $\sigma_{f_{\rm tot}} \propto \sqrt{f_{\rm tot}}$, so that $w_i \propto 1/\sigma_{f_{\rm tot},i}$.} We quantify the goodness-of-fit and the typical fitting errors using two primary metrics: the weighted reduced chi-squared ($\chi_\nu^2$) and the Root Mean Square Error (RMSE), the latter providing an intuitive measure of the absolute fitting error in the same units as the stellar mass fraction.

We present three representative examples of satellite tidal tracks in Fig.~\ref{fig:fitting_example1}. Both models are fitted to the actual simulation points using the identical weighting scheme described above. The middle panel of Fig.~\ref{fig:fitting_example1} illustrates a satellite where the best-fitting exponent is $b = 0.99$. In this scenario, Eq.~\eqref{eq:He26} naturally reduces to the \citetalias{Smith2016} formulation, and the two fitting curves are nearly identical. However, the top and bottom panels display satellites with best-fitting parameters of $b = 0.5$ ($b < 1$) and $b = 1.77$ ($b > 1$), respectively. In these cases, it is evident that without introducing the exponent $b$, the single-parameter model (blue dashed lines) fundamentally fails to capture the distinct curvature of the tidal tracks. 

Fig.~\ref{fig:red_chi2} shows the probability density distribution of the weighted reduced chi-squared, $\log_{10}(\chi_\nu^2)$, for our entire selected sample. The distribution is strongly peaked at low values, with a median of $\log_{10}(\chi_\nu^2) = -2.12$ (corresponding to a typical RMSE of $\sim 2\%$). This indicates that the vast majority of the simulated tidal tracks are exceptionally well-described by Eq.~\eqref{eq:He26}. We find that extreme outliers, defined here as having $\chi_\nu^2 > 0.1$ ($\log_{10}\chi_\nu^2 > -1$), account for only $2.5\%$ of the total fitted sample. To understand the origins of these poor fits, we highlight two typical outlier examples in Fig.~\ref{fig:fitting_example2}.

The top panel of Fig.~\ref{fig:fitting_example2} shows a satellite with a highly complex evolutionary path. Such discontinuous dynamical evolution deviates from the assumed smooth, monotonic nature of the analytical fitting formula. The bottom panel illustrates another common cause for high $\chi_\nu^2$ values: a lack of sufficient temporal sampling. In this example, the track is resolved with only four valid snapshots (the minimum required by our selection criteria). We find that a large fraction of the $2.5\%$ outlier population originates from these poorly sampled tracks, where the sparse data points either fail to provide meaningful constraints or overly penalize the $\chi_\nu^2$ statistic due to minor numerical fluctuations.

Despite these rare outliers, the overall statistical performance confirms that Eq.~\eqref{eq:He26} serves as a robust and highly accurate empirical description of satellite stellar mass evolution.

\section{model validation}\label{sec:ftotbin}
\begin{figure}
    \centering
    \includegraphics[width=0.9\linewidth]{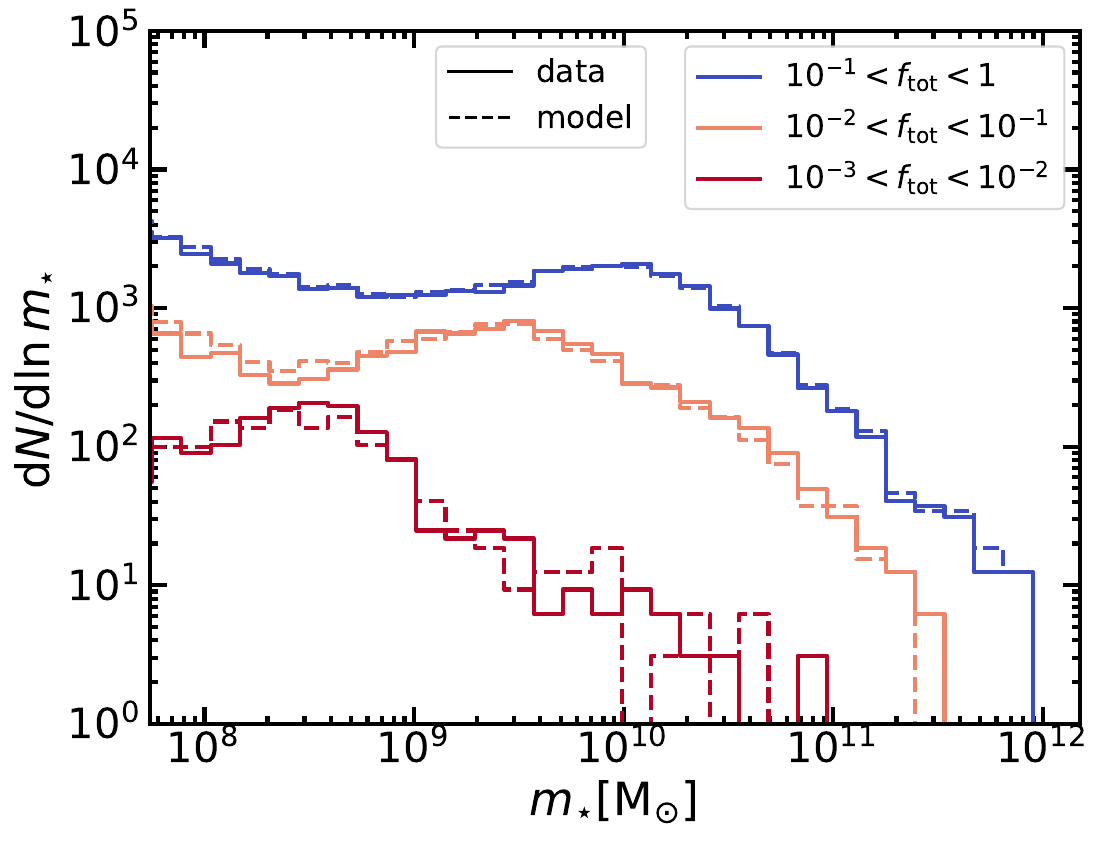}
    \caption{The stellar mass function for resolved satellites shown in Fig.~\ref{fig:ad}, but is split into subsets based on the bound mass fraction $f_{\rm tot}$. The solid lines represent the simulation results, and the dashed lines represent the results predicted by the tidal track model.}
    \label{fig:ftotbin}
\end{figure}

\added{To evaluate the robustness of our tidal track model and justify its extrapolation to the lost population, we perform an internal validation test using the resolved satellite sample. We partition the resolved satellites into three distinct sub-samples based on their remaining bound mass fractions ($f_{\rm tot}$). Fig.~\ref{fig:ftotbin} presents a comparison between the evolved satellite stellar mass functions predicted by our tidal track model and those directly measured from the \colibre\ simulation data across these different $f_{\rm tot}$ bins. In the weakly stripped regime ($10^{-1} < f_{\rm tot} < 1$), the model perfectly reproduces the simulated stellar mass function, establishing a reliable baseline. As we move towards the heavily stripped regime ($10^{-3} < f_{\rm tot} < 10^{-2}$), our model predictions still maintain agreement with the simulation data. The fact that the model's performance does not degrade in the lowest resolved bound mass fraction bin demonstrates the generalizability of our model prescriptions. This successful validation provides a solid physical and statistical foundation, justifying our extrapolation of the model to the extremely stripped, lost population where $f_{\rm tot} \ll 10^{-3}$.}

%


\bsp	
\label{lastpage}
\end{document}